\documentclass{report}

\usepackage{latexsym,amssymb,epsf}
\usepackage{setspace,cite} 
\usepackage{equations,fancyhdr,graphicx}
\usepackage{subfigure}
\usepackage{a4}

\newcommand{\beq}{\begin{equation}}
\newcommand{\eeq}{\end{equation}}
\newcommand{\bea}{\begin{eqnarray}}
\newcommand{\eea}{\end{eqnarray}}

\newcommand{\CB}{{\cal B}}
\newcommand{\CA}{{\cal A}}
\newcommand{\CO}{{\cal O}}

\newcommand{\half}{{1\over 2}}
\newcommand{\threehalves}{{3\over 2}}

\newcommand{\pa}{\partial}

\newcommand{\obar}{\overline}

\begin{document}
 
\newpage
\pagenumbering{arabic}
\setcounter{page}{1} \pagestyle{fancy} 
\renewcommand{\chaptermark}[1]{\markboth{\chaptername%
\ \thechapter:\,\ #1}{}}
\renewcommand{\sectionmark}[1]{\markright{\thesection\,\ #1}}


\newpage
\thispagestyle{empty}

\pagestyle{empty} 
\begin{center} 
    {\LARGE UNIVERSITY OF OXFORD} \\
\vspace{3cm} 
    {\Huge{\bf Glueballs in Flatland}} \vspace{12pt} \\ 

\vspace{2cm} 
    by \\
\vspace{2cm} 
    {\LARGE Robert W. Johnson} \\
\vspace{3cm} 
    A thesis submitted for the degree of \\
    Doctor of Philosophy \\
    Department of Theoretical Physics \\
    Trinity 2001 \\
\end{center}

\singlespacing

\begin{abstract}

The pure gauge theory in 2+1 dimensions is explored, through both a
phenomenological model and a lattice calculation.  The Isgur-Paton
model is extended to include a curvature term and various mixing
mechanisms.  The method of inferential statistics is used to extract
the parameters of best fit and to compare the likelihoods of the
various models when compared to existing lattice data.  The conventional 
assignment of spin 0 to the pseudoscalar state is called into question by
the proximity of a spin 4 state in the model, which motivates calculating
the mass of the spin 4 state on the lattice.
Novel lattice operators are constructed from a matrix of effective
Greens functions which attempt to overcome the lattice rotational
ambiguities.  Correlation functions are presented for the channels
with even J, and effective masses extracted.  The resulting masses
compare well with the extended Isgur-Paton model.

\vspace*{5cm}


\end{abstract}

\doublespacing

\pagenumbering{roman}
\setcounter{page}{1} \pagestyle{plain}

\tableofcontents

\listoffigures
\listoftables

\chapter*{Acknowledgments} 
\addcontentsline{toc}{chapter} 
		 {\protect\numberline{Acknowledgments\hspace{-96pt}}} 
 
Many thanks to Jack Paton and Brian Buck for illuminating and educational conversations.  Gratefully I appreciate funding from the Rhodes Trust.  Thanks to my friends and family for their support through these long, lean years.  Most of all, I thank my supervisor, Mike Teper, without whose guidance and perseverance in the face of my musical distractions this thesis could not exist.

\pagestyle{fancy}

\newpage

\addtolength{\headheight}{3pt}    
\fancyhead{}
\fancyhead[LE]{\sl\leftmark}
\fancyhead[LO,RE]{\rm\thepage}
\fancyhead[RO]{\sl\rightmark}
\fancyfoot[C,L,E]{}

\pagenumbering{arabic}

\chapter{Introduction} \label{chap:intro}

Among the many phenomena currently being explored in the world's particle accelerators, few are as interesting as the search for the glueball.  Several candidates have presented themselves~\cite{Close:2001ga} at around 1.6 GeV, but mixing with nearby quarkonia states~\cite{Lee:1999kv} or other non-perturbative effects has kept the situation as clear as mud.  In order to identify the glueball from the potential candidates, a clear understanding of QCD's nonperturbative spectrum is required.  Unfortunately, solving QCD completely is still a long ways off.  Much progress has been made, however, using various techniques.  Among the most promising is the study of nonperturbative QCD using lattice regularization~\cite{Creutz:1983}.  Once theory has made an unequivocal prediction for the QCD spectrum, experimentalists will be able to verify the existence of the glueball.

While some progress has been made in studying dynamical quarks on the lattice~\cite{Allton:2001sk}, the pure gauge theory without quarks is much better under control and sufficiently interesting to warrant study.  Without quarks around, the only bound states that exist must be purely gluonic:  glueballs~\cite{Isgur:1985bm}, gluelumps~\cite{Karl:1999sz,Michael:1998sm}, and perhaps even more exotic states~\cite{Paria:1995zt}.  Lattice methods have produced remarkably accurate estimates of the glueball spectrum~\cite{Teper:1999te,Michael:2001qz}, as well as estimates of the color-electric flux profile~\cite{Wong:1994kz,Green:1997be}.  Unfortunately, the best estimate of the lightest scalar glueball's mass~\cite{Teper:1999te}, about 1.6 GeV, is smack between the two most promising experimental candidates~\cite{Close:2001ga}.  Thus a more complete understanding of gluonic dynamics needs to be developed, so that we may build a model for glueballs which makes experimentally testable predictions.

Several models for glueballs currently exist, each with its own strengths and weaknesses.  The bag model~\cite{Chodos:1974} receives much attention in the literature~\cite{Karl:1999wq,Karl:1999sz,Juge:1998nd}, but unfortunately is ignorant of the consequences of the no-hair theorem, namely that a spherical shell of nonvanishing flux cannot exist~\cite{Robson:1980}.  Constituent gluon models also exist~\cite{Close:1988}, but their development is hampered by the intrinsically nonperturbative aspect of QCD.  The most promising candidate for a model of glueballs is the flux tube model, first applied to QCD by Isgur and Paton~\cite{Isgur:1985bm}.  Chapter~\ref{chap:ipmodel} reviews the traditional Isgur-Paton model as it applies to D=2+1 QCD.  In this thesis, we work with one less spatial dimension for simplicity -- there is still plenty of structure to be quite interesting.  As is often the case, a system reduced to 2+1 dimensions behaves as though it were the cross section of the full 3+1 system~\cite{Teper:1997tq}, and so understanding glueballs in 2+1 dimensions should impart much insight into the full theory.

Proceeding to Chapter~\ref{chap:extensions}, we modify the traditional Isgur-Paton Hamiltonian to include a contribution from a curvature term.  This term would arise from a finite coefficient of elasticity for the flux tube, as is the case with flux tubes in a type-II superconductor~\cite{'tHooft:1981ht,Mandelstam:1973,Baker:2000,Lampert:1999ws}.  The traditional Isgur-Paton model does not include states with negative charge conjugation eigenvalue in 2+1 dimensions, and so in Chapter~\ref{chap:mixing} we explore several mechanisms which give rise to mixing between oppositely orientated flux tubes to produce both the positive and negative charge conjugate sectors of the spectrum.  Chapter~\ref{chap:analysis1} presents the results of the calculations, and Chapter~\ref{chap:N2andchi2} explores the behavior of the parameters and the merit function for the various models.

One feature of the flux tube model is the presence of a divergent term in the potential for small radius~\cite{Shevchenko:2000du}.  Taking inspiration from a recent study of flux rings in the dual Ginzburg-Landau model~\cite{Koma:1999sm}, Chapter~\ref{chap:analysis2} explores how including an effective string tension which goes to zero at small radius modifies the spectrum.

When comparing the Isgur-Paton model's spectrum to existing lattice data~\cite{Teper:1999te}, some surprising features are observed.  The model seems to imply that the conventional assignment of spin to certain lattice operators may be in error.  Traditional lattice operators belong to one of the symmetry channels of the cubic lattice, which may be compatible with several values of spin {\it{modulo}} 4.  Usually the lowest value of spin is assigned to the lightest state in the channel, but the presence of spin 4 states in the extended Isgur-Paton model at the mass of the pseudoscalar ($J^{PC}=0^{-+}$) from the lattice casts doubt on this particular assignment, and so in Chapter~\ref{chap:lattice} we begin developing an alternative construction of lattice operators that should be free of spin ambiguities.  Chapter~\ref{chap:lattcalc} details the calculation and presents a new algorithm for inverting an Hermitian matrix with the specific band structure encountered here.  The results using this new method are presented in Chapter~\ref{chap:lattres}.

We conclude this thesis with Chapter~\ref{chap:conclusion}, reviewing the modifications to the traditional Isgur-Paton model and their implications for the glueball spectrum and comparing the results with the new lattice operators.  In Appendix~\ref{chap:bayes} we present an overview of inferential statistics as used in this work.  Appendix~\ref{chap:IPreview} reviews the details of the Isgur-Paton model and demonstrates how states are identified within its context.  Finally, representative copies of the programs used may be found in Appendix~\ref{chap:programs}.

For future reference, here we present the lattice data from~\cite{Teper:1999te} in Table~\ref{table:lattmass}.

\begin{table}[h]
\begin{center}
\begin{tabular}{|l|l|l|l|l|l|}\hline
\multicolumn{6}{|c|}{$m_G/\surd\sigma$} \\ \hline
state & SU(2) & SU(3) & SU(4) & SU(5) & SU($\infty$) \\ \hline
$0^{++}$         & 4.718(43) & 4.329(41) & 4.236(50) & 4.184(55) & 4.065(55) \\
$0^{++\ast}$     & 6.83(10)  & 6.52(9)  & 6.38(13) & 6.20(13)  & 6.18(13) \\
$0^{++\ast\ast}$ & 8.15(15)  & 8.23(17) & 8.05(22) & 7.85(22)  & 7.99(22) \\ 
$0^{-+}$         & 9.95(32)  & 9.30(25) & 9.31(28) & 9.19(29) & 9.02(30) \\
$2^{++}$         & 7.82(14)  & 7.13(12) & 7.15(13) & 7.19(20)  & 6.88(16) \\
$2^{++\ast}$     &           &          & 8.51(20) & 8.59(18)  &          \\
$2^{-+}$         & 7.86(14)  & 7.36(11) & 6.86(18) & 7.18(16) & 6.89(21)  \\
$2^{-+\ast}$     &           & 8.80(20) & 8.75(28) & 8.67(24) & 8.62(38) \\
$1^{++}$         & 10.42(34) & 10.22(24) & 9.91(36) & 10.26(50) & 9.98(25) \\
$1^{-+}$         & 11.13(42) & 10.19(27) & 10.85(55)& 10.28(34) & 10.06(40) \\ \hline
$0^{--}$         &           & 6.48(9)  & 6.271(95) & 6.03(18) & 5.91(25) \\ 
$0^{--\ast}$     &           & 8.15(16) & 7.86(20) & 7.87(25) & 7.63(37) \\ 
$0^{--\ast\ast}$ &           & 9.81(26) & 9.21(30) & 9.51(41) & 8.96(65)\\
$0^{+-}$         &           & 10.52(28) & 10.35(50) & 9.43(75) & 9.47(116) \\
$2^{--}$         &           & 8.75(17) & 8.22(32) & 8.24(21)  & 7.89(35) \\
$2^{--\ast}$     &           & 10.31(27) & 9.91(41) & 9.79(45) & 9.46(66) \\
$2^{+-}$         &           & 8.38(21)  & 8.33(25) & 8.02(40)  & 8.04(50)\\
$2^{+-\ast}$     &           & 10.51(30) & 10.64(60) & 9.97(55) & 9.97(91) \\
$1^{--}$         &           & 9.86(23)  & 9.50(35) & 9.65(40) & 9.36(60) \\
$1^{+-}$         &           & 10.41(36) & 9.70(45) & 9.93(44) & 9.43(75) \\   \hline
\end{tabular}
\caption{Glueball masses from the lattice in units of the string tension.}
\label{table:lattmass}
\end{center}
\end{table}

\chapter{The Isgur-Paton Model of Glueballs} \label{chap:ipmodel}

The traditional Isgur-Paton model stems from the strong coupling limit
of (lattice) QCD~\cite{Isgur:1985bm}.  In this scenario, the color-electric flux between
two point sources does not spread uniformly over space but is
restricted to the region of a flux tube between the sources.  As
remarked earlier, such a flux tube picture also comes about from the
dual Meissner effect in a Ginzburg-Landau type superconductor~\cite{Lampert:1999ws,'tHooft:1981ht,Mandelstam:1973}.  Much numerical evidence~\cite{Trottier:1993jv,Green:1997be,Michael:1998sm} points to the existence of
this flux tube, even for moderate values of the coupling.  The flux
tube is quantized between the static quarks upto some small cutoff
using an adiabatic Born-Oppenheimer type approximation.  In the
limit of infinite quark mass, the model provides a good reckoning for
mesons~\cite{Isgur:1985bm,Michael:1998sm}.

Gauge invariant bound states also exist in the gauge theory without
constituent quarks.  The simplest topologically is a single loop of
color flux.  If the $q$ and $\bar q$ of the flux tube meson are
brought sufficiently close as to annihilate, the remaining loop of
flux might remain as a purely gluonic bound state.  This closed flux
tube is going to have some finite width on the order of
$\sqrt{\sigma}$, where $\sigma$ is the string tension in units of
energy/length.  Numerical estimates of this width~\cite{Wong:1994kz} and the
distribution of the chromoelectric flux within~\cite{Trottier:1993jv} exist and agree
with the rough estimate.  To simplify matters, though, we will suppose
the flux tube to be described by a loop of string of vanishing thickness with
coordinates given in polar form as
\beq
\vec r = \vec r (r, \vec \Phi)
\eeq
where $\vec \Phi$ represents the angular variable(s) for either 2 or 3
spatial dimensions.  This loop is going to have complicated dynamics
akin to the Nambu-Goto string~\cite{Nambu:1979bd}:
\beq
S_{\mathrm{Nambu-Goto}} \propto \sigma \int d^2\zeta \sqrt{g}
\eeq
But we desire a simpler model with more intuitive dynamics.  Without
the static quarks around to provide a natural adiabatic limit, we are
a little more cavalier in performing the Born-Oppenheimer
approximation.  Nevertheless, we make the approximation, hoping,
despite the evidence of previous calculations~\cite{Moretto:1993zj}, that
the magnitude of the resulting contributions will justify the gross
approximation that is made.  The fluctuations of the string are
divided into an overall radial ``breathing'' mode of low conjugate
momentum and transverse vibrations of higher momentum.  In applying
the semi-classical approach to quantizing this system, we separate off
the presumably ``fast'' transverse vibrations and quantize these as
phonons on the string providing a radial potential
$V_{phonon}(\vec r) = V_{phonon}(r)$ within which the loop
fluctuates in its breathing mode.  (Details of this procedure can be
found in~\cite{Moretto:1993tm}.)  Also arising from the angular quantization of the
loop is the L\"uscher universal correction term
\beq
V_{L\ddot{u}scher} = \frac{\gamma_0}{r}
\eeq
which depends on the number of spatial dimensions D~\cite{Luscher:1980fr}.
\footnote{The reason for the nought on $\gamma_0$ is that
later $\gamma$ will represent a free parameter with a constant
correction of $\gamma_0$}  Evidence for the actual existence of this
contribution (in the case of a straight string within a meson) has
been presented in~\cite{Soloviev:1995}.  When we write down the expressions for the
energy on our way to quantization, we need to include the self-energy
of the string,
\beq
\mu = 2 \pi \sigma r
\eeq which is easily seen to be the product of the string tension and
the total (average) length of the string.

This string of flux is related to the flux-ring solution of the dual Ginzburg-Landau (dGL) theory~\cite{Lampert:1999ws}.  In that theory, the non-perturbative QCD vacuum is represented by a type-II superconductor, and vortex field solutions form around a line on which the monopole field vanishes.  Having no kinetic term, the dGL flux-ring collapses to its center and thus cannot represent a stable bound state.  Introducing a kinetic term for the flux ring meets with varying degrees of success at solving the relativistic Schroedinger equation that results.  Progress has been made recently, and soon perhaps the detailed dynamics of the dGL flux-ring can be matched with the gross features of the flux-tube model~\cite{Koma:1999sm}.

The IP model approaches the kinetic term in a non-relativistic way.  Consequently, we can expect the model's predictions to be off by some amount due to ``relativistic effects.'' This systematic error will be later handled by increasing the errors on the ``experimental'' data points taken from the lattice by 5\% in quadrature with the statistical errors, so that we may say that the model ``fits within 10\%'' of the lattice data.  Details will be provided later of the effect of this fudge factor.

For a classical bosonic string~\cite{Artru:1983} of circular radius $r$ and mass $\mu$, we write its kinetic energy as
\beq KE = P_r^2 / 2 \mu , \eeq
where $\vec{P_r}$ is the momentum in the $\hat{r}$ direction.  Substituting the expression for $\mu$, we have
\beq KE = P_r^2 / 4 \pi \sigma r . \eeq
This operator, as written, will not be Hermitian upon quantization (because it is not symmetric).  Following standard techniques~\cite{Courant:1937}, the independent variable becomes
\beq r = \rho ^{\frac{2}{3}} \eeq
so that, when $P_r \rightarrow i \hbar \frac{\pa}{\pa r}$ and $\hbar$ is set equal to unity,
\bea KE &=& \frac{-1}{4 \pi \sigma} \frac{P_r^2}{r} \\
&=& \frac{-1}{4 \pi \sigma} \frac{P_r}{r^{1/2}} \frac{P_r}{r^{1/2}} \\
&\rightarrow& \frac{-1}{4 \pi \sigma} \frac{\threehalves \rho^{1/2} \frac{\pa}{\pa \rho}}{\rho^{1/3}}\frac{\threehalves \rho^{1/2} \frac{\pa}{\pa \rho}}{\rho^{1/3}} \\
&=& \frac{-1}{4 \pi \sigma} \frac{9}{4} \frac{\pa^2}{\pa \rho^2} , \eea
which is Hermitian.  When normalizing the wavefunction, we must take care to include the appropriate measure induced by the change of variables:
\beq \delta_{gf} = \int d\rho \tilde{g}(\rho) \tilde{f}(\rho) = \threehalves \int r^{1/2} dr g(r) f(r) . \eeq

The quantization of the transverse vibrations into phonons is easily visualized.  Vibrations are classified according to their number of nodes $m$ and their polarization, left-going or right-going.  We write the occupation number of a particular mode as $n_m^\pm \in \bf{Z}^+$.  Each mode contributes a term $\sim m(n_m^+ + n_m^-) /r$ to the potential, so the total contribution is
\beq V_{phonons} = \sum_{m=2}^\infty m(n_m^+ + n_m^-)\frac{1}{r} = M / r , \eeq
where $M$ is the phonon mode number.  This sum starts at $m=2$ because infinitesimal $m=1$ modes are equivalent to translations and rotations, and thus are excluded in the center of mass frame.  The quantization of the string itself produces the universal bosonic correction $\gamma_0 / r$ first identified by L\"uscher~\cite{Luscher:1980fr}, and so, including the contribution from the self-energy, we write the potential as
\beq PE = V(r) = \frac{M + \gamma_0}{r} + 2 \pi \sigma r . \eeq
We can now write down our Schr\"odinger equation (in the $\rho$ variable)
\bea H f &=& (T + V) f \\
&=& \{ \frac{-9}{16 \pi \sigma} \frac{\pa^2}{\pa \rho^2} + V(\rho^{2\over3}) \}f = E_f f \eea
We recognize this equation as a Sturm-Liouville problem with self-adjoint form~\cite{Courant:1937}
\beq \{ \pa_\rho^2 + \lambda_f - \frac{16 \pi \sigma}{9}V(\rho^{2\over3}) \} f = 0 , \eeq
where $\lambda_f = \frac{16 \pi \sigma}{9} E_f$.

The traditional Isgur-Paton model looks on the potential $\sim 1/r$ as unphysical for $r \rightarrow 0$.  The potential $V(r) = (M+\gamma_0)/r + 2\pi\sigma r$ diverges at small r, and indeed we should expect a string model for the flux tube to break down when the radius is on the order of the thickness of the flux tube.  Thus Isgur and Paton multiplied the offending term by a suppression factor
\beq F(r) = (1 - e^{-f \sqrt{\sigma} r}) \eeq
with $f$ a free parameter determined by fitting the model to the available lattice spectrum.  As there are no other adjustable parameters around to influence the spectrum, it was appealing to have something adjustable to minimize $\chi^2$.  However, after studying the behavior of the eigenfunctions numerically and analytically, it seems the eigenfunctions fall off sufficiently fast as $r \rightarrow 0$ for the singularity at $r = 0$ to pose no problem.

There are other ways to handle the suppression factor.  The factor could be applied to the whole potential, for instance.  Doing so is like supposing that the model ceases to exist as $r \rightarrow 0$, or that the wavefunction is suppressed by such factor.  New evidence coming from a study of the dual Ginzburg-Landau flux-ring~\cite{Koma:1999sm} implies that $\sigma \rightarrow 0$ as $r \rightarrow 0$, that is, the string (self-energy) disappears as the potential becomes infinite.  This approach produces some nearly intractable mathematics--some avenues are explored in Chapter~\ref{chap:analysis2}.

\chapter{The Isgur-Paton Model in 2D -- Including Elasticity} \label{chap:extensions}

\section{Canonical variables}

Let's look in detail at the Isgur-Paton model in 2 spatial dimensions.  Viewing the flux tube as a classical string with
\beq r = r(\phi)\; , \; \phi \in [0,2\pi) \eeq
we expand
\beq r = r_0 + \sum_{m=2} (a_m \cos{m\phi} + b_m \sin{m\phi}) \eeq
and quantize with respect to $\phi$ to arrive at the one dimensional problem
\beq T = \frac{P_r^2}{4\pi\sigma r} \; , \; V = F(r)(\frac{M+\gamma_0}{r} + 2\pi\sigma r) . \label{eqn:TV} \eeq
Setting $F(r)$ to unity for now, we examine the kinetic term as we apply the prescription for taking a classical system over to quantum mechanics~\cite{Schiff:1955}.  When $P_r \rightarrow i\hbar\frac{\pa}{\pa r}$, we are faced with a quandary over what to do with the initial $1/r$ coming from the inertial mass.  One approach would be to multiply through by $\mu$ and work with the system
\beq P_r^2 + 2\mu[V - E] = 0 , \eeq
converting the generalized eigenvalue problem to one that is diagonal by means of a Cholesky decomposition~\cite{Press:1992}.  Our approach is to find the canonical variables for the system.  This step actually should be performed at the classical level, before discussion of quantum mechanics enters the picture~\cite{Schiff:1955}, as we want the Hamiltonian to equal the total energy, but we can just as readily address the problem after $P_r \rightarrow i\hbar\frac{\pa}{\pa r}$.  We want to transform the independent variable $r$ into some canonical variable $\rho$ such that the kinetic energy becomes proportional to the square of the canonical momentum, ie:
\beq T(r) \rightarrow T(\rho) \propto \frac{\pa^2}{\pa \rho^2}  . \eeq

Looking at the troublesome kinetic term,
\beq T = \frac{P_r^2}{2\mu} \rightarrow \frac{-1}{4\pi\sigma}\frac{\frac{\pa}{\pa r}\frac{\pa}{\pa r}}{r} \; , \; \mathrm{where} \; \hbar \equiv 1 . \label{eqn:kterm} \eeq
Symmetric forms of this operator are (upto a constant):
\bea T &\sim& r^{-\half} \frac{\pa^2}{\pa r^2} r^{-\half} \\
 &\sim& \frac{\pa}{\pa r} \left( \frac{1}{r} \frac{\pa}{\pa r}\right) . \eea 
Evaluating these two operators leads to an ambiguous term in the potential.
The first operator leads to the equation (for $V=0$)
\bea E &=& r^{-\half} \frac{\pa^2}{\pa r^2} r^{-\half} \\
E r^\half &=& \frac{\pa^2}{\pa r^2} r^{-\half} \\
 &=& \frac{\pa}{\pa r}\left(r^{-\half} \frac{\pa}{\pa r} - \half r ^{-\threehalves} \right) \\
 &=& r^{-\half} \frac{\pa^2}{\pa r^2} - r^{-\threehalves}\frac{\pa}{\pa r} + {3\over 4} r^{-{5\over 2}} \\
E r &=& \frac{\pa^2}{\pa r^2} - \frac{1}{r}\frac{\pa}{\pa r} + {3\over 4} \frac{1}{r^2} \label{Eq:star} \eea
As this operator equation acts on $\psi$, let $\psi = fg$, then
\beq f'' g + 2 f' g' + f g'' - {1\over r} (f' g + f g') + {3\over 4}{1\over r^2} f g = E r f g \eeq
Annihilating terms in $g'$, 
\beq {f'\over f} = {1\over {2r}} \Rightarrow f = r^\half, {f''\over f} = {-1\over 4} {1\over r^2} \eeq
thus
\bea g'' + g \left( {3\over 4}{1\over r^2} - \half {1\over r^2} - {1\over 4}{1\over r^2} \right) &=& E r g \\
g'' &=& E r g \eea
The second operator gives
\bea E &=& \frac{\pa}{\pa r}{1\over r}\frac{\pa}{\pa r} \\
 &=& {1\over r} \frac{\pa^2}{\pa r^2} - {1\over r^2}\frac{\pa}{\pa r} \eea
so
\beq E r = \frac{\pa^2}{\pa r^2} - {1\over r}\frac{\pa}{\pa r} . \eeq
Comparing to Eqn \ref{Eq:star}, we note a difference of ${3\over 4}{1\over r^2}$. Proceeding, we find
\beq g'' - {3\over 4}{1\over r^2} g = E r g , \eeq
and we are still left with a generalized eigenproblem.  This ambiguity arises from the choice of whether to sandwich the 1/r factor between the derivatives or {\it{vice versa}}.

The historical treatment~\cite{Isgur:1985bm} of this term is equivalent to using the first analytical form of the kinetic operator, as no contribution to the potential arises from the change of variables.
Attacking the independent variable directly in Equation~\ref{eqn:kterm} yields
\beq \frac{\pa}{\pa r}{1\over r}\frac{\pa}{\pa r} \rightarrow \frac{\frac{\pa}{\pa r}}{r^\half}\frac{\frac{\pa}{\pa r}}{r^\half} , \eeq
so that conceptually $P_r = \frac{\frac{\pa}{\pa r}}{r^\half}$.  For our canonical variable $\rho$, $T \propto \frac{\pa^2}{\pa \rho^2}$, so let
\beq \frac{\pa}{\pa r} \rightarrow \frac{\pa \rho}{\pa r} \frac{\pa}{\pa \rho} , \eeq
then
\beq P_r \rightarrow \frac{\frac{\pa \rho}{\pa r} \frac{\pa}{\pa \rho}}{r^\half} , \eeq
so $\frac{\pa \rho}{\pa r} = r^\half$ gives
\beq r^\half dr = d\rho \Rightarrow \rho = {2\over 3}r^{3\over 2} , \eeq
but we want not to change the scale, so we relax our requirement to $\frac{\pa \rho}{\pa r} \propto r^\half$ so that
\beq \rho = r^{3\over 2}, \rho(r=1) = 1 . \eeq
Then
\bea T &\rightarrow& \frac{-1}{4\pi\sigma}\left( {3\over 2} \right)^2 \frac{\pa^2}{\pa \rho^2} \\
 &=& \frac{-9}{16\pi\sigma}\frac{\pa^2}{\pa \rho^2} \eea
and our Hamiltonian becomes
\beq \frac{-9}{16\pi\sigma}\frac{\pa^2}{\pa \rho^2} + V(\rho(r)) = E . \eeq
Absorbing the initial constant into the eigenvalue and potential,
\beq \left[ \frac{\pa^2}{\pa \rho^2} + \lambda - \tilde{V} \right] \tilde{f} = 0 \eeq
where $\rho$ goes from 0 to $\infty$.  Dropping the tildes, we follow~\cite{Miller:1982mj} and solve by first letting $t \equiv {1\over {1+\rho}}$ on the interval $[0,1]$ so that
\bea \frac{\pa t}{\pa \rho} = -t^2 &,& \frac{\pa^2 t}{\pa \rho^2} = 2t^3 \\
\frac{\pa^2}{\pa \rho^2} \rightarrow t^4 \frac{\pa^2}{\pa t^2} + 2t^3 \frac{\pa}{\pa t} \eea
and then $\Phi(t) \equiv t f$
gives
\bea \frac{\pa^2}{\pa \rho^2} f &\rightarrow& \left( t^4 \frac{\pa^2}{\pa t^2} + 2t^3 \frac{\pa}{\pa t} \right) {\Phi\over t} \\
 &=& t^3  \frac{\pa^2}{\pa t^2}\Phi . \eea
Our Hamiltonian goes to
\beq \frac{\pa^2}{\pa t^2}\Phi + \lambda {\Phi\over t^4} - V {\Phi\over t^4} = 0 . \eeq
Taking $t$ now to be discrete
\beq t_j = j h, h = {1\over {n+1}} \; , \; j \in [0,n+1] \eeq
and $\Phi(t) \rightarrow {\Phi_j \equiv \Phi(t_j)}$, with the boundary conditions
\beq \Phi(0) = \Phi(1) = 0 \Rightarrow \Phi_0 = \Phi_{n+1} = 0 . \eeq
The second order approximation
\beq \frac{\pa^2}{\pa t^2}\Phi_j \approx {1\over {12 h^2}} \left( -\Phi_{j-2} + 16 \Phi_{j-1} - 30 \Phi_j + 16 \Phi_{j+1} - \Phi_{j+2} \right) \eeq
gives us
\beq S_b \mathbf{\Phi} + {12\over {j^4 h^2}} \left[ V - \lambda \right] \mathbf{\Phi} = S \mathbf{\Phi} = 0 . \eeq
The boundary conditions on $\Phi'$ show up on the extreme of the diagonal of the band matrix $S_b$.  As we are in 2 dimensions, $\Phi_{n+2} = -\Phi_n$ and $\Phi_{-1} = \Phi_1$.  To get the usual form
\beq A \mathbf{x} = \lambda \mathbf{x} , \eeq
use the diagonal matrix $K$ so that
\beq K S K \left( K^{-1} \Phi \right) = 0 \eeq
then
\beq A - \lambda = {{h^2}\over 12} K S K = {{h^2}\over 12} j^4 S_b + V - \lambda \eeq
is the matrix to diagonalize for the eigenstates.  ($K$ is the diagonal matrix $K_{jj} = j^2$.)  Orthogonality is given by
\beq \delta_{gf} = \int d\rho g f \rightarrow (-) \int dt t^{-4} \Psi \Phi \rightarrow \sum {1\over {h^4}} \mathbf{y x} . \eeq

If we include the suppression factor $F(r)$ as in~\cite{Isgur:1985bm}, we can calculate the spectrum as a function of the parameter $f$ in 
\beq F(r) = 1 - e^{-f r},\label{eqn:f} \eeq
where $\sqrt{\sigma} \equiv 1$ sets the scale in the theory.  While the lattice community is moving away from the $\sqrt{\sigma}$ convention when reporting data, we will continue in $\sqrt{\sigma}$ units.  The spectrum shown in Figure~\ref{fig:b0ncdirN2} reproduces that of~\cite{Isgur:1985bm,Moretto:1993tm}.

\section{Elasticity}

Looking now at the potential in detail, let's consider the properties a flux tube should have.  Evidence from the lattice supports the L\"{u}scher term for straight flux tubes~\cite{Soloviev:1995} and from type-II superconductor flux tubes~\cite{Baker:2000} a contribution from the curvature.  Nonrelativistically the curvature term is
\beq V_{curvature} = \gamma / r . \eeq
For a circular loop, we can think of the L\"{u}scher term as a finite additive renormalization to the coefficient of elasticity $\gamma \rightarrow \gamma - {13\over 12}$ as we quantize the angular variable $\phi$.  Incidentally, the string tension also receives a renormalization which we absorb into $\sigma \rightarrow \sigma_{renorm}$.

\begin{figure}[!b]
\centering
\includegraphics[width=\textwidth]{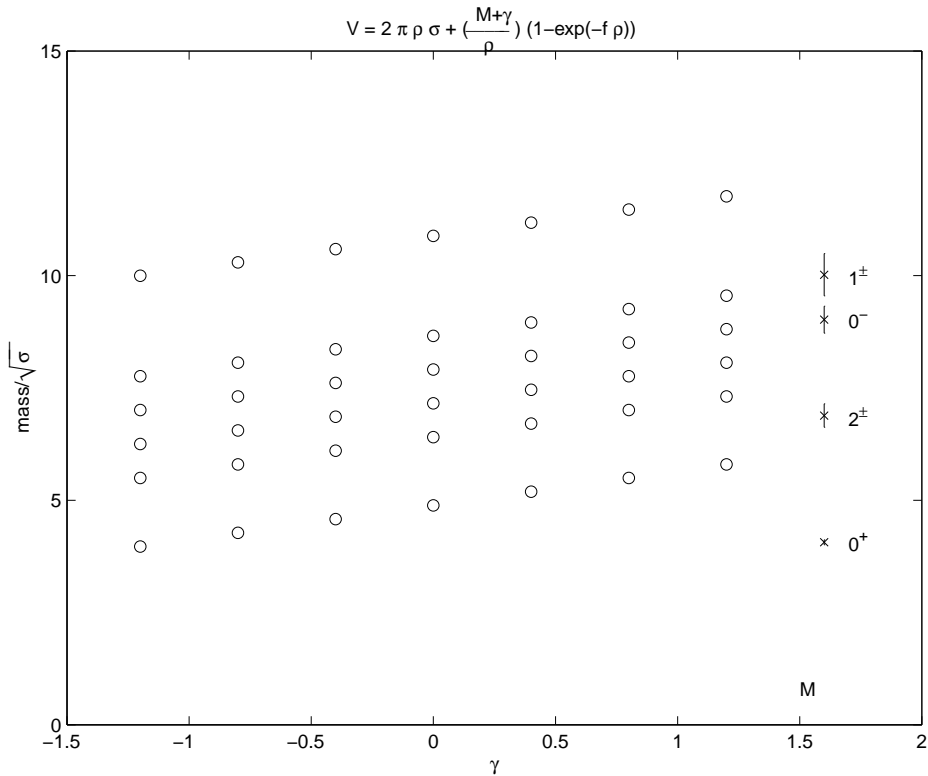}
\caption{The spectrum of the Isgur-Paton model as a function of the parameter $\gamma$.  For comparison, lattice masses are displayed on the right.}
\label{fig:ipvsg}
\end{figure}

Our potential, Equation~\ref{eqn:TV}, is now a function of the two parameters $\gamma$ and the $f$ from Equation~\ref{eqn:f}, as well as the variable $\rho$. We could now do a comparison of the spectrum to lattice data, defining a merit function which we will call $\chi^2$ (see Equation~\ref{eqn:MD}) as a function of the two parameters $f$ and $\gamma$.  At first we use the statistical errors as given by the lattice data, later adjusting the weightings by manipulating the errors.  The behavior of the spectrum as a function of $\gamma$ at $f=1$ is shown in Figure~\ref{fig:ipvsg}.  The lattice masses for SU(2) are given on the right hand side.  Its effect is roughly linear, and the signal for no curvature term would be $\gamma_{fit} = -{13\over 12} \sim -1$.  Preliminary work not reported here~\cite{Johnson:1997ap}, suggests that the suppression $F(r)$ is mostly superfluous, and as we will be adding another parameter later to incorporate splitting between the charge conjugate sectors, we might as well warm up on a 1 parameter minimization.

So far the model can produce states with $C=+$ and exhibits parity doubling for states with $J>0$.  Looking at the lattice masses as a function of $1/N^2$~\cite{Teper:1997tq}, we see that the $0^+$ in N=2 is a smooth continuation of the $0^{++}$ for higher N.  This fact will be important later as we consider mixing mechanisms to produce the $C=-$ sector.  So we select the N=2 data for our initial comparison, which has only $C=+$ and should test the viability of the Isgur-Paton model to explain the lattice spectrum.

\begin{figure}[!t]
\centering
\includegraphics[width=\textwidth]{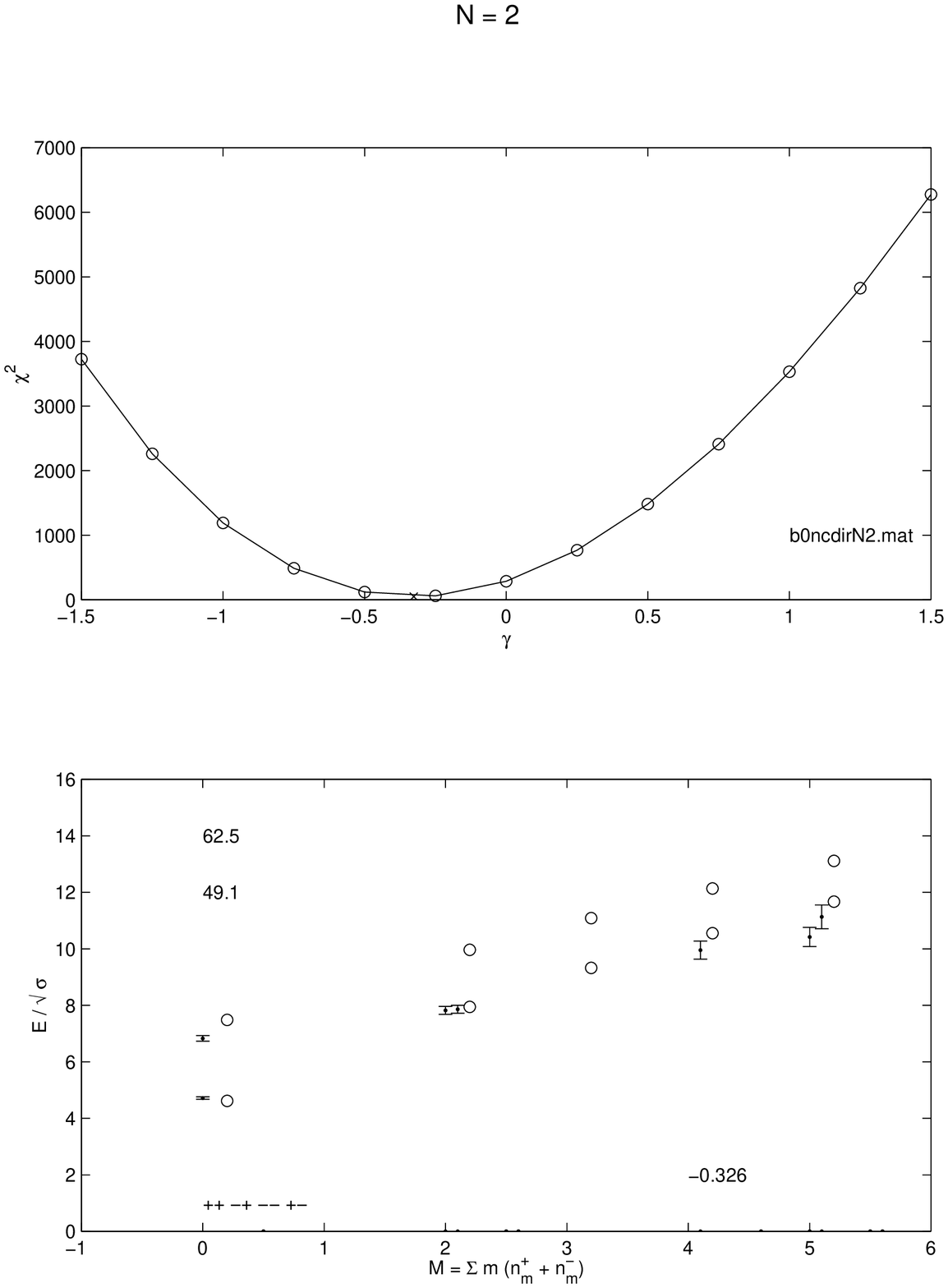}
\caption{The posterior and spectrum for N=2.  The model's predictions are shown by 'o', and the corresponding lattice states by '.' with errorbars.  The x-axis is labelled by the phonon number M.  The relationship between M and the spin J is given in Appendix~\ref{chap:IPreview}, Table~\ref{table:MtoJ}.  See Subsection~\ref{subsec:fig} for more details.}
\label{fig:b0ncdirN2}
\end{figure}

Some ambiguity exists in the assignment of spin to the lattice operators~\cite{Morningstar:1999rf}; the variational method should select out the state with the greatest overlap with the selected symmetry channel and excitation, but convention dictates that the lowest compatible spin be assigned.  For example, on a cubic lattice the continuous rotational symmetry is broken to the discrete $O(2)$ group--rotations are only defined upto a phase of $\pi/2$, so that spin is defined upto mod4.  Practically, what is assigned spin 0 could be spin 4; furthermore, spin 1 could be confused with spin 3, as it is really $|J| \mathrm{mod} 4$ for $J=0,\pm1,\pm2,
\mathrm{etc.}$ that is determined.  Eyeballing the lattice data for N=2 in Table~\ref{table:lattmass} and the spectrum in Figure~\ref{fig:b0ncdirN2}, we see that the lowest $0^+, 2^+, \mathrm{and} 1^+$ are in the right ballpark, but the $0^-$ is nowhere near.  The lightest state in the Isgur-Paton model with $0^-$ has M=8 (see Appendix~\ref{chap:IPreview} for details) and a mass around 15, but there is a state $4^-$ near the lattice mass, suggesting that that operator is coupling to J=4 and not J=0.  To be safe, we will only include the well identified states $0^+$ and $2^{\pm}$ when determining $\chi^2$.

For details on why we can use $\chi^2$ and what it really means, see Appendix~\ref{chap:bayes}.  Succinctly, maximum likelihood with uniform priors and normal errors implies minimizing
\beq \chi^2 \equiv \sum_j \frac{(M_j - D_j)^2}{\sigma_j^2}, \label{eqn:MD} \eeq where the sets ${M_j} \; \mathrm{and} \; {D_j}$ are the model's predictions and the data for comparison, and the ${\sigma_j}$ are the weightings of the data points by the statistical errors.
While our statistical errors are normal, we expect a systematic discrepancy arising from the breakdown of the string picture as $r \rightarrow 0$, so that the smallest state with the tightest statistical weighting, the $0^+$, should be least well represented by the model.  We will later deal with this difficulty in the next section by introducing a fudge parameter $\beta$, but for now we will stick to using the statistical errors and ignore the absolute value of $\chi^2$, which will be rather large.  Our problem reduces to one dimensional minimization.  Following~\cite{Bretthorst:1988,Sivia:1996}, we write the log of the posterior
\bea L &=& prob(\gamma|{D_k}) \label{eqn:logpost} \\
 &=& \mathrm{const} - \half \chi^2 . \eea
Taking a Taylor series about a local minimizer $\gamma_0$, we write
\beq L = L(\gamma_0) + \half \frac{\pa^2 L}{\pa \gamma^2}|_{\gamma_0} (\gamma - \gamma_0)^2 + ... \eeq
Stopping at the quadratic term represents a Gaussian distribution centered at $\gamma_0$ with variance
\beq \sigma_\gamma^2 = - \left[\frac{\pa^2 L}{\pa \gamma^2}|_{\gamma_0} \right]^{-1} = 2\left[ \frac{\pa^2 \chi^2}{\pa \gamma^2}|_{\gamma_0} \right]^{-1} . \eeq
Brute force is a perfectly acceptable method to find the global minimum, so we compute $\chi^2(\gamma)$ as shown in the top half of Figure~\ref{fig:b0ncdirN2}, and start the numerical minimization routine at the minimum of the grid.  To get the variance we use the numerical derivative~\cite{Press:1992}
\beq f' \approx \frac{f(\gamma+h)-f(\gamma-h)}{2h} , \eeq
where the discrete step h is chosen carefully to reduce error.  For $\chi^2$ as in Equation~\ref{eqn:MD}, 
\bea \frac{\pa \chi^2}{\pa \gamma} &=& 2 \sum_k \frac{M_k-D_k}{\sigma_k^2} \frac{\pa M_k}{\pa \gamma} \\
\frac{\pa^2 \chi^2}{\pa \gamma^2} &=& 2 \sum_k \frac{1}{\sigma_k^2}\left[ \left( \frac{\pa M_k}{\pa \gamma}\right)^2 + (M_k - D_k) \frac{\pa^2 M_k}{\pa \gamma^2}\right] . \eea
Near the minimum $\gamma_0$ we can ignore the second term and write
\beq \sigma_\gamma^2 = \left[ \sum_k \left(\frac{\pa M_k}{\pa
\gamma}|_{\gamma_0}\right)^2\right]^{-1} . \eeq

\subsection{Explaing the figures}
\label{subsec:fig}

The results are shown in line 1 of Table~\ref{table:b0ncdir} and the bottom half of Figure~\ref{fig:b0ncdirN2}.  We should explain the layout of this figure as it will be repeated throught this thesis.  The x-axis is given by the total phonon number M, and is related to the spin J.  For M = 1 or 2, the spin is equivalent.  At M = 4 we compare with the lattice state $0^{-+}$, and at M = 5 we compare with the lattice states with J = 1.  Plotted as a '.' with errorbars are the masses from Table~\ref{table:lattmass}.  The ordering of the PC eigenvalues for the lattice states are given in the lower left corner.  As we are currently in SU(2), the $C=-$ sector is not present.  Alongside the corresponding C sector are the predictions of the model, plotted as an 'o'.  The smallest $\chi^2$ from the grid and the $\chi^2$ from the fit are presented in the upper left corner.  In the lower right corner we display the values of the parameters, only $\gamma$ in this instance.  

We easily see that the minimum is found at $\gamma_0 = -.326(141)$.  With only the four lightest states at M = 0 and 2, the heavier states are pure predictions of the model.  The validity of the adiabatic approximation is called into question looking at the excited $0^+$.  For states with a radius on the order of the width of the flux tube, the adiabatic separation of modes is a rather drastic approximation.  That so, the agreement which the model displays is rather impressive, especially out at M=5 with the J=1 from the lattice.  Looking at M=4, we find a state with J=$4^\pm$ quite close to the lattice $0^-$.  This feature persists throughout the calculations.

\clearpage

\section{Adjusting the weightings}

Knowing that we have nonnormal systematic effects, we might be interested to what level the model is capable of fitting the lattice spectrum.  We face the dilemma that the state with the greatest weighting in $\chi^2$, the $0^+$, is so small that a string picture for the state does not really seem applicable.  In order to more evenly spread the weighting of the data points, we add in quadrature to the statistical error a percentage $\beta$ of the mass of the data point $D_k$, ie
\beq \sigma_k^2 \rightarrow \sigma_k^2 + \beta^2 D_k^2 . \eeq
The intended effect is to deweight the lightest (and smallest) states so that they do not unduly influence the model.
So that $\beta = \infty$ is not the trivial solution, we must also add a regularizing term, thus the quantity to minimize becomes
\beq \CA + \CB = \sum_k \left[ \frac{(M_k - D_k)^2}{\sigma_k^2 + \beta^2 D_k^2} + \beta^2 D_k^2 \right] . \eeq
To find the extremal value for $\beta$, we take the derivative
\beq \frac{\pa}{\pa \beta} (\CA + \CB) = 2 \beta \sum_k D_k^2 \left[ 1 - \left( \frac{M_k - D_k}{\sigma_k^2 + \beta^2 D_k^2}\right)^2 \right] , \eeq
and hence
\beq 0 = \sum_k D_k^2 \left[ 1 - \left( \frac{M_k - D_k}{\sigma_k^2 + \beta^2 D_k^2}\right)^2 \right] \label{eqn:forbeta} \eeq
is our equation to solve for $\beta$.  While messy, if we fix the ${M_k}$ to those given by the best fit parameters of Equation~\ref{eqn:logpost}, we can solve the equation numerically for $\beta$.  Technically we should perform a simultaneous minimization over all the parameters~\cite{Bretthorst:1988}, but this approach will produce a satisfactory qualitative answer.  Using the data in Table~\ref{table:b0ncdir} and Equation~\ref{eqn:forbeta}, $\beta \sim .05$, thus we can say the model fits to within 10\% ($\pm \sigma_\mathrm{new}$ errorbars) of the lattice data.

\begin{figure}[!t]
\centering
\includegraphics[width=\textwidth]{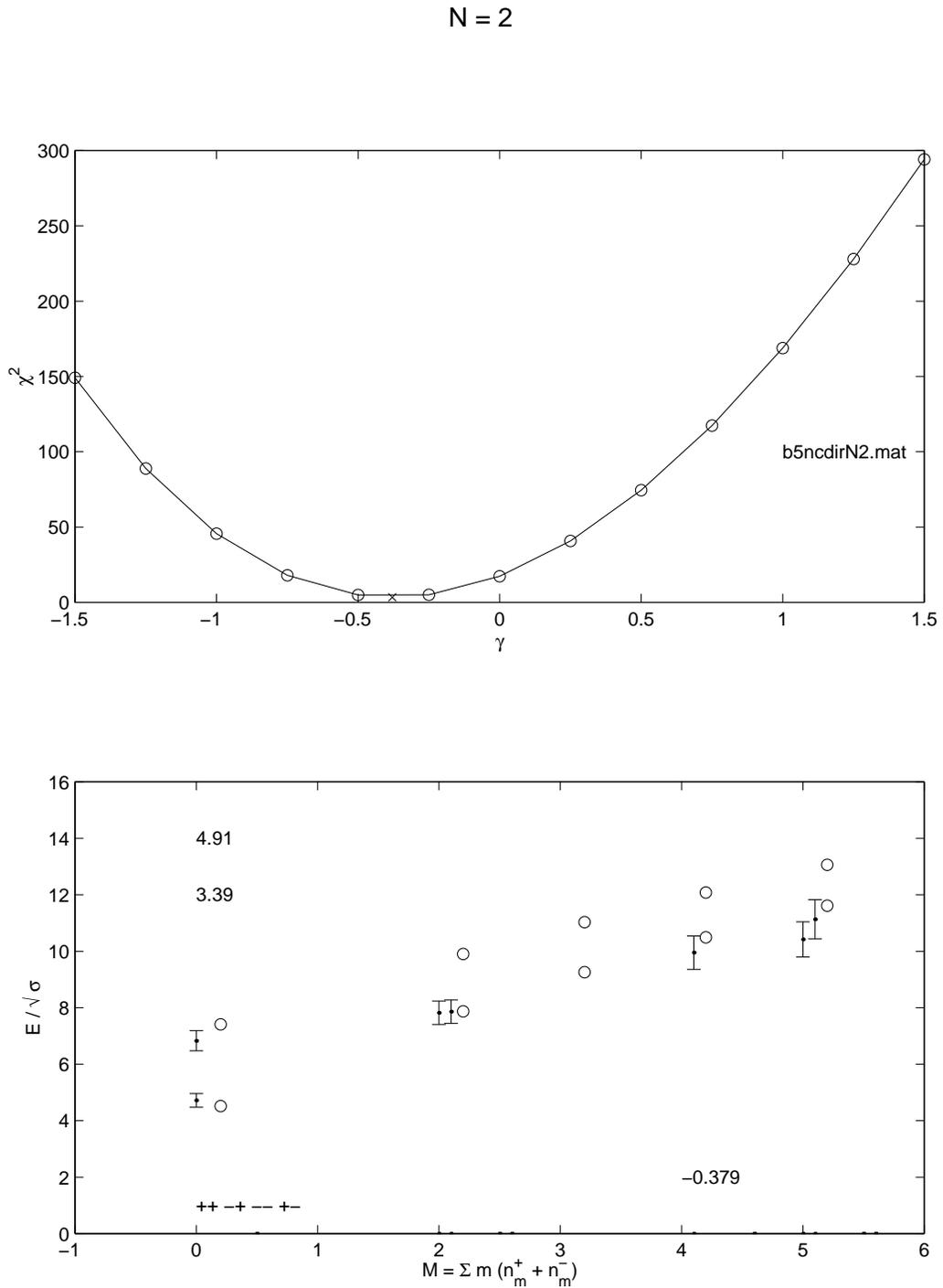}
\caption{The posterior and spectrum for N=2 with adjusted weightings.}
\label{fig:b5ncdirN2}
\end{figure}

These new errorbars correct for the systematic discrepancy we expect at small values of the radius, where the model is less applicable.
Using these new values for the ${\sigma_k}$, we again perform the minimization of $\chi^2$ to find $\gamma_0$.  Figure~\ref{fig:b5ncdirN2} shows the new posterior and the resulting spectrum.  The best fit $\gamma_0$ has not changed much, and the value $\chi^2_\mathrm{fit} = 3.39$ is consistent with 4 data points - 1 parameter.  We see that the errorbars on the $0^{++}$, while still the smallest, are now of the same order of magnitude as the others.  Later we will perform parallel fits at $\beta = 0$ and $\beta = .05$ and compare the results.  Essentially the effect of $\beta$ is to affect the normalization of $\chi^2$ in $\gamma$ space.  Over at M=5, the lattice $1^+$ looks a little low (possibly closer to the $3^+$), and the M=4 is still predicted at the ``$0^-$'' mass.

\chapter{Mixing Mechanisms and the C~=~--~Sector} \label{chap:mixing}

\begin{figure}[t]
\centering
\subfigure[$\psi_L$]{
	\label{fig:psi_LR:L}
	\includegraphics{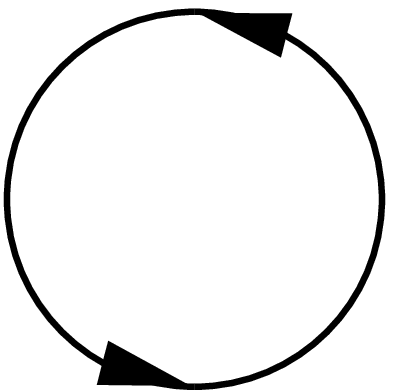}}
\hspace{1in}
\subfigure[$\psi_R$]{
	\label{fig:psi_LR:R}
	\includegraphics{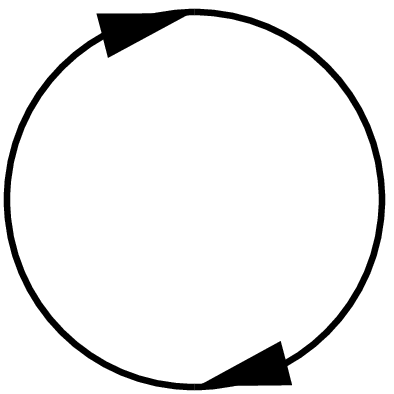}}
\caption{The orientations of the Isgur-Paton flux tube.}
\label{fig:psi_LR}
\end{figure}

We would like to apply the Isgur-Paton model to N$>$2, particularly N=3 and the extrapolation to N=$\infty$.  However, these gauge theories posses states with charge conjugation (C) eigenvalue of minus~\cite{Teper:1999te}.  For N$>$2, the flux tube has an orientation associated with the color electric field, giving rise to two orientations L and R which cannot be connected by a rotation in two dimensions.  Taking $\psi_{L,R}$ to represent these states of opposite orientation (see Figure~\ref{fig:psi_LR}), the linear combination $\pm(\psi_L - \psi_R)$ has $C=-$.  These states are degenerate and unique.  To go from the degenerate basis $[\psi_L,\psi_R]$ to $[\psi_+,\psi_-]$, where the subscript on $\psi$ gives the C eigenvalue, we need a mechanism to transform a $\psi_L$ into a $\psi_R$ (and vice versa).  A transition through an intermediate state with coefficient $\alpha$ immediately suggests itself, which we call indirect mixing.  We also consider a direct mechanism, whereby at small radius the flux tube picture breaks down and a more bag like picture is appropriate, allowing a loop of one orientation to flip to the other, with magnitude $\alpha$.  We write the Hamiltonian for the enlarged basis as
\beq H_{\mathrm{mix}} = H_0 + H' = \left[ \begin{array}{cc} H_1 & 0 \\ 0 & H_2 \end{array}\right] + \alpha \left[ \begin{array}{cc} 0 & 1 \\ 1 & 0 \end{array}\right] \eeq
where $H_{1,2}$ are selected according to the details of the mixing mechanism and operate on the relevant subspace of the full Hilbert space of states.

In the direct mechanism, $H_{1,2}$ are simply $H_{L,R}$ which are really each just copies of the extended Isgur-Paton Hamiltonian $H_{IP}$, where 
\beq H_{IP} = \frac{-9}{16\pi\sigma}\frac{\pa^2}{\pa \rho^2} + V(\rho(r)) , \label{eqn:Hip} \eeq
thus we can write our problem as
\beq H_{\mathrm{mix}} \Psi = \left( \left[\begin{array}{cc} H_{IP} & 0 \\ 0 & H_{IP} \end{array}\right] + \alpha \left[\begin{array}{cc} 0 & 1 \\ 1 & 0 \end{array}\right] \right)\left[\begin{array}{c}\psi_L \\ \psi_R \end{array}\right] , \eeq
For small $\alpha$, we can regard H' as a perturbation~\cite{Schiff:1955}, and solve to first order.  Then 
\beq \lambda = E_{IP} \pm \sqrt{E_{IP}^2 - (E_L E_R - \alpha ^2)} = E_{IP} \pm \alpha . \eeq
Thus the C=$\pm$ sectors are split evenly away from their mean equal to $E_{IP}$.  The problem is that the $0^+$ ought to be a smooth continuation of the $0^{++}$, see~\cite{Teper:1997tq}, not of the average of the $0^{++}$ and $0^{--}$. Also, as written, we have not given $\alpha$ any dependence on r.  Nonetheless, we consider the model viable and calculate its spectrum.

For the two indirect mechanisms considered, we suppose there exists a state of higher mass which interpolates between $\psi_L$ and $\psi_R$.  The Hamiltonian is
\beq H_{\mathrm{mix}} \Psi = \left( \left[\begin{array}{cc} H_{IP} & 0 \\ 0 & H_I \end{array}\right] + \alpha \left[\begin{array}{cc} 0 & 1 \\ 1 & 0 \end{array}\right] \right) \left[\begin{array}{c}\psi_\pm  \\ \psi_I \end{array}\right] \eeq
and the perturbed mass is given by
\beq \lambda = \half (E_{IP} + E_I) \pm \half \sqrt{(E_{IP} - E_I)^2 + 4 \alpha^2} . \eeq
The state $\psi_I$ couples to $\psi_\pm = {1\over \sqrt{2}}(\psi_L \pm \psi_R)$ depending on the details of the model.  The unperturbed eigenstate retains the mass of the Isgur-Paton model, $E_{IP}$.

\begin{figure}[t]
\centering
\subfigure[$\psi_A$]{
	\label{fig:psi_AV:A}
	\includegraphics{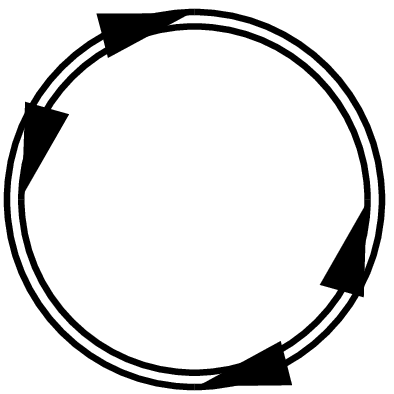}}
\hspace{1in}
\subfigure[$\psi_V$]{
	\label{fig:psi_AV:V}
	\includegraphics{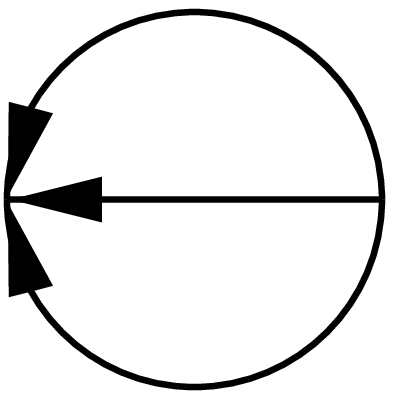}}
\caption{Flux tube states for indirect mixing.}
\label{fig:psi_AV}
\end{figure}

The vertex mixing mechanism posits a state consisting of the original flux tube plus a diameter, see Figure~\ref{fig:psi_AV:V}.  When N=3, this diameter is a single flux tube meeting the circle at two ``baryonic vertices''; for N$>$3 we suppose the diameter to consist of $N-2$ flux tubes together.  The baryonic vertices could contribute to the state's mass directly~\cite{Momen:1997eq,Michael:1998sm}, but we ignore such contribution for simplicity.  To write its Hamiltonian, we include the self energy of the additional flux tubes and a contribution from the angular momentum
\beq H_V = KE + (2\pi + 2(N-2)) \sigma r + \frac{J^2}{2I} . \eeq
Here, J is the spin of the state whose mass we are calculating, and I is the classical moment of intertia of a ring with N-2 colinear diameters.
Specifically,
\beq I = (2\pi + 2(N-2)/3)r^2 . \eeq

The angular momentum also affects the C eigenvalue.  Writing
\beq \psi_J = \sum_{n=0}^3 e^{iJn\theta_0} \psi_{\theta = 0} \; , \mathrm{where} \; \theta_0 = \pi/2 , \eeq
states of even J are invariant under C, while states of odd J transform as $C=-$.  Thus, for example, the $0^{++}$ mixes with $\psi_V$ and is lowered relative to the unmixed $0^{--}$, while the $1^{++}$ is given by the unmixed spectrum and the $1^{+-}$ is lowered.  The resulting structure might explain the anomaly at J=1 in the lattice data:  while data for even J shows the $C=-$ sector to be heavier than the $C=+$, for J=1 the $C=-$ is nearly degenerate with the $C=+$.  As $N\rightarrow \infty$, so does the number of flux tubes in the diameter, and the mass should diverge.  Therefore we can only do the comparison for the three values of N for which we have data.  Later we can extrapolate to $N=\infty$ the values of the parameters if we wish.  Since there is no analog of $\psi_V$ in N=2, the $0^+$ is at the $E_{IP}$ mass, and so is not a smooth continuation from the mixed $0^{++}$ but of the unperturbed $0^{--}$.

For adjoint mixing, the intermediate state is another Isgur-Paton
model flux tube, where the flux is in the adjoint representation, Figure~\ref{fig:psi_AV:A}.  An adjoint flux tube exists from $N=\infty$ down to N=2~\cite{Michael:1998sm,Faber:1998rp,Darmohval:1999ht}, and so the $0^+$ is a linear combination of $\psi_{IP}$ and $\psi_A$, and thus is a smooth continuation of the $0^{++}$.  The Hamiltonian $H_A$ resembles that for $H_{IP}$, Equation~\ref{eqn:Hip}, except for the replacement of $\sigma_F$ by $\sigma_A$
\beq H_{A} = H_{IP}(\sigma_F \rightarrow \sigma_A) . \eeq
To keep our parameter count down, we accept the hypothesis of Casimir scaling~\cite{Shevchenko:2000du,Bali:2000un,Faber:1998rp} to give $\sigma_A$ in terms of $\sigma_F$.  Accordingly, the ratio of the string tension in representation j to the fundamental string tension equals the ratio of the 2nd Casimir operators~\cite{Trottier:1993jv}, thus
\beq {\sigma_A \over \sigma_F} = {C_A\over C_F} = \frac{2N^2}{N^2 -1} \eeq
is the value we take for each N.

At this stage, we are in order to compute the log of the posterior (Appendix~\ref{chap:bayes}) for each model
\bea L &=& \ln{prob(\gamma,\alpha|\{D_k\},\{\sigma_k\},I)} \\
 &\approx& L_0 - \half \chi^2 \eea
in the Gaussian approximation.  Now that we have two parameters, we expand
\beq L = L(\gamma_0,\alpha_0) + \half \left[ \frac{\pa^2 L}{\pa \gamma^2}|_{\gamma_0,\alpha_0}(\gamma - \gamma_0)^2 + \frac{\pa^2 L}{\pa \alpha^2}|_{\gamma_0,\alpha_0}(\alpha - \alpha_0)^2 + 2 \frac{\pa^2 L}{\pa \gamma \pa \alpha}|_{\gamma_0,\alpha_0}(\gamma - \gamma_0)(\alpha - \alpha_0) \right] + \ldots \eeq
Isolating the quadratic part, we write the curvature matrix
\beq [C] = \left[\begin{array}{cc} \frac{\pa^2 L}{\pa \gamma^2} & \frac{\pa^2 L}{\pa \gamma \pa \alpha} \\ \frac{\pa^2 L}{\pa \gamma \pa \alpha} & \frac{\pa^2 L}{\pa \alpha^2}\end{array}\right] \; , \eeq
whence the covariance matrix is
\beq -[C]^{-1} = \left[\begin{array}{cc} \sigma_\gamma^2 & \sigma_{\gamma\alpha}^2 \\
\sigma_{\gamma\alpha}^2 & \sigma_\alpha^2 \end{array}\right] . \eeq
As 
\beq \frac{\pa L}{\pa \gamma} = - \half \frac{\pa \chi^2}{\pa \gamma} , \eeq
then
\beq C_{jk} = - \half \frac{\pa^2 \chi^2}{\pa j \pa k} \; , \mathrm{where } \; j,k \in \{\gamma,\alpha\} , \eeq
and
\beq \frac{\pa^2 \chi^2}{\pa j \pa k} = 2 \sum_i {1\over \sigma_i^2} \left[ \frac{\pa M_i}{\pa j} \frac{\pa M_i}{\pa k} - (D_i - M_i)\frac{\pa^2 M_i}{\pa j \pa k} \right] . \eeq
Near the minimum of $\chi^2$, where the expansion is happening, we can ignore the nonlinear term and write
\beq [C] = - \sum_i {1\over \sigma_i^2} \left[\begin{array}{cc} \left(\frac{\pa M_i}{\pa \gamma}\right)^2 & \frac{\pa M_i}{\pa \gamma}\frac{\pa M_i}{\pa \alpha} \\ \frac{\pa M_i}{\pa \gamma}\frac{\pa M_i}{\pa \alpha} & \left(\frac{\pa M_i}{\pa \alpha}\right)^2 \end{array} \right] \eeq
which is evaluated by
\beq \frac{\pa^2 f}{\pa \gamma \pa \alpha} \approx [f(\gamma+h,\alpha) - f(\gamma,\alpha) - f(\gamma, \alpha+h) + f(\gamma,\alpha)]/h^2 \eeq
for some appropriately chosen h~\cite{Press:1992}.  To find $(\gamma_0,\alpha_0)$, we again apply brute force to a grid in parameter space to map $\chi^2(\gamma,\alpha)$ and begin a simplex~\cite{Press:1992} minimization from the minimum on the grid.

\chapter{Analysis I} \label{chap:analysis1}

\section{Fitting the data}

Let's begin by using the statistical errors, i.e. $\beta=0$:  these fits are driven mostly by the lightest state, the $0^{++}$.  For direct mixing, as the flux tube loses its orientation for N=2, we should start by looking at Figure~\ref{fig:b0ncdirN2} from the preceding chapter.  By using only the $0^+$ and the $2^+$ states when fitting the parameter, states at higher M are predictive.  We see that the $1^-$ agrees within $1\sigma$, but the $1^+$ is as close to a state with M=3 as it is to M=5.  The question of lattice spin assignments is further called into question by the $0^-$.  Here, the model predicts a state with M=J=4 quite close to the lattice $0^-$.  The lightest state with $0^-$ quantum numbers in the Isgur-Paton model has M=8 and has mass around 16 in our units, much too heavy to match the lattice estimate.  The first excitation of $0^+$ is included in the fit.  The model's prediction is too high by about $4\sigma$.  We should remember that the model's adiabatic approximation separating the radial from the phonon modes has little justification and will contribute to the systematic error.

\begin{figure}[!t]
\centering
\includegraphics[width=\textwidth]{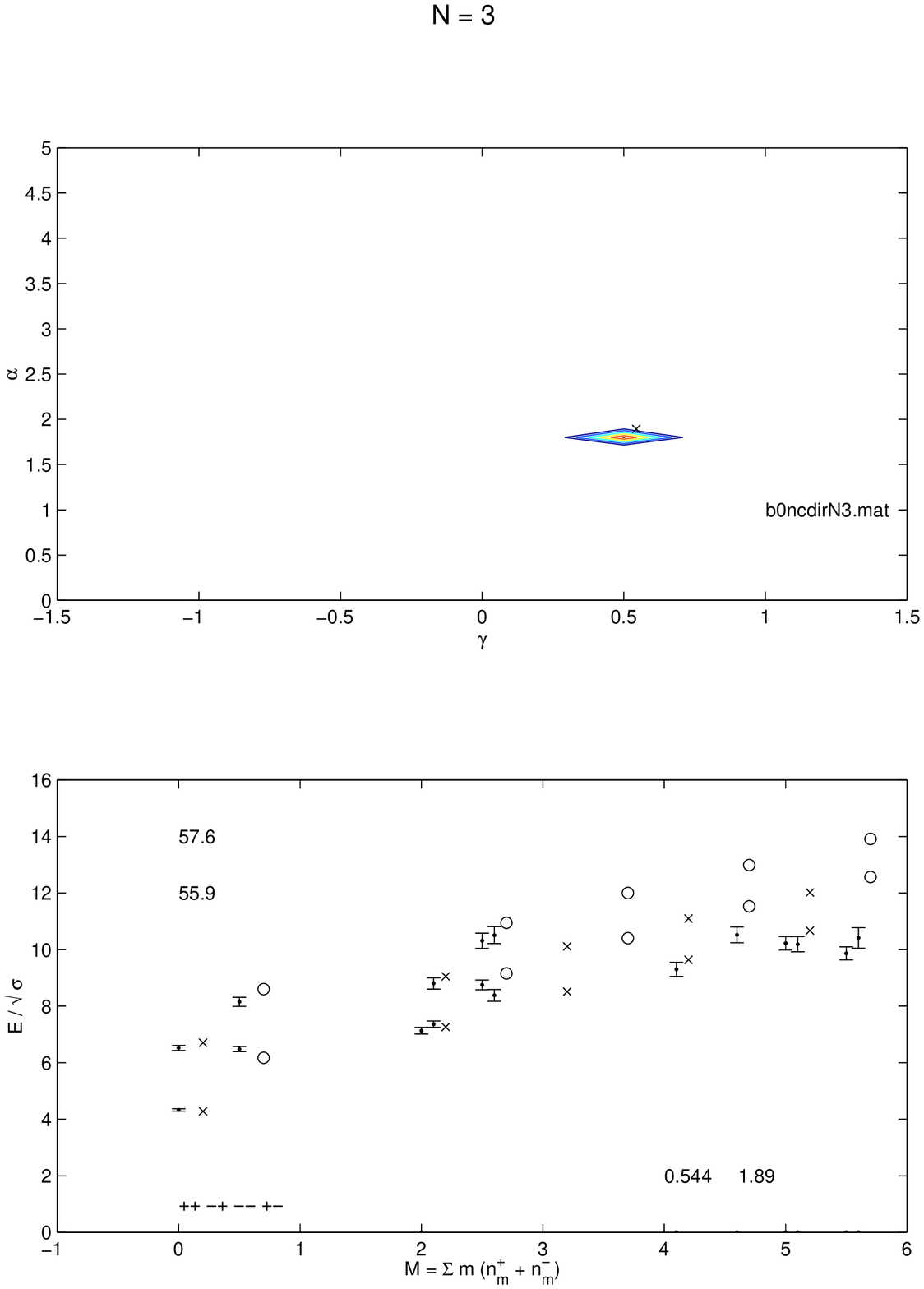}
\caption{The direct mixing posterior and spectrum for N=3.  The model's predictions are shown by 'o', and the corresponding lattice states by '.' with errorbars.  The x-axis is labelled by the phonon number M.  The relationship between M and the spin J is given in Appendix~\ref{chap:IPreview}, Table~\ref{table:MtoJ}.  See Subsection~\ref{subsec:fig} for more details.}
\label{fig:b0ncdirN3}
\end{figure}

When N=3 we encounter our first 2 dimensional parameter estimation problem.  The posterior density function (pdf) is shown in Figure~\ref{fig:b0ncdirN3}.  Despite the coarse graining of parameter space, we see a clearly defined maximum in the pdf.  The best $\chi^2$ is at $\gamma=0.54, \alpha=1.89$.  Examining the best fit spectrum, we see a qualitatively good fit.  Except for the $2^{+-}$, everything used in the fit is within $2\sigma$ of the lattice values.  As the mixing is just a constant, no properties of the states can influence the magnitude of the mixing, which might account for the discrepancies at higher M.  These states are larger and thus less likely to collapse to the ball pictured for this mechanism.  The discrepancy in the $C=-$ sector continues for the M=4 and 5 states.  Here we see good predictive agreement for the $C=+$ sector (allowing for the spin assignment at M=4), but the $C=-$ sector does not agree at higher M.  If we are concerned with an 0/4 ambiguity, then we also have to consider a 1/3 ambiguity.  In that case, there is a J=M=$3^{\pm-}$ at the right location in the spectrum, but then the $3^{++}$ should couple to what is assigned to J=1.

What should we make of the mess in the heavier part of the spectrum?  First, we should stress that these states are not included when determining $\chi^2$ and hence the parameters -- they are free predictions of the model, which is what raises the question of ambiguities in the conventional assignment of spin to certain lattice operators.  A preliminary exploration~\cite{Johnson:1998ev} showed that for the ++ sector the J=0 should be distinguished from the J=4.  But states with more complicated quantum numbers and higher spin require operators of more complicated construction, and the situation regarding which states a (finite) operator couples to is much less clear~\cite{Morningstar:1999rf}.  It is possible that the $1^{\pm+}$ operator is good while the $1^{\pm-}$ is being misled by the J=3, but the issue requires further investigation.  Exploratory calculations on a new technique of constructing lattice operators for arbitrary spin are discussed in Chapter~\ref{chap:lattice}.

\begin{figure}[!t]
\centering
\includegraphics[width=\textwidth]{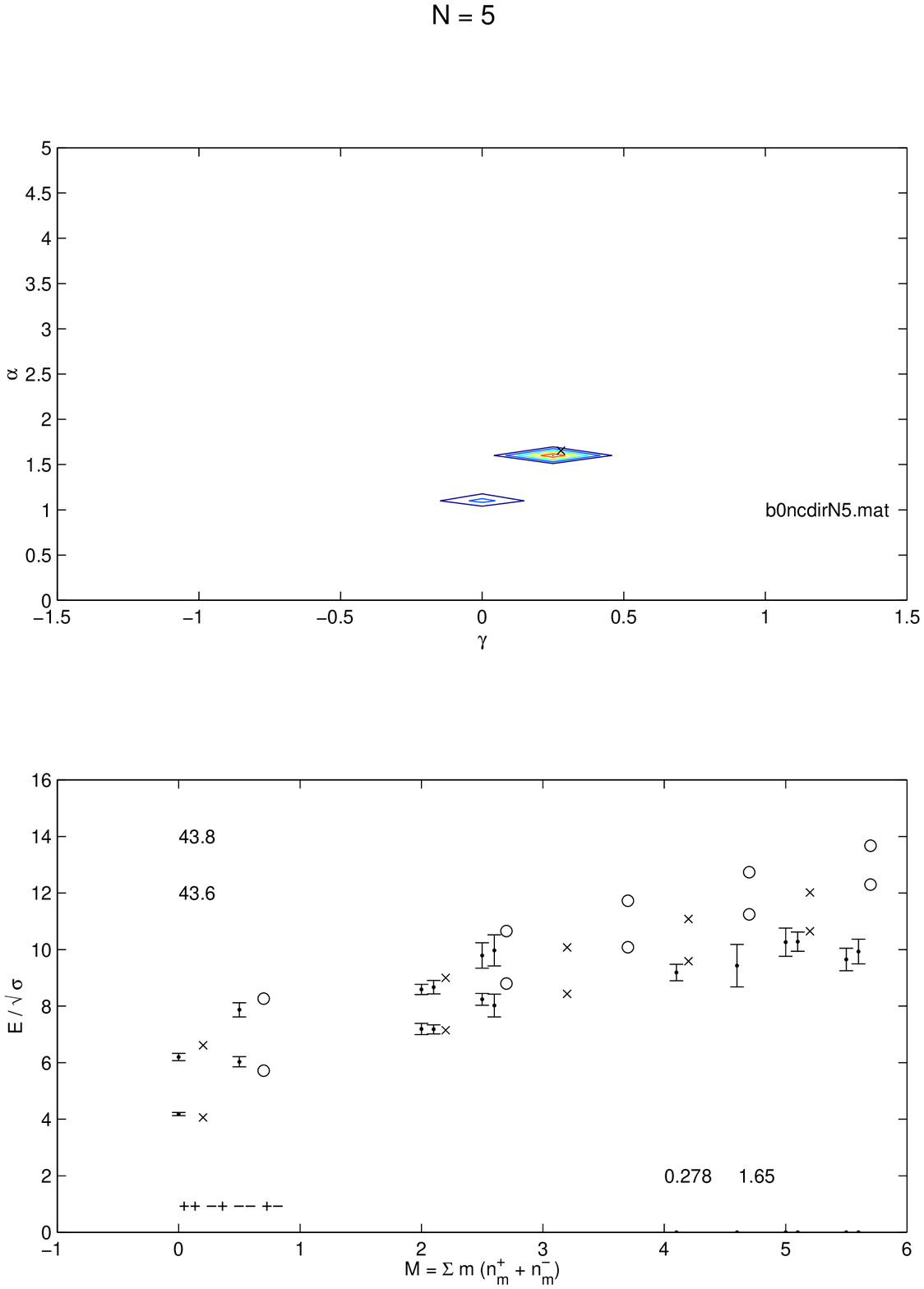}
\caption{The direct mixing posterior and spectrum for N=5.}
\label{fig:b0ncdirN5}
\end{figure}

For N=4 and 5, the qualitative features of the spectrum do not change dramatically.  The low lying states reproduce the lattice spectrum reasonably well, and the higher states continue to follow the spin ambiguity predictions.  Turning to the posterior, Figure~\ref{fig:b0ncdirN5}, at N=5 we start seeing evidence of a second peak near $\gamma=0$.  This peak is not included when we take the local quadratic approximation to compute the errors on $\gamma$ and $\alpha$, but its presence should be noted.  The main peak has also shifted closer to the $\gamma=0$ axis.

\begin{figure}[!t]
\centering
\includegraphics[width=\textwidth]{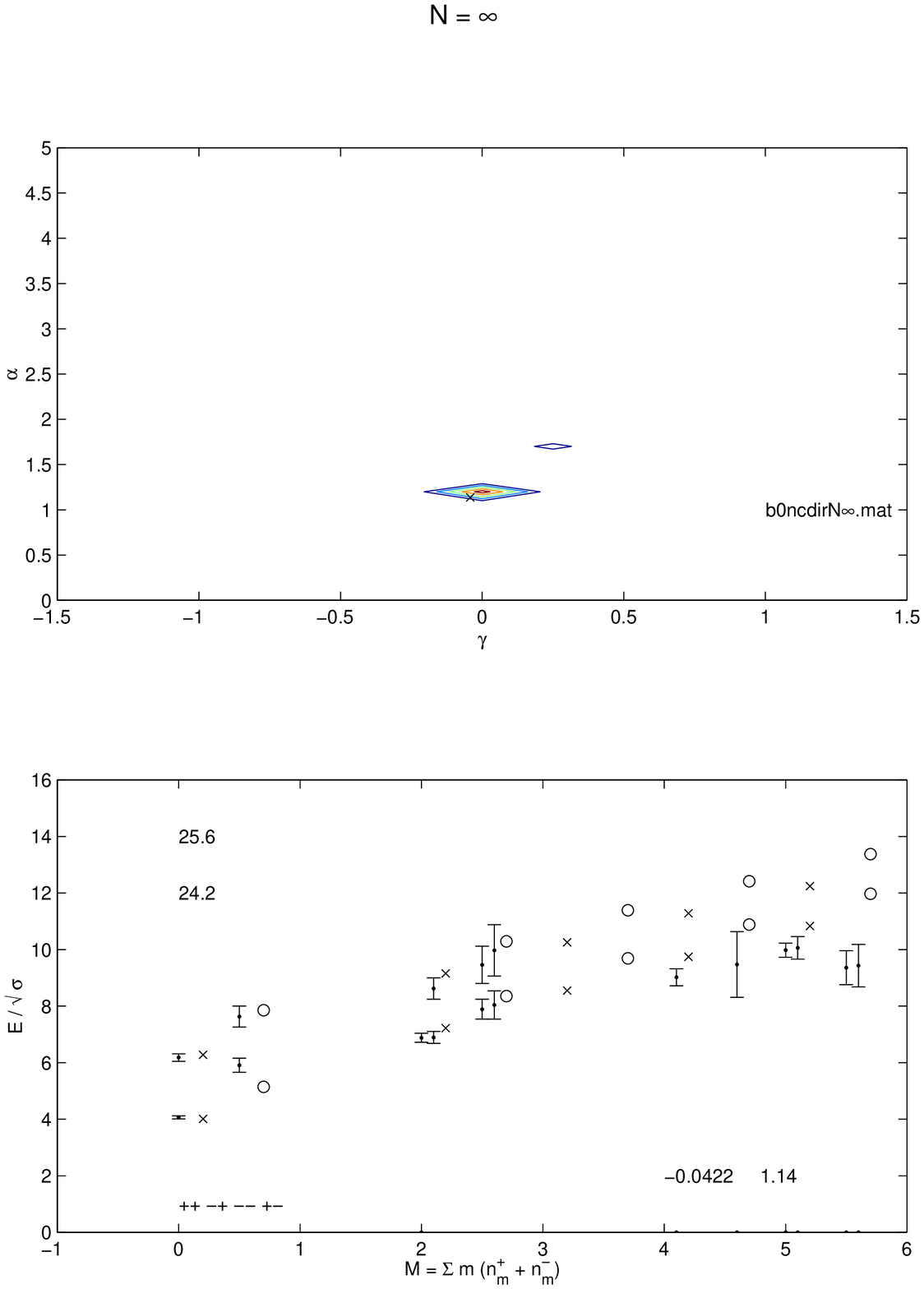}
\caption{The direct mixing posterior and spectrum for N=$\infty$.}
\label{fig:b0ncdirN6}
\end{figure}

The lattice spectrum has been extrapolated to N=$\infty$ using a correction linear in (1/$N^2$)~\cite{Teper:1997tq}.  Repeating our fitting procedure with these lattice values, we get the posterior and spectrum of Figures~\ref{fig:b0ncdirN6}.  Now the second peak has swamped the first to dominate the posterior, though there is a little of the first peak left.  The best fit gives $\gamma=-.042(190)$ with an error consistent with $\gamma=0$.  Were $\gamma$ to equal zero would require a delicate balance of the elasticity and the L\"{u}scher correction to cancel.  We will look at the behavior of $\gamma$ as a function of (1/$N^2$) later in the analysis and compare its extrapolation to N=$\infty$ with $\gamma$ fit to the extrapolated spectrum.

\begin{figure}[!t]
\centering
\includegraphics[width=\textwidth]{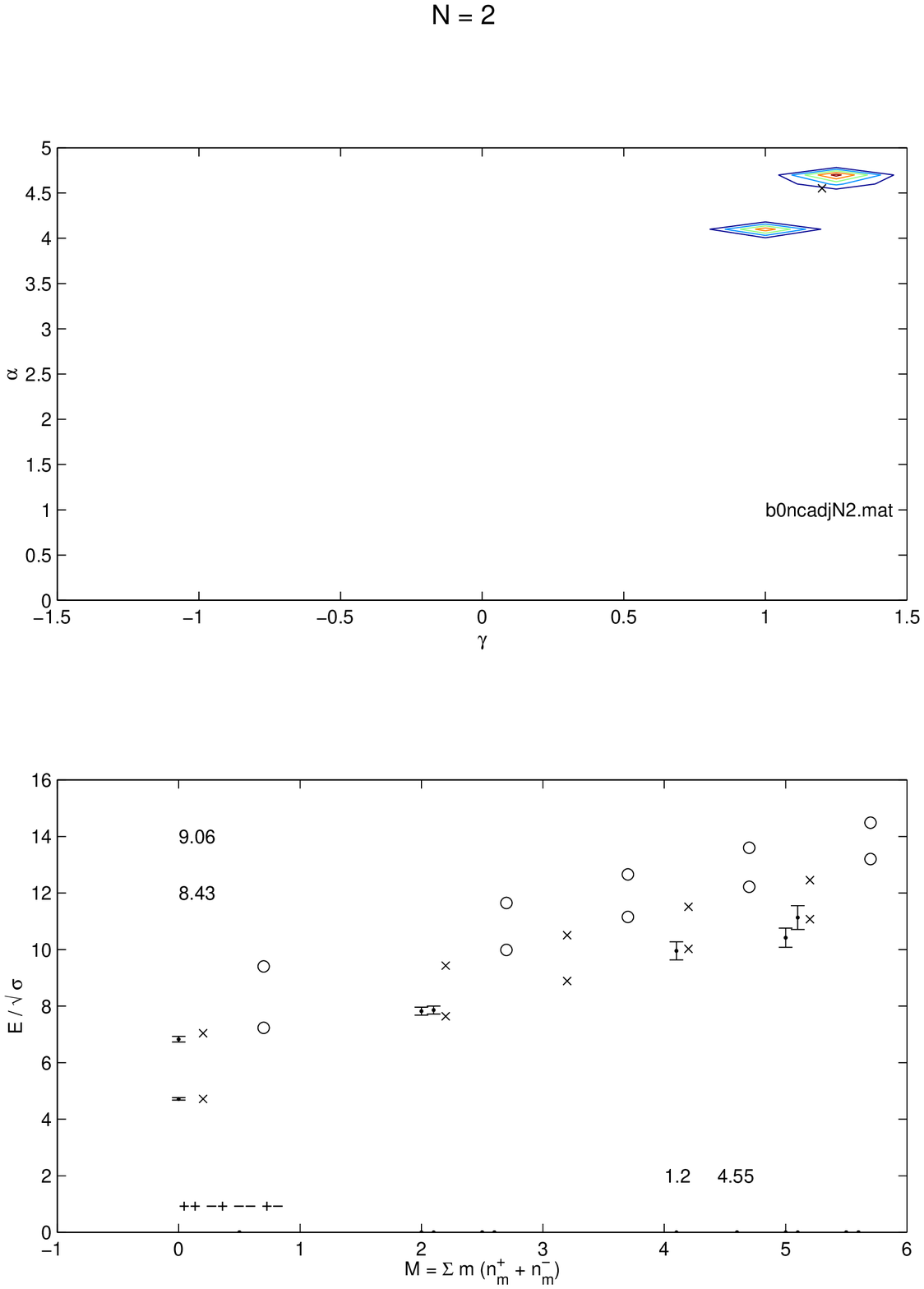}
\caption{The adjoint mixing posterior and spectrum for N=2.}
\label{fig:b0ncadjN2}
\end{figure}

Looking now at the adjoint mixing mechanism, we can start with the N=2, as there still exists an adjoint loop mixing with the fundamental loop, which is plotted in the figure.  Our prior nearly misses the peak, Figure~\ref{fig:b0ncadjN2}, but as its location moves into the prior for later N, we should feel confident we are seeing the global minimum.  The low spin spectrum fits quite well -- all states are within 1.5$\sigma$.  The higher spin states also display good agreement.

At N=3, the model starts showing signs of difficulty in fitting the lattice data.  Including both radial excitations and the $C=-$ sector contributes many more states to the fit.  Still, we see a well defined peak in $\chi^2$ space and can determine $\gamma_0 = .576$ and $\alpha_0 = 3.56$.  Looking at its best fit spectrum, the model is in general qualitative agreement with the lattice data.  The most obvious discrepancy is the $2^{\pm-}$.  At higher spins, the model consistently is too heavy.  In the $C=-$ sector, the oddly low $1^{\pm-}$ does compare well with the $3^{\pm-}$.  The magnitude of the radial splitting seems a little off as well, again perhaps because of the adiabatic approximation.
As $N\rightarrow\infty$, nothing dramatic happens.  The values of the best fit parameters, Table~\ref{table:ParametersB0}, do change somewhat.  The best fit $\gamma$ are still positive, and $\alpha\sim 3$.  Except for the lowest J=0 state in either C sector, the model's predictions are too heavy, especially at higher spin.

\begin{figure}[!t]
\centering
\includegraphics[width=\textwidth]{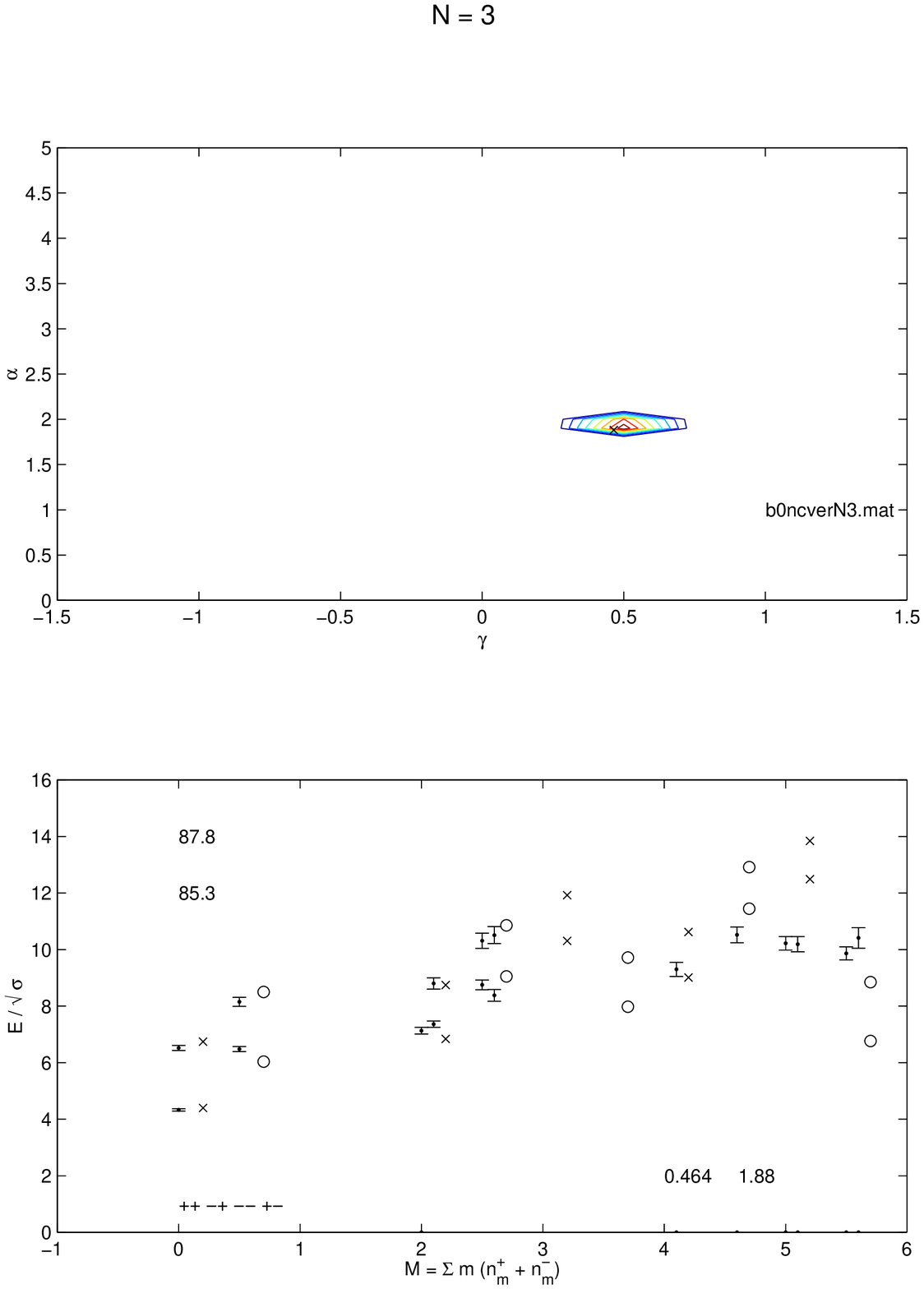}
\caption{The vertex mixing posterior and spectrum for N=3.}
\label{fig:b0ncverN3}
\end{figure}

\begin{figure}[!t]
\centering
\includegraphics[width=\textwidth]{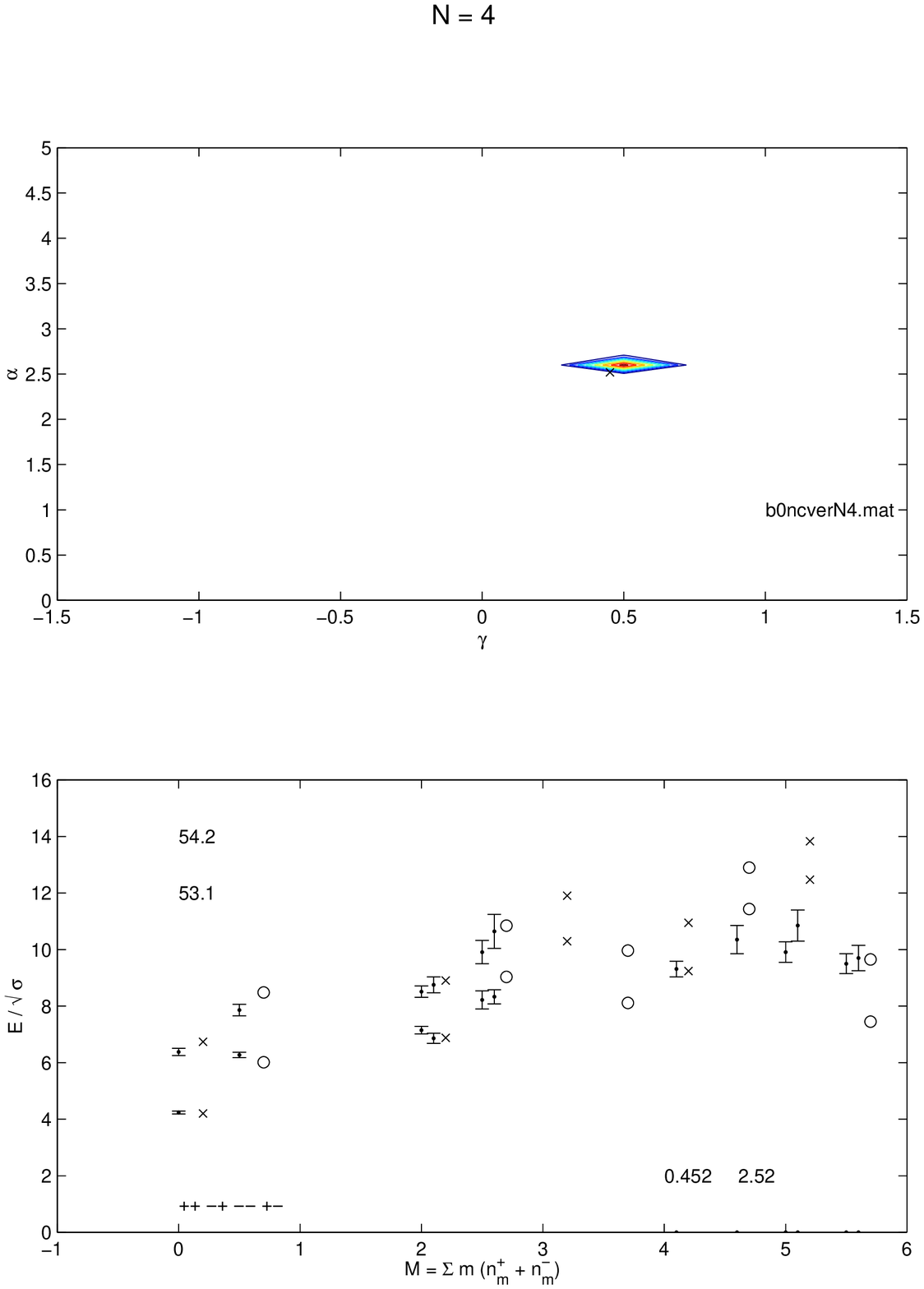}
\caption{The vertex mixing posterior and spectrum for N=4.}
\label{fig:b0ncverN4}
\end{figure}

\begin{figure}[!t]
\centering
\includegraphics[width=\textwidth]{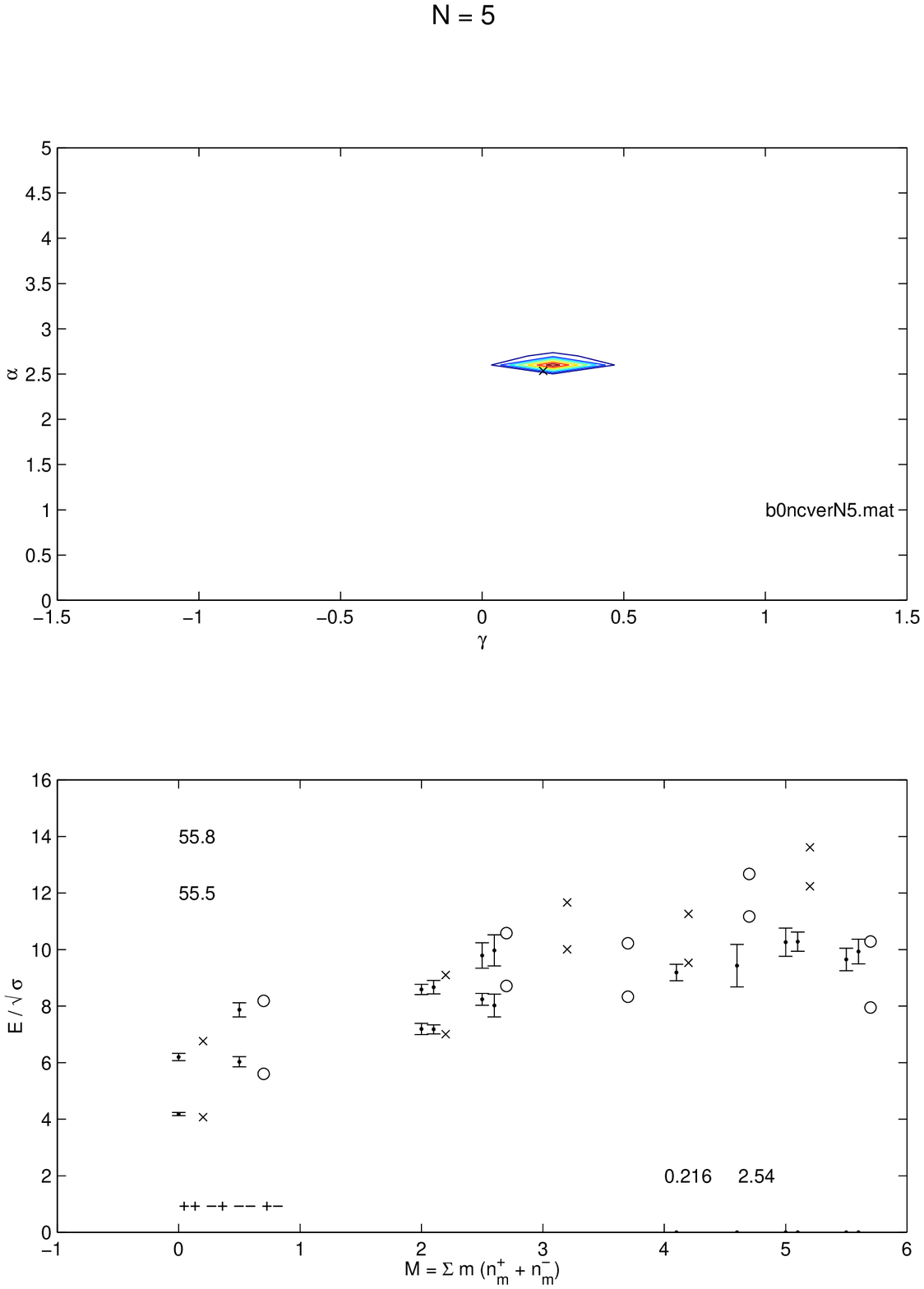}
\caption{The vertex mixing posterior and spectrum for N=5.}
\label{fig:b0ncverN5}
\end{figure}

For the vertex model we have only three cases to look at.  From N=3 to
N=5, we find a positive $\gamma$ and a mixing $\alpha\sim 2$,
Figures~\ref{fig:b0ncverN3} through \ref{fig:b0ncverN5}.  For J=0 and 2, the model generally reproduces the features of the lattice spectrum.  The robustness of the Isgur-Paton model is shown in the consistency of results across mixing mechanisms.  At higher spin, things are a mess, except for $4^{-+}$.  That is, if we persist in comparing the lattice J=1 state to M=5.  Allowing for more spin ambiguity, we might compare these states to M=3.  In that case, the $3^{\pm+}$ would be in good agreement, while the $3^{\pm-}$ might also be as well; the lowest state $3^{\pm-}$ in the model is mostly $\psi_V$, while the lattice operator is not constructed to couple to such a state~\cite{Teper:1999}.  At this point, such remarks are only conjecture -- a much more thorough technique needs to be developed, as in Chapter~\ref{chap:lattice}.  As $N\rightarrow\infty$, the features remain the same.  One state that does not fit well is the $0^{+-}$, which the lattice claims is well below the model's prediction for the $4^{\pm-}$.

\newpage

\begin{figure}[!t]
\centering
\includegraphics[width=\textwidth]{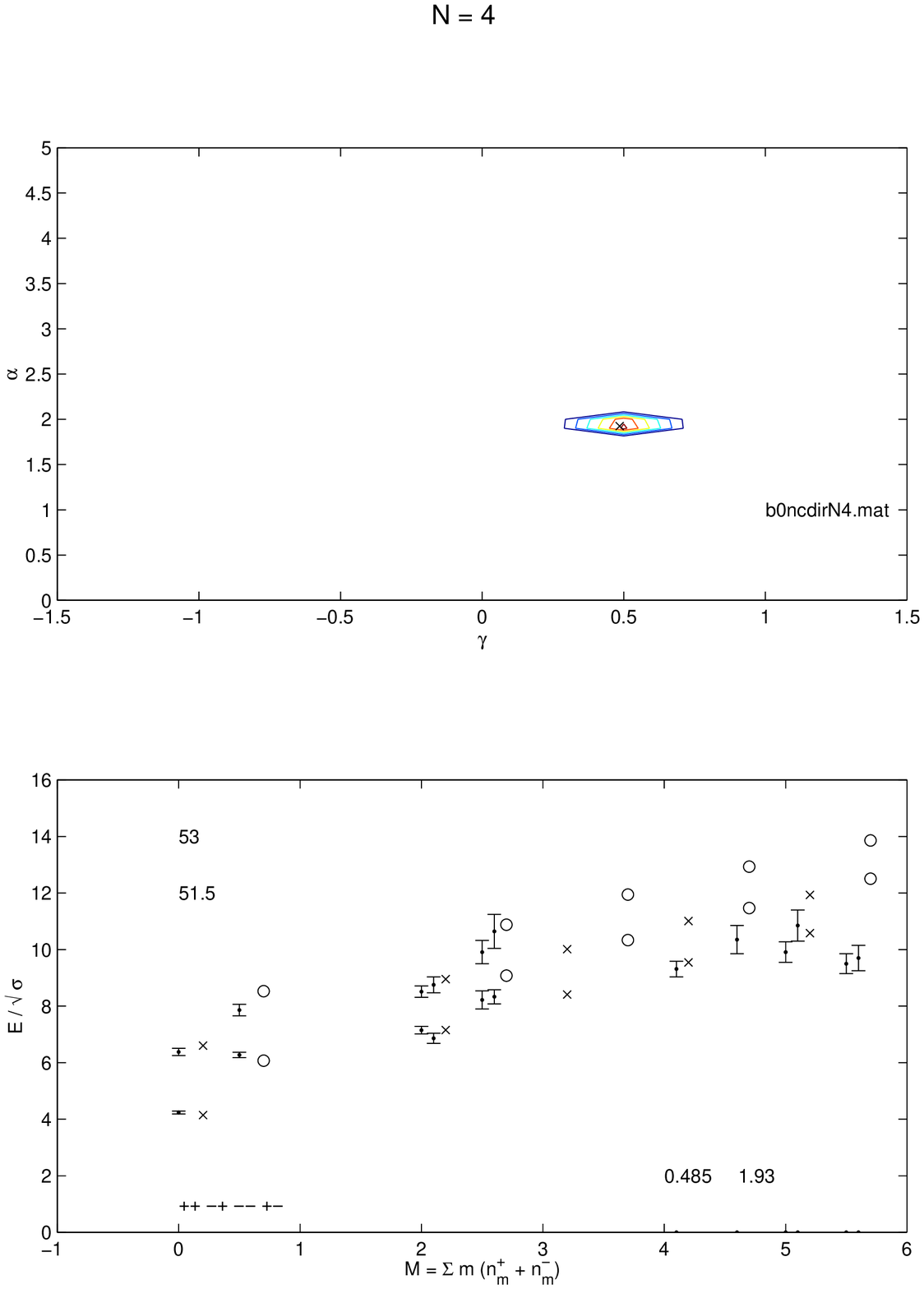}
\caption{The direct mixing posterior and spectrum for N=4.}
\label{fig:b0ncdirN4}
\end{figure}

\begin{figure}[!t]
\centering
\includegraphics[width=\textwidth]{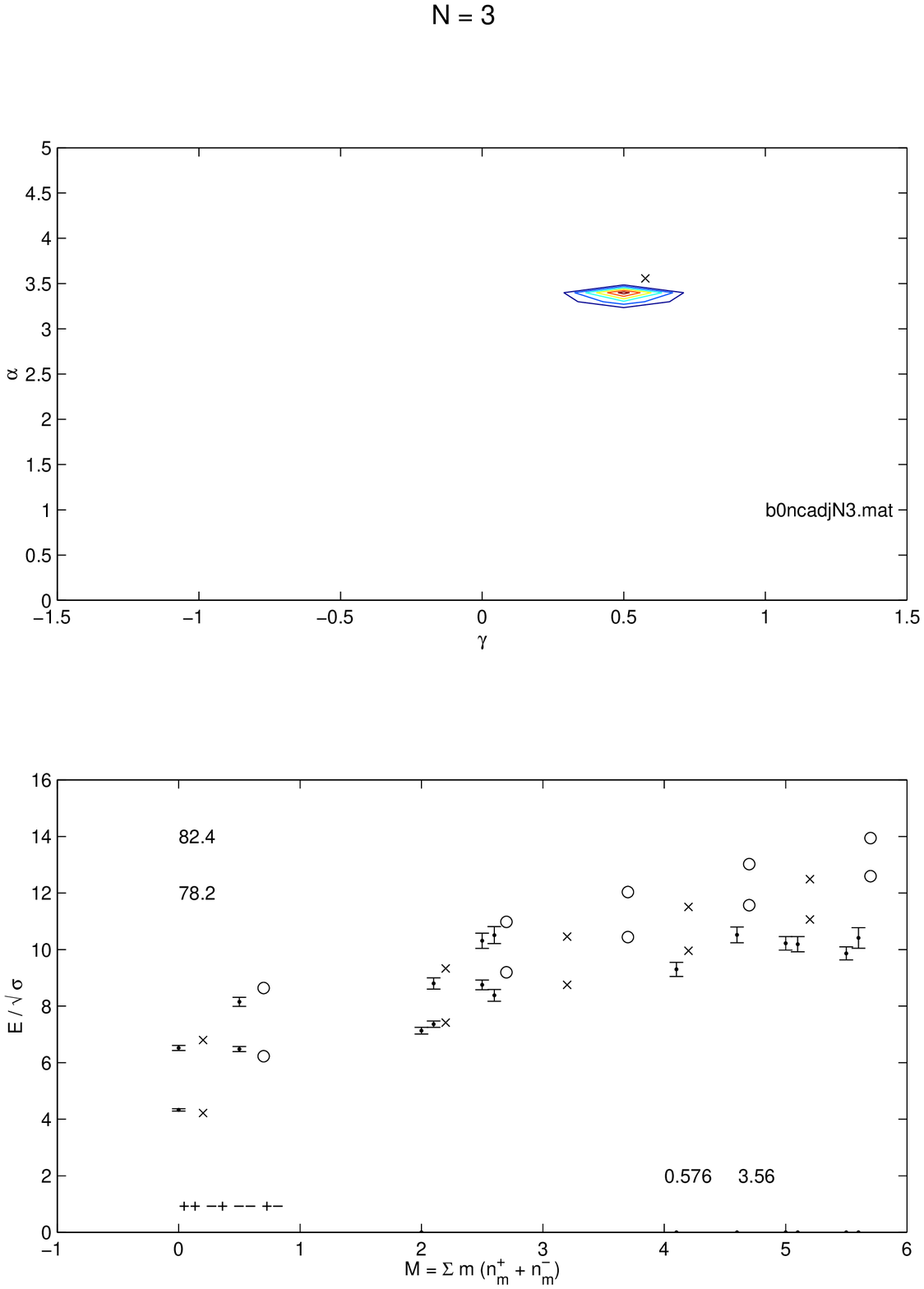}
\caption{The adjoint mixing posterior and spectrum for N=3.}
\label{fig:b0ncadjN3}
\end{figure}

\begin{figure}[!t]
\centering
\includegraphics[width=\textwidth]{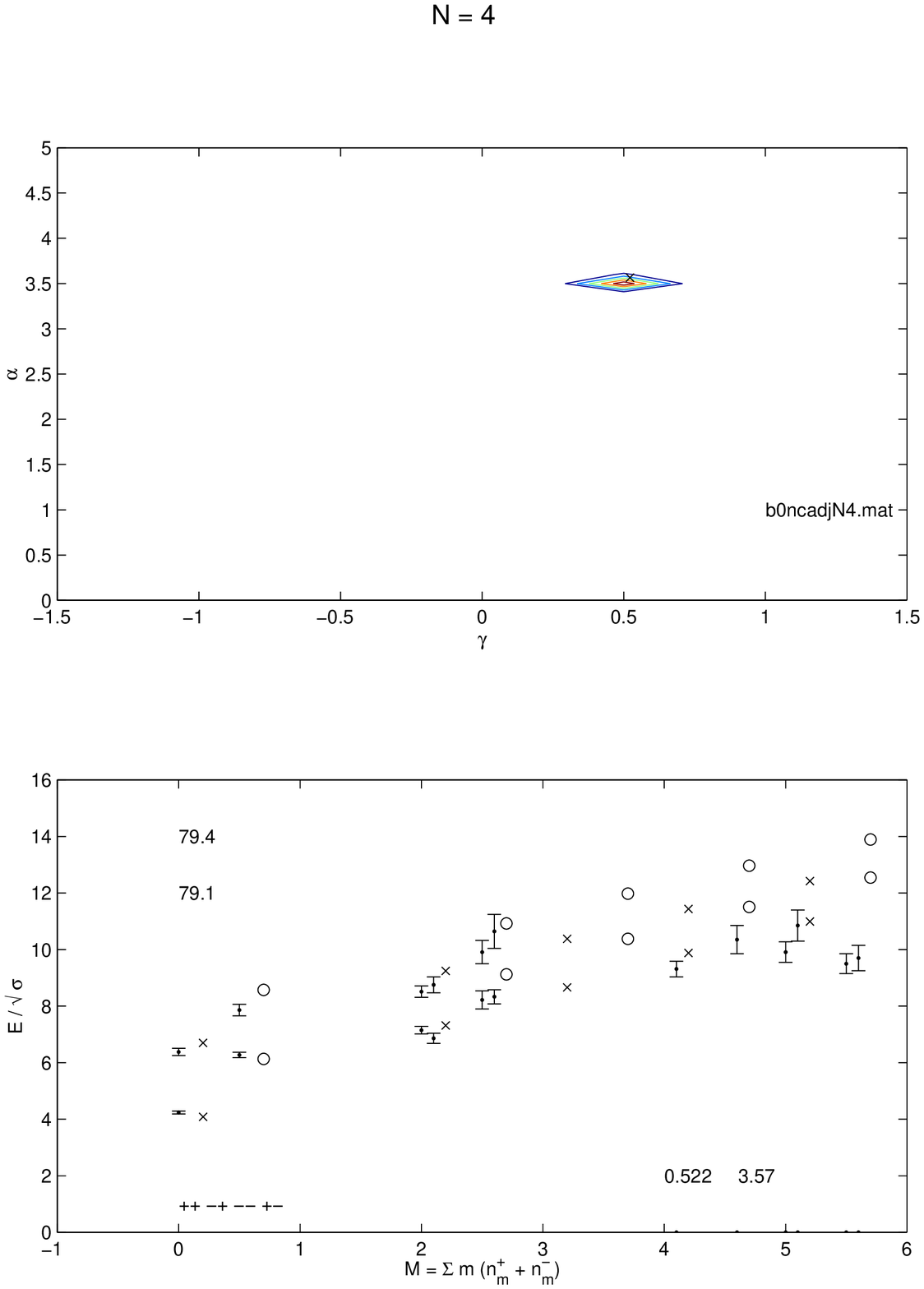}
\caption{The adjoint mixing posterior and spectrum for N=4.}
\label{fig:b0ncadjN4}
\end{figure}

\begin{figure}[!t]
\centering
\includegraphics[width=\textwidth]{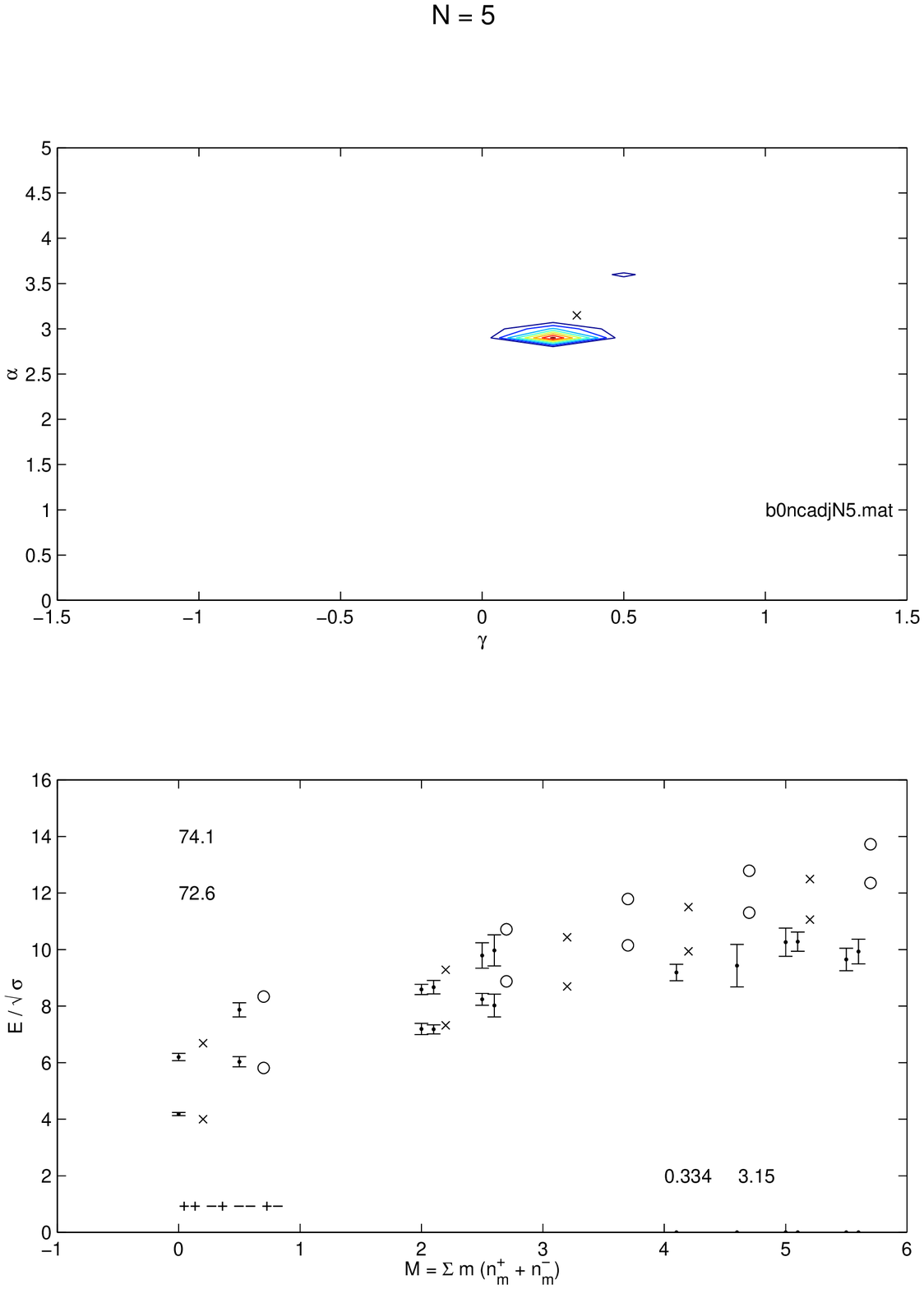}
\caption{The adjoint mixing posterior and spectrum for N=5.}
\label{fig:b0ncadjN5}
\end{figure}

\begin{figure}[!t]
\centering
\includegraphics[width=\textwidth]{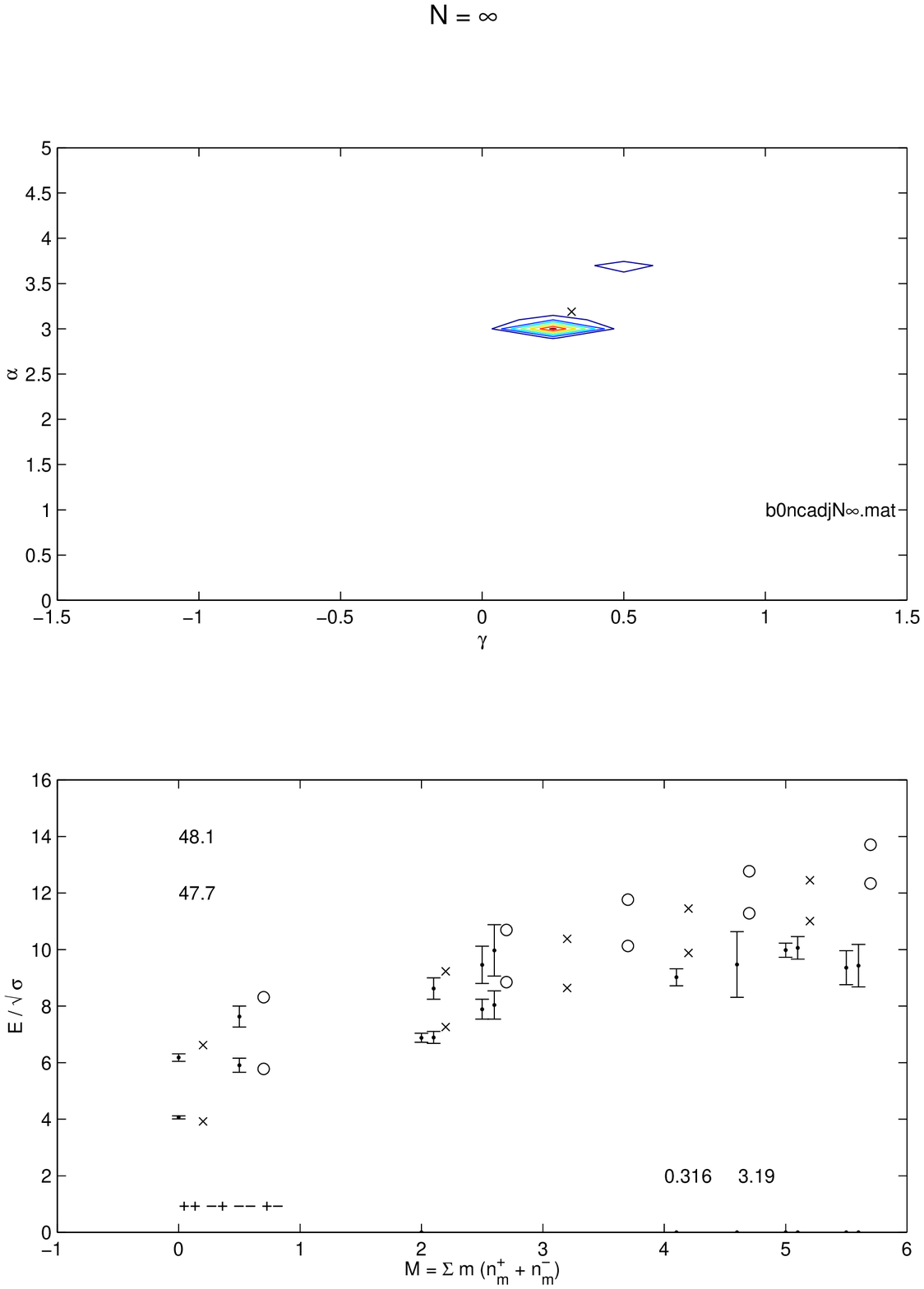}
\caption{The adjoint mixing posterior and spectrum for N=$\infty$.}
\label{fig:b0ncadjN6}
\end{figure}

\newpage

\begin{table}[!h]
\centering
\begin{tabular}{|lc|cc|cc|cc|cc|cc|} \hline
N & direct & $0^{++}$ & $0^{--}$ & $2^{\pm+}$ & $2^{\pm-}$ &
$3^{\pm+}$ & $3^{\pm-}$ & $4^{\pm+}$ & $4^{\pm-}$ & $1^{\pm+}$ &
$1^{\pm-}$ \\\hline

2 & $m_{J^{PC}}$ & 4.618 & & 7.946 & & 9.324 & & 10.56 & & 11.67 &  \\
  & $m_{J^{PC}}^\star$ & 7.482 & & 9.963 & & 11.09 & & 12.13 & & 13.11 &  \\\hline

3 & & 4.278 & 6.171 & 7.261 & 9.154 & 8.509 & 10.4 & 9.638 & 11.53 &
10.67 & 12.57 \\
  & & 6.707 & 8.60 & 9.051 & 10.94 & 10.11 & 12.00 & 11.09 & 12.99 &
12.02 & 13.91 \\\hline

4 & & 4.146 & 6.072 & 7.151 & 9.076 & 8.407 & 10.33 & 9.542 & 11.47 &
10.58 & 12.51 \\
  & & 6.601 & 8.526 & 8.955 & 10.88 & 10.02 & 11.94 & 11.01 & 12.93 &
11.93 & 13.86 \\\hline

5 & & 4.062 & 5.715 & 7.145 & 8.798 & 8.431 & 10.08 & 9.589 & 11.24 &
10.65 & 12.3 \\
  & & 6.612 & 8.265 & 8.998 & 10.65 & 10.07 & 11.73 & 11.08 & 12.73 &
12.02 & 13.67 \\\hline

$\infty$ & & 4.009 & 5.144 & 7.219 & 8.354 & 8.552 & 9.687 & 9.747 &
10.88 & 10.83 & 11.97 \\
  & & 6.279 & 7.853 & 9.155 & 10.29 & 10.26 & 11.39 & 11.28 & 12.42 &
12.24 & 13.38 \\\hline

\end{tabular}
\caption{The spectrum for direct mixing at various N.  For each N, the
first mass is the lowest state in that channel, and the second mass is
the radial excitation.}
\label{table:b0ncdir}
\end{table}

\begin{table}[!h]
\centering
\begin{tabular}{|lc|cc|cc|cc|cc|cc|} \hline
N & adjoint & $0^{++}$ & $0^{--}$ & $2^{\pm+}$ & $2^{\pm-}$ &
$3^{\pm+}$ & $3^{\pm-}$ & $4^{\pm+}$ & $4^{\pm-}$ & $1^{\pm+}$ &
$1^{\pm-}$ \\\hline

2 & $m_{J^{PC}}$ & 4.717 & & 7.643 & & 8.888 & & 10.02 & & 11.07 & \\
  & $m_{J^{PC}}^\star$ & 7.04 & & 9.432 & & 10.51 & & 11.51 & & 12.46 \\\hline

3 & &
4.217  &  6.225  &  7.417  &  9.196  &  8.754  &  10.44  &  9.961  &
  11.57  &  11.06  &   12.6 \\ & & 
6.801  &  8.64   &  9.329  &  10.98  &  10.46  &  12.03  &  11.51  &
  13.02  &  12.49  &  13.94 \\\hline

4 & & 
4.083  &  6.134  &  7.316  &  9.124  &  8.664  &  10.38  &  9.879  &
11.51  &  10.99  &  12.54 \\ & & 
6.697  &  8.572  &  9.244  &  10.92  &  10.38  &  11.98  &  11.43  &
12.97  &  12.42  &  13.89 \\\hline

5 & & 
3.999  &  5.813  &  7.322  &  8.874  &  8.698  &  10.15  &  9.934  &
11.3  &  11.06 &   12.36 \\ & & 
6.687  &  8.336  &  9.283  &  10.71  &  10.43  &  11.79  &   11.5  &
12.79  &   12.5 &   13.72 \\\hline

$\infty$ & & 
3.919  &   5.78  &  7.258  &  8.849  &  8.639  &  10.13  &  9.879  &
11.28  &  11.01 &   12.34 \\ & & 
6.621  &  8.312  &  9.226  &  10.69  &  10.38  &  11.77  &  11.45  &
12.77  &  12.45 &   13.71 \\\hline

\end{tabular}
\caption{The spectrum for adjoint mixing at various N.  For each N, the
first mass is the lowest state in that channel, and the second mass is
the radial excitation.}
\label{table:b0ncadj}
\end{table}

\begin{table}[!h]
\centering
\begin{tabular}{|lc|cc|cc|cc|cc|cc|} \hline
N & vertex  & $0^{++}$ & $0^{--}$ & $2^{\pm+}$ & $2^{\pm-}$ &
$3^{\pm+}$ & $3^{\pm-}$ & $4^{\pm+}$ & $4^{\pm-}$ & $1^{\pm+}$ &
$1^{\pm-}$ \\\hline

3 & $m_{J^{PC}}$ &
4.4  &  6.036  &  6.839  &  9.049  &  10.31  &  7.975  &  9.007  &
11.45   & 12.49   &  6.765 \\
  & $m_{J^{PC}}^\star$ &
6.742 &     8.5  &  8.748  &  10.86  &  11.92  &  9.719  &  10.62  &
12.91  &  13.84  &  8.849 \\\hline

4 & & 
4.203  &  6.014  &  6.878  &  9.031  &  10.29  &  8.113  &  9.238  &
11.43  &  12.47  &  7.448 \\ & &
6.735  &  8.484  &  8.905  &  10.84  &  11.91  &  9.961  &  10.94  &
12.9  &  13.83  &  9.649 \\\hline

5 & & 
4.074  &  5.606  &  7.006  &  8.713  &  10.01  &  8.331  &  9.531  &
11.17  &  12.24  &  7.947 \\ & &
6.762  &  8.185  &  9.102  &  10.58  &  11.66  &  10.22  &  11.26  &
12.67  &  13.61  &  10.28 \\\hline

\end{tabular}
\caption{The spectrum for vertex mixing at various N.  For each N, the
first mass is the lowest state in that channel, and the second mass is
the radial excitation.}
\label{table:b0ncver}
\end{table}

\clearpage

\section{Adjusting the weightings}

How do things change when $\beta=.05$?  Actually, not by very much.  By adding $\pm5$\% to the error, we de-weight the states with the tightest statistical errors, which are the low lying states of small radius on the order of the thickness of the flux tube, allowing the radial excitations to take on more of a role in determining the best fit parameters.  We are also trying to quantify what we can expect of a nonrelativistic model with a drastic adiabatic approximation when compared to our best estimates of the pure glue spectrum from the lattice.

\begin{figure}[!t]
\centering
\includegraphics[width=\textwidth]{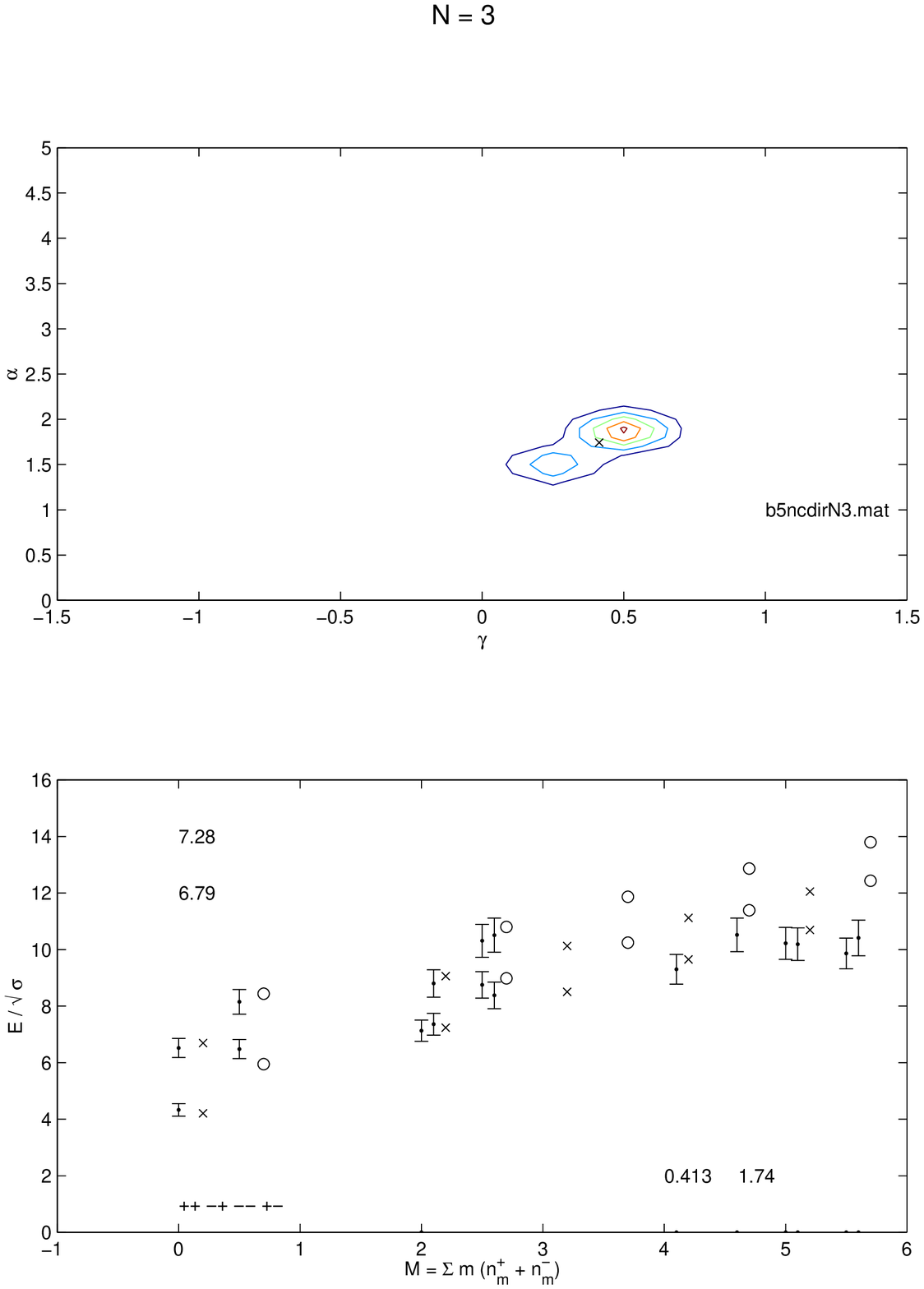}
\caption{The direct mixing posterior and spectrum for N=3.}
\label{fig:b5ncdirN3}
\end{figure}

\begin{figure}[!t]
\centering
\includegraphics[width=\textwidth]{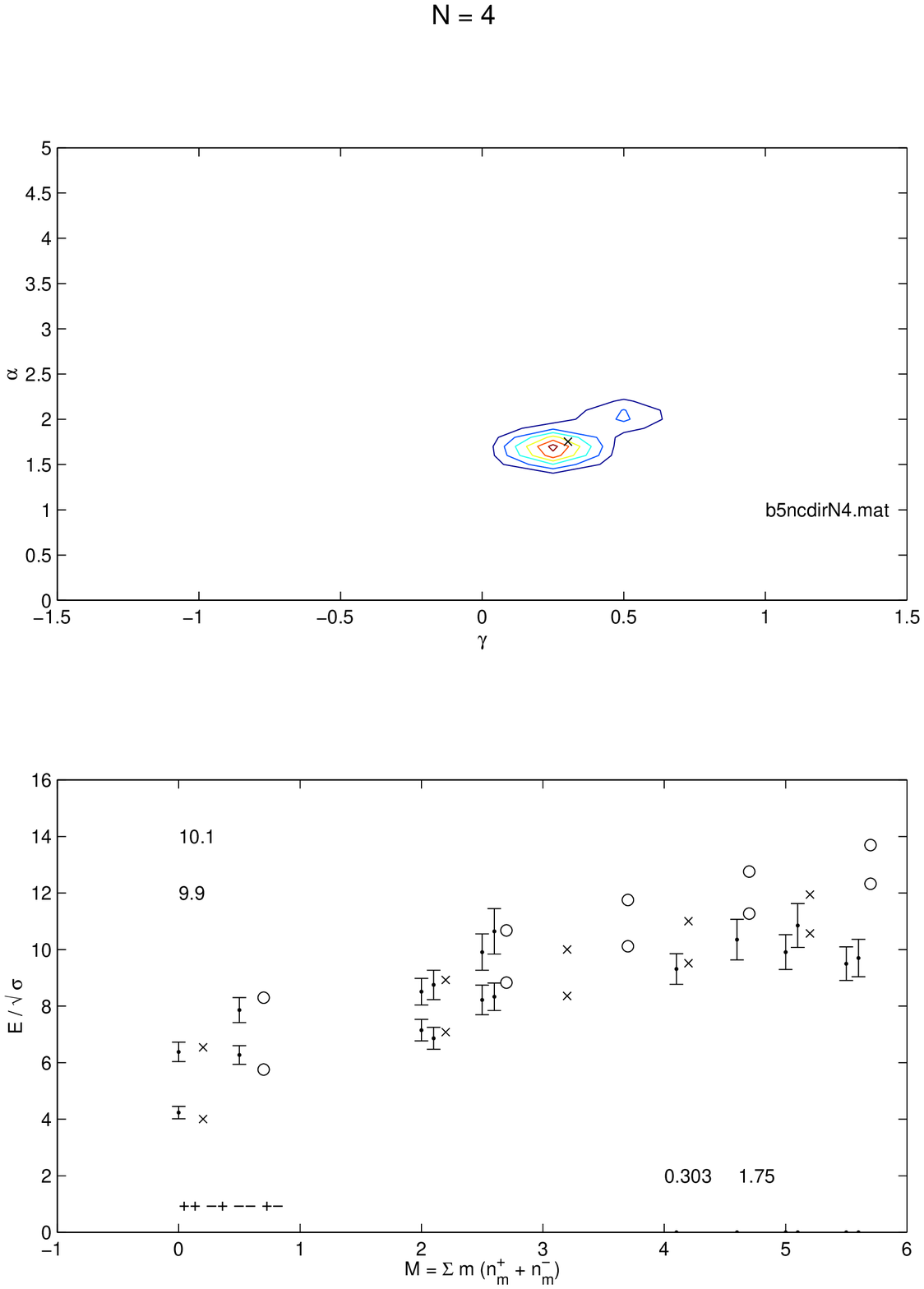}
\caption{The direct mixing posterior and spectrum for N=4.}
\label{fig:b5ncdirN4}
\end{figure}

\begin{figure}[!t]
\centering
\includegraphics[width=\textwidth]{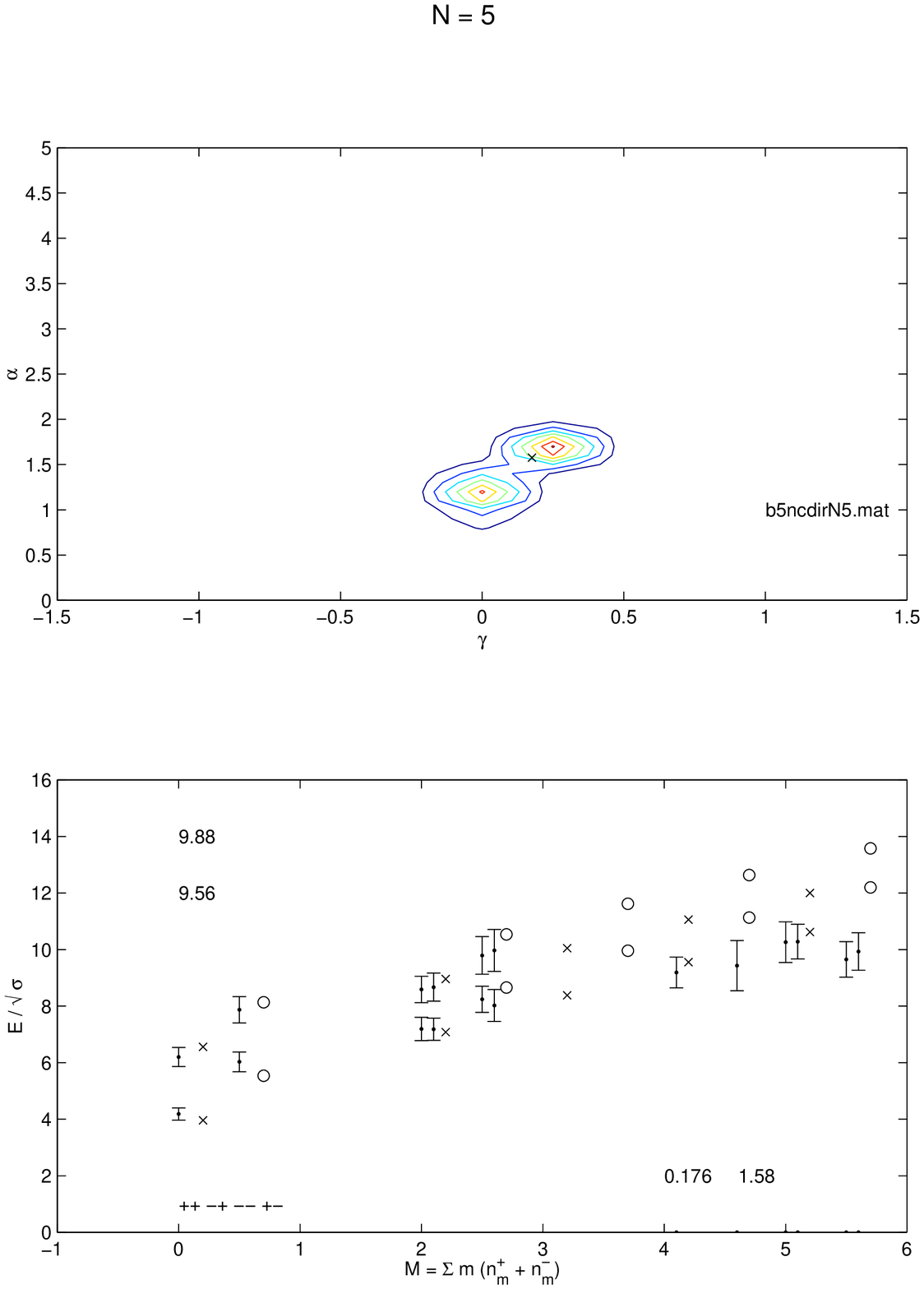}
\caption{The direct mixing posterior and spectrum for N=5.}
\label{fig:b5ncdirN5}
\end{figure}

\begin{figure}[!t]
\centering
\includegraphics[width=\textwidth]{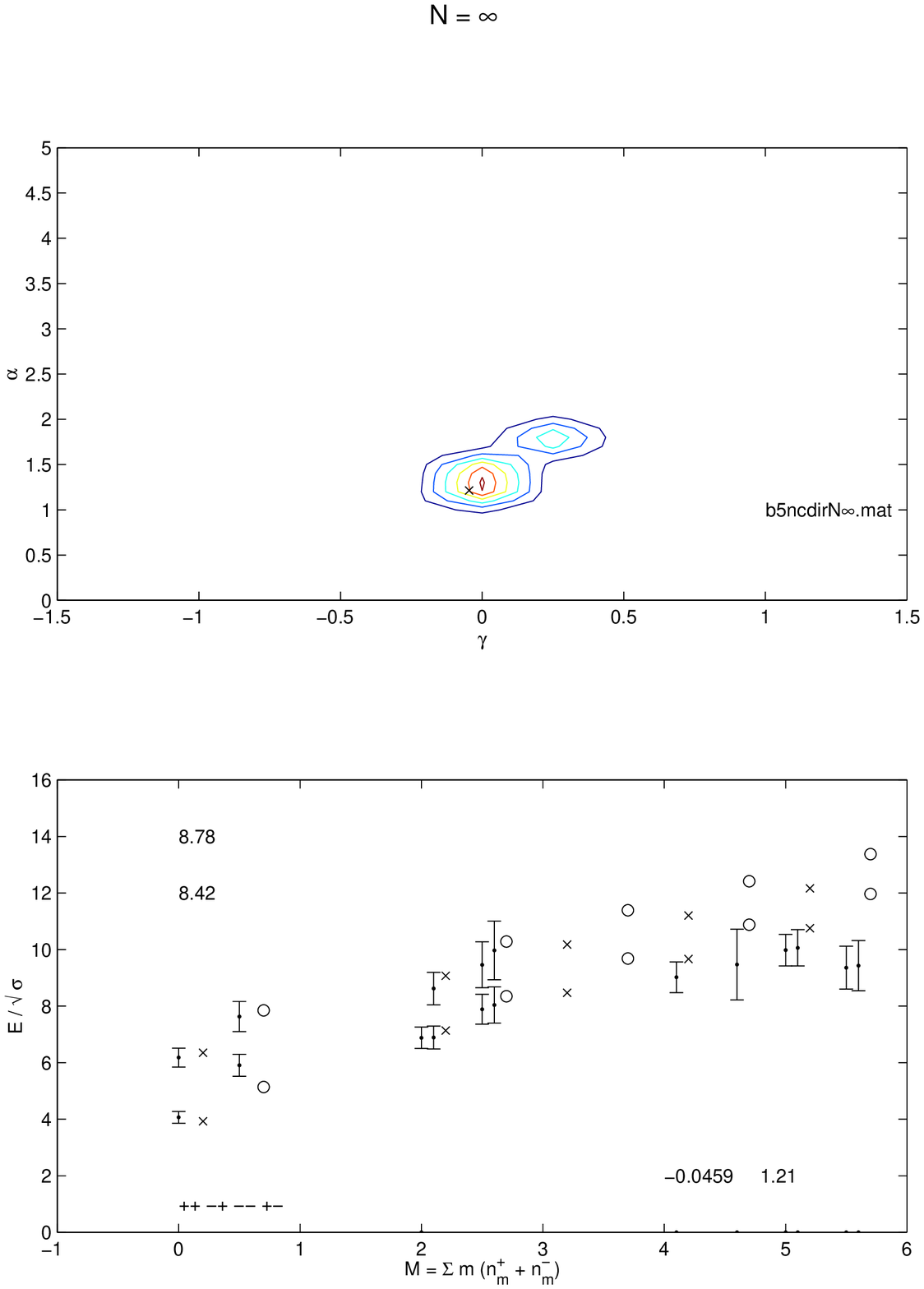}
\caption{The direct mixing posterior and spectrum for N=$\infty$.}
\label{fig:b5ncdirN6}
\end{figure}

Again considering direct mixing first, we look at the log of the
posterior as a function in parameter space,
Figure~\ref{fig:b5ncdirN3}.  We note a broader, smoother peak located
at roughly the same location $(\gamma_0, \alpha_0)$.  Consistency in
the value of the best fit parameters implies that we have not done
anything too drastic by manipulating the weightings.  A broader
posterior actually implies a more conservative estimate of the
parameters, in the sense of having a larger errorbar (see~\cite{Bretthorst:1988 ,Sivia:1996}).  The
spectrum, for N=3, Figure~\ref{fig:b5ncdirN3}, shows good agreement
across the lower states -- except for the $0^{--}$ and the $2^{-+}$,
all the predictions are within the expanded errorbars.  The higher
states are again consistent with ambiguous spin assignments at M=4 and
5.  As $N\rightarrow\infty$, Figures~\ref{fig:b5ncdirN4} through \ref{fig:b5ncdirN6}, the features of the model are essentially the same.  The prediction for the $0^{--}$ is too low, which could also be explained if the tighter errorbar of the $0^{++}$ were pulling the whole spectrum towards the state which it should least fit.  At N=$\infty$, where we are fitting parameters to an extrapolation of the lattice data, our fit for $\gamma_0 = -.046(295)$ is negative, but within its error is consistent with zero.  We will more closely examine the behavior of $\gamma$ and $\alpha$ as functions of N in the next chapter.

\begin{figure}[!t]
\centering
\includegraphics[width=\textwidth]{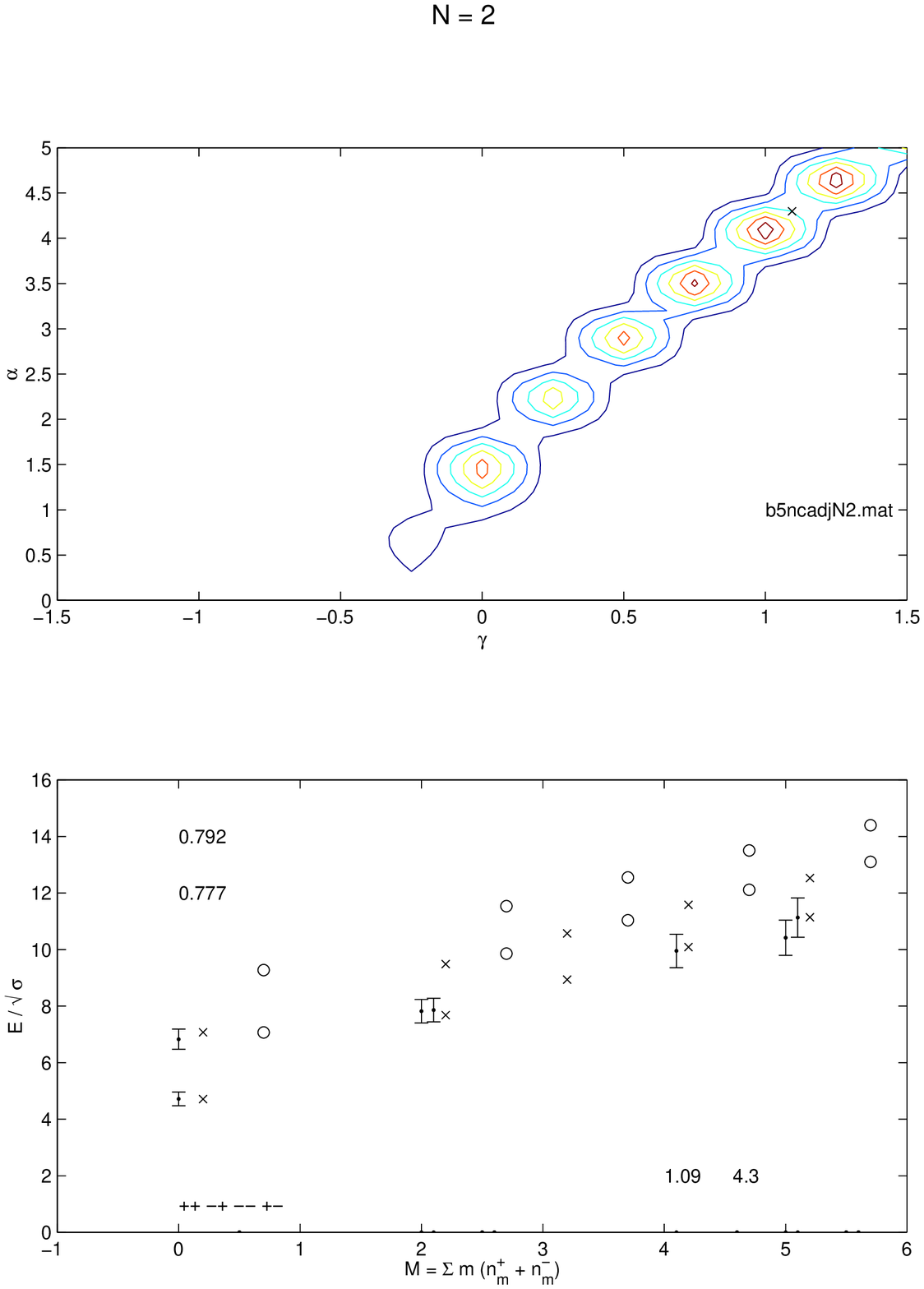}
\caption{The adjoint mixing posterior and spectrum for N=2.}
\label{fig:b5ncadjN2}
\end{figure}

For the adjoint mixing mechanism at N=2, we see a very elongated posterior (and a miniscule value for $\chi^2$).  The peak is still in about the same location, however, which is good.  The length of the peak tells us that a wide range of parameter values are consistent with the lattice data.  When N=3, our peak takes on a more traditional shape at $\gamma_0 = .44(36) \;,\; \alpha_0 = 3.33(60)$.  Looking at the spectrum, Figure~\ref{fig:b5ncadjN2}, we see good agreement for J=0 and 2 and the usual mess for the higher spins.  As $N\rightarrow\infty$, the $1^{\pm-}$ states are nowhere close, suggesting either they are spin 3 or that the model does not have the necessary structure to explain the data.

As before, we have only three cases for vertex mixing.  For each N,
the peak is well defined, giving parameter values in  Table~\ref{table:ParametersB5}.  The spectrum again demonstrates the mechanism's complex structure.
At M=5, the direction of the C splitting is in the right direction to
follow the lattice assignments, but far too large in magnitude.  These
ambiguities highlight the need for better determination of the spin
assignments.

\begin{table}[!h]
\centering
\begin{tabular}{|lc|cc|cc|cc|cc|cc|} \hline
N & direct & $0^{++}$ & $0^{--}$ & $2^{\pm+}$ & $2^{\pm-}$ &
$3^{\pm+}$ & $3^{\pm-}$ & $4^{\pm+}$ & $4^{\pm-}$ & $1^{\pm+}$ &
$1^{\pm-}$ \\\hline

2 & $m_{J^{PC}}$ &
4.519  & &          7.869  & &          9.255  & &          10.49  & &
         11.61 &  \\
  & $m_{J^{PC}}^\star$ &
7.413   & &         9.902   & &         11.03  & &           12.08 & &
13.06 & \\\hline

3 & &
4.206  &  5.948  &  7.237   &  8.98  &  8.504  &  10.25  &  9.647  &
11.39  &  10.69  &  12.43 \\ & &
6.693  &  8.435  &  9.058   &  10.8  &  10.12  &  11.87  &  11.12  &
12.86  &  12.05  &  13.79 \\\hline

4 & &
4.005  &  5.758  &  7.079   & 8.832  &  8.361  &  10.11  &  9.516  &
11.27  &  10.57 &   12.32 \\ & &
6.543  &  8.296  &  8.926   & 10.68  &     10  &  11.75  &     11  &
12.76  &  11.94 &   13.69 \\\hline

5 & &
3.959  &  5.535  &  7.082   & 8.658  &  8.383  &  9.958  &  9.552  &
11.13  &  10.62 &   12.19 \\ & & 
6.558  &  8.134  &  8.961   & 10.54  &  10.05  &  11.62  &  11.06  &
12.63  &     12 &   13.58 \\\hline

$\infty$ & &
3.925  &  5.137  &  7.137  &  8.349  &   8.47  &  9.682  &  9.666  &
10.88  &  10.75 &   11.97 \\ & & 
 6.35  &  7.849  &  9.073  &  10.29  &  10.17  &  11.39  &   11.2  &
12.41  &  12.16 &   13.37 \\\hline

\end{tabular}
\caption{The spectrum for direct mixing with adjusted weightings at various N.  For each N, the
first mass is the lowest state in that channel, and the second mass is
the radial excitation.}
\label{table:b5ncdir}
\end{table}

\begin{table}[!h]
\centering
\begin{tabular}{|lc|cc|cc|cc|cc|cc|} \hline
N & adjoint & $0^{++}$ & $0^{--}$ & $2^{\pm+}$ & $2^{\pm-}$ &
$3^{\pm+}$ & $3^{\pm-}$ & $4^{\pm+}$ & $4^{\pm-}$ & $1^{\pm+}$ &
$1^{\pm-}$ \\\hline

2 & $m_{J^{PC}}$ &
4.714      & &       7.68      & &      8.938      & &      10.08
& &      11.14      &   \\
  & $m_{J^{PC}}^\star$ &
7.075      & &      9.486      & &      10.57      & &      11.58
& &     12.53       &   \\\hline

3 & & 
4.086  &  5.987  &  7.357  &   9.01  &  8.717  &  10.27   &  9.94  &
11.41  &  11.06  &  12.46 \\ & & 
6.732  &  8.463  &  9.298  &  10.83  &  10.44  &  11.89   &  11.5  &
12.88  &  12.49 &   13.81 \\\hline

4 & & 
3.869  &   5.82   &   7.2  &   8.88  &   8.58  &  10.16   &  9.82  &
11.31  &  10.95 &   12.36 \\ & & 
6.568  &  8.341  &  9.169  &  10.72  &  10.32  &  11.79   & 11.39  &
12.79  &  12.39  &  13.73 \\\hline

5 & & 
3.826  &  5.577  &  7.227  &  8.691  &  8.628  &  9.987  &  9.882  &
11.15  &  11.02 &   12.22 \\ & & 
6.573  &  8.164  &  9.218  &  10.56  &  10.38  &  11.65  &  11.46  &
12.66  &  12.47 &    13.6 \\\hline

$\infty$ & & 
3.733  &  5.422  &  7.186  &   8.57  &  8.602  &  9.879  &  9.867  &
11.06  &  11.01 &   12.13 \\ & & 
6.511  &  8.052  &  9.193  &  10.46  &  10.37  &  11.55  &  11.46  &
12.57  &  12.47 &   13.52 \\\hline

\end{tabular}
\caption{The spectrum for adjoint mixing with adjusted weightings at various N.  For each N, the
first mass is the lowest state in that channel, and the second mass is
the radial excitation.}
\label{table:b5ncadj}
\end{table}

\begin{table}[!h]
\centering
\begin{tabular}{|lc|cc|cc|cc|cc|cc|} \hline
N & vertex & $0^{++}$ & $0^{--}$ & $2^{\pm+}$ & $2^{\pm-}$ &
$3^{\pm+}$ & $3^{\pm-}$ & $4^{\pm+}$ & $4^{\pm-}$ & $1^{\pm+}$ &
$1^{\pm-}$ \\\hline

3 & $m_{J^{PC}}$ &
4.456  &  5.825   &  6.95  &  8.884  &  10.16  &  8.106   &  9.15  &
11.31  &  12.36 &   6.678 \\
  & $m_{J^{PC}}^\star$ & 
6.75   & 8.345   & 8.869   & 10.72   & 11.79   & 9.863   & 10.77   &
12.79   & 13.73  &   8.85 \\\hline

4 & & 
4.111  &  5.701  &  6.879  &  8.787  &  10.07  &  8.147  &  9.296  &
11.23  &  12.29  &   7.28 \\ & & 
6.682  &  8.254  &  8.931  &  10.64  &  11.72  &  10.01  &  11.01  &
12.72  &  13.66  &  9.609 \\\hline

5 & & 
3.992  &  5.469  &  6.962  &  8.607  &  9.912  &    8.3  &   9.51  &
11.09  &  12.16  &  7.848  \\ & & 
6.702  &  8.086  &  9.071  &  10.49  &  11.58  &   10.2  &  11.24  &
12.6  &  13.54  &  10.24   \\\hline

\end{tabular}
\caption{The spectrum for vertex mixing with adjusted weightings at various N.  For each N, the
first mass is the lowest state in that channel, and the second mass is
the radial excitation.}
\label{table:b5ncver}
\end{table}

\newpage

\begin{figure}[!t]
\centering
\includegraphics[width=\textwidth]{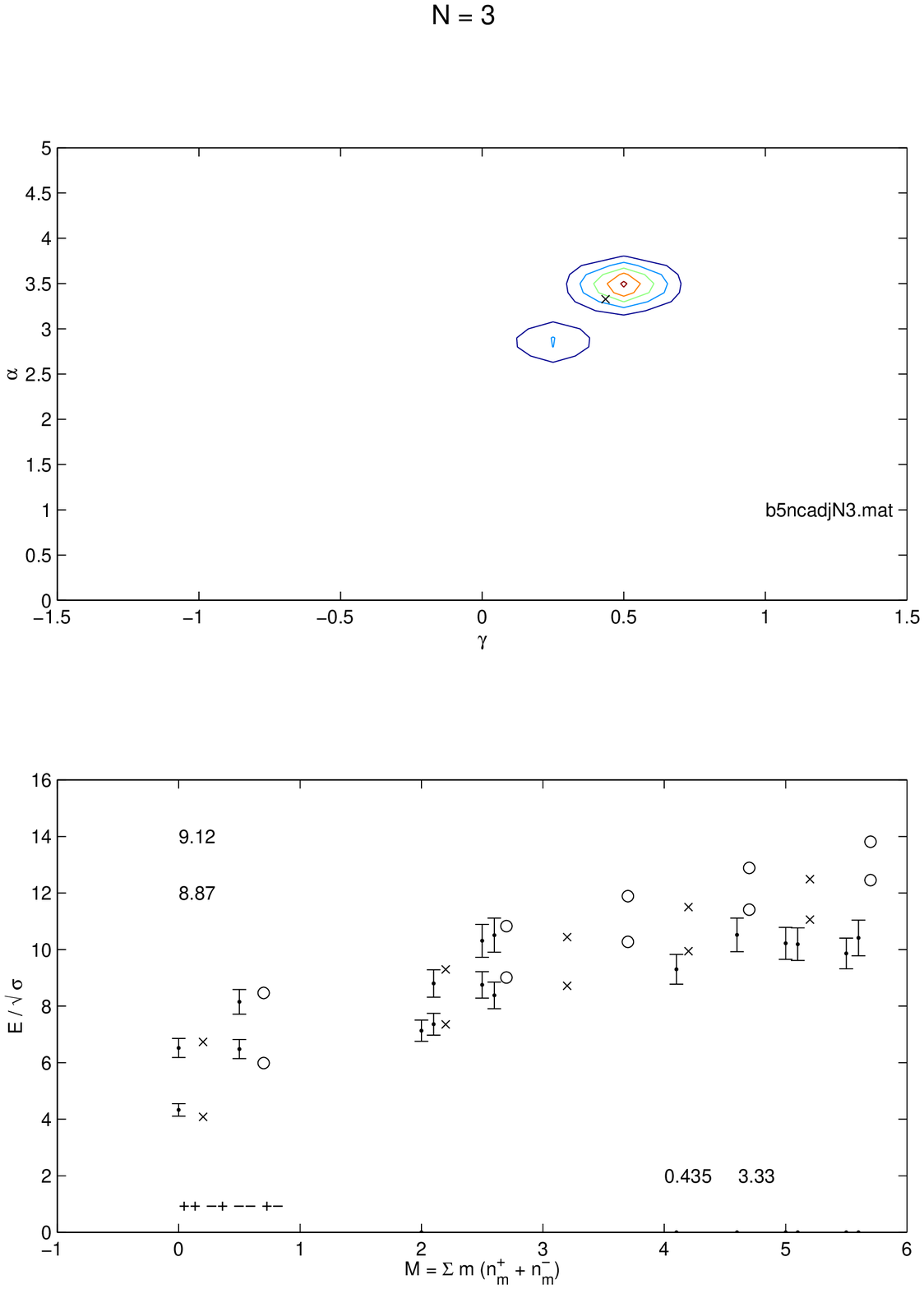}
\caption{The adjoint mixing posterior and spectrum for N=3.}
\label{fig:b5ncadjN3}
\end{figure}

\begin{figure}[!t]
\centering
\includegraphics[width=\textwidth]{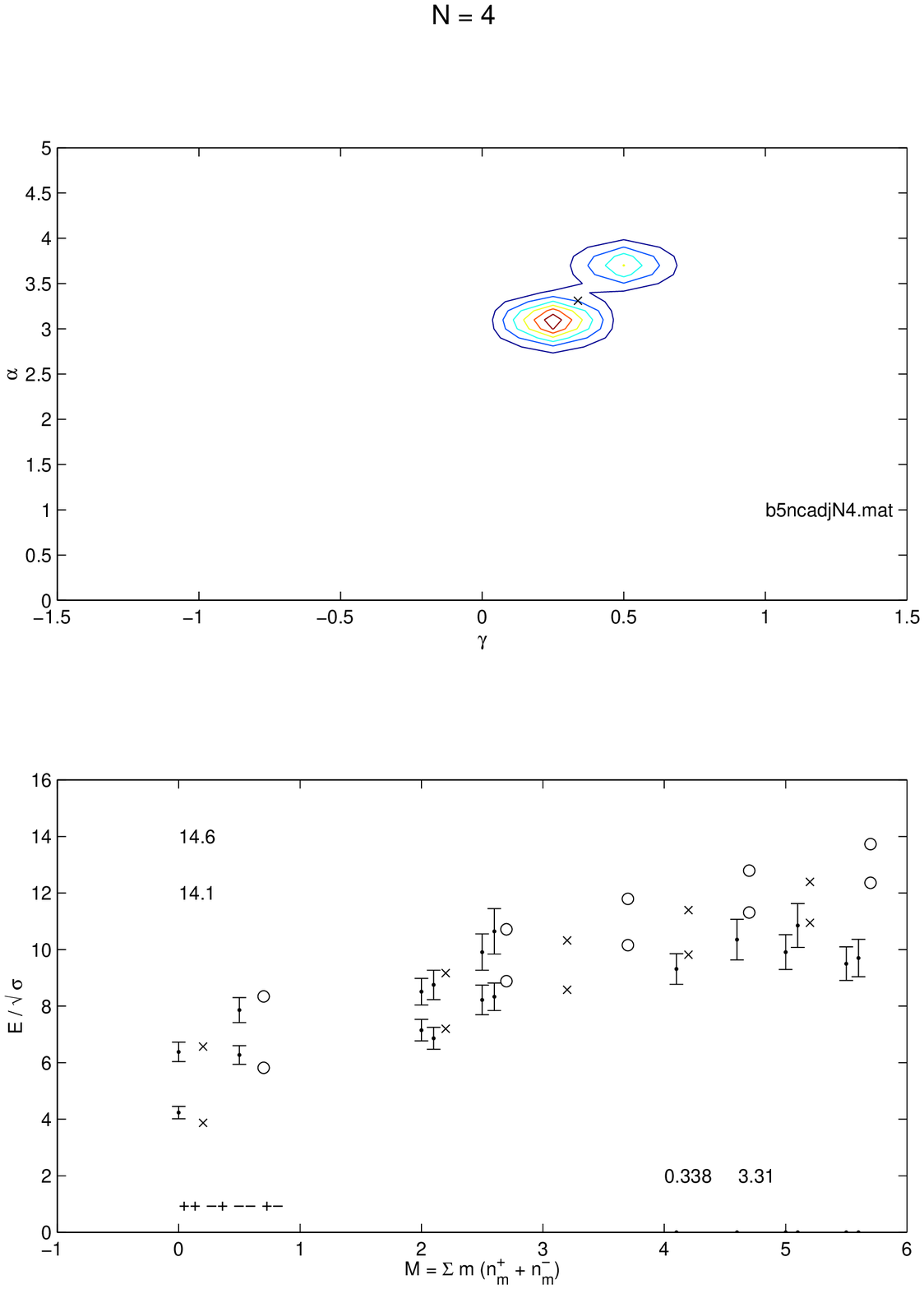}
\caption{The adjoint mixing posterior and spectrum for N=4.}
\label{fig:b5ncadjN4}
\end{figure}

\begin{figure}[!t]
\centering
\includegraphics[width=\textwidth]{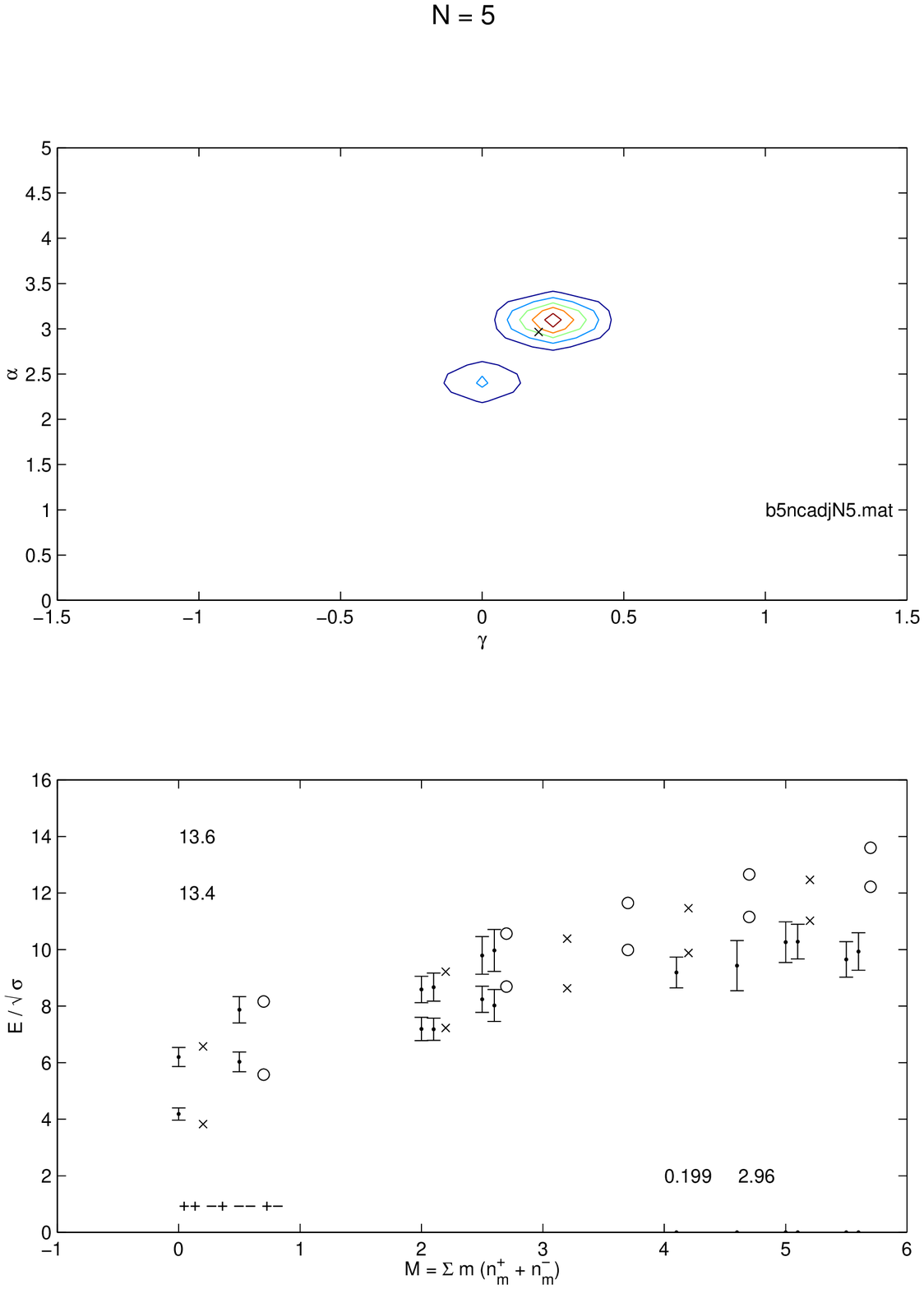}
\caption{The adjoint mixing posterior and spectrum for N=5.}
\label{fig:b5ncadjN5}
\end{figure}

\begin{figure}[!t]
\centering
\includegraphics[width=\textwidth]{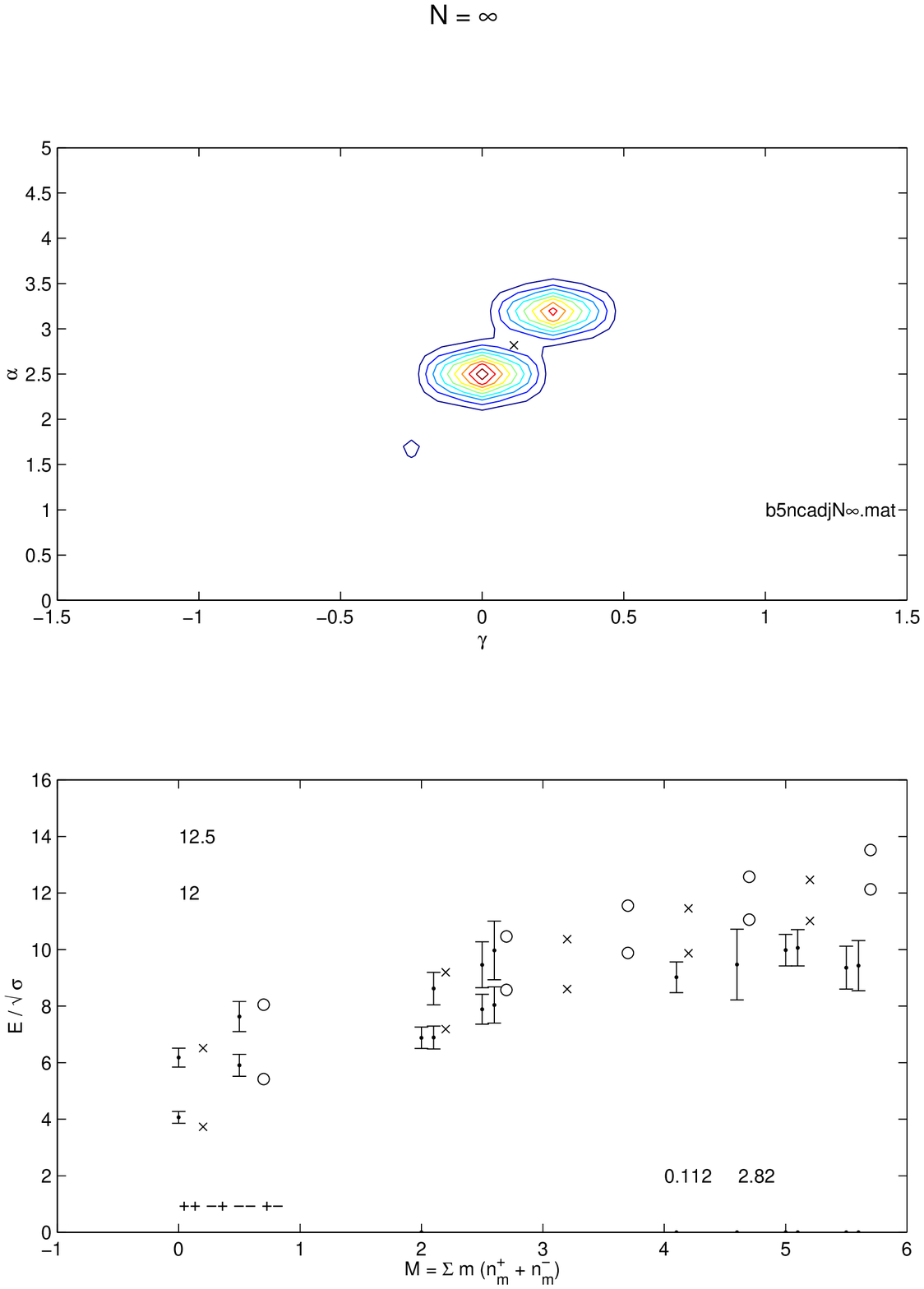}
\caption{The adjoint mixing posterior and spectrum for N=$\infty$.}
\label{fig:b5ncadjN6}
\end{figure}

\begin{figure}[!t]
\centering
\includegraphics[width=\textwidth]{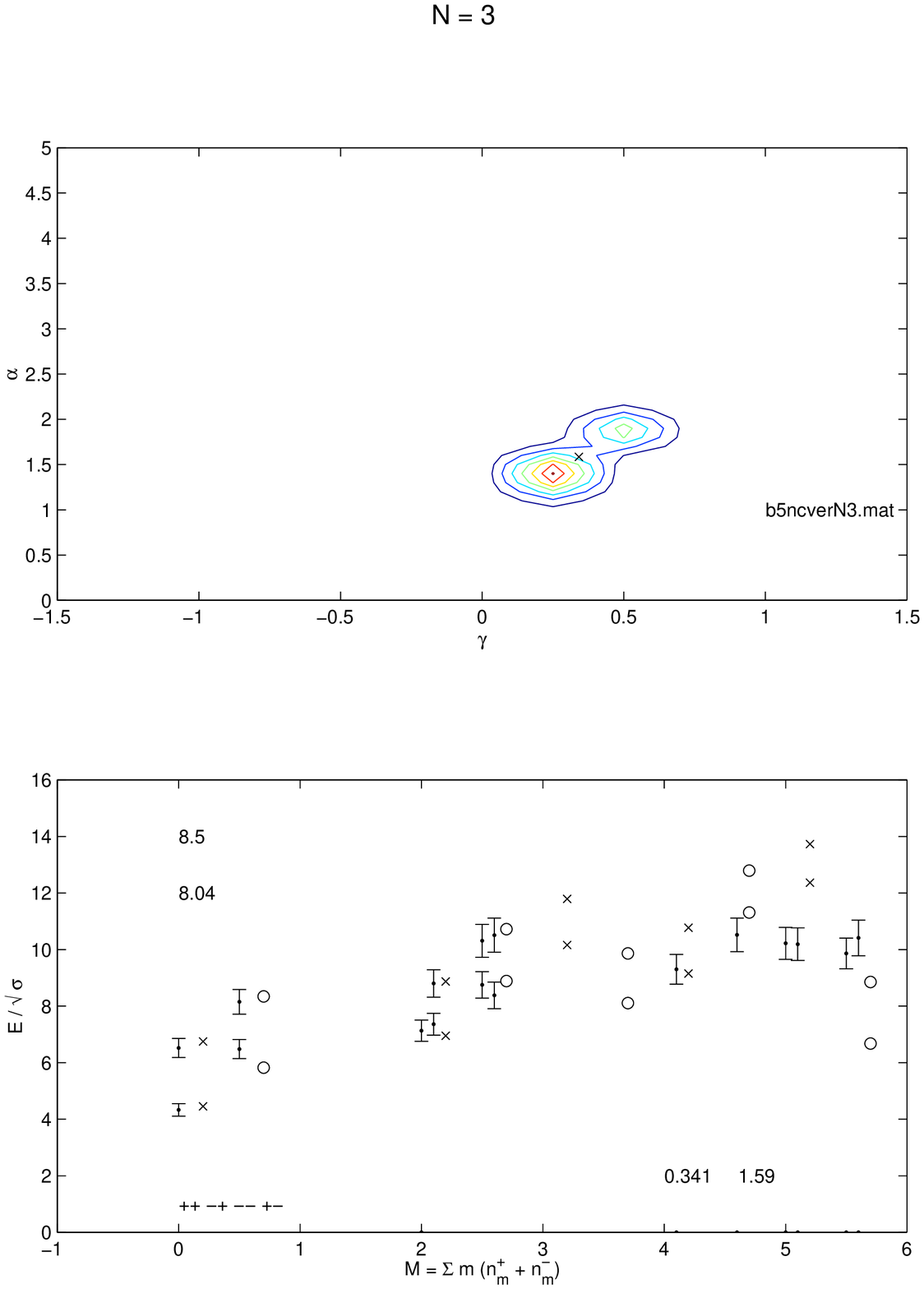}
\caption{The vertex mixing posterior and spectrum for N=3.}
\label{fig:b5ncverN3}
\end{figure}

\begin{figure}[!t]
\centering
\includegraphics[width=\textwidth]{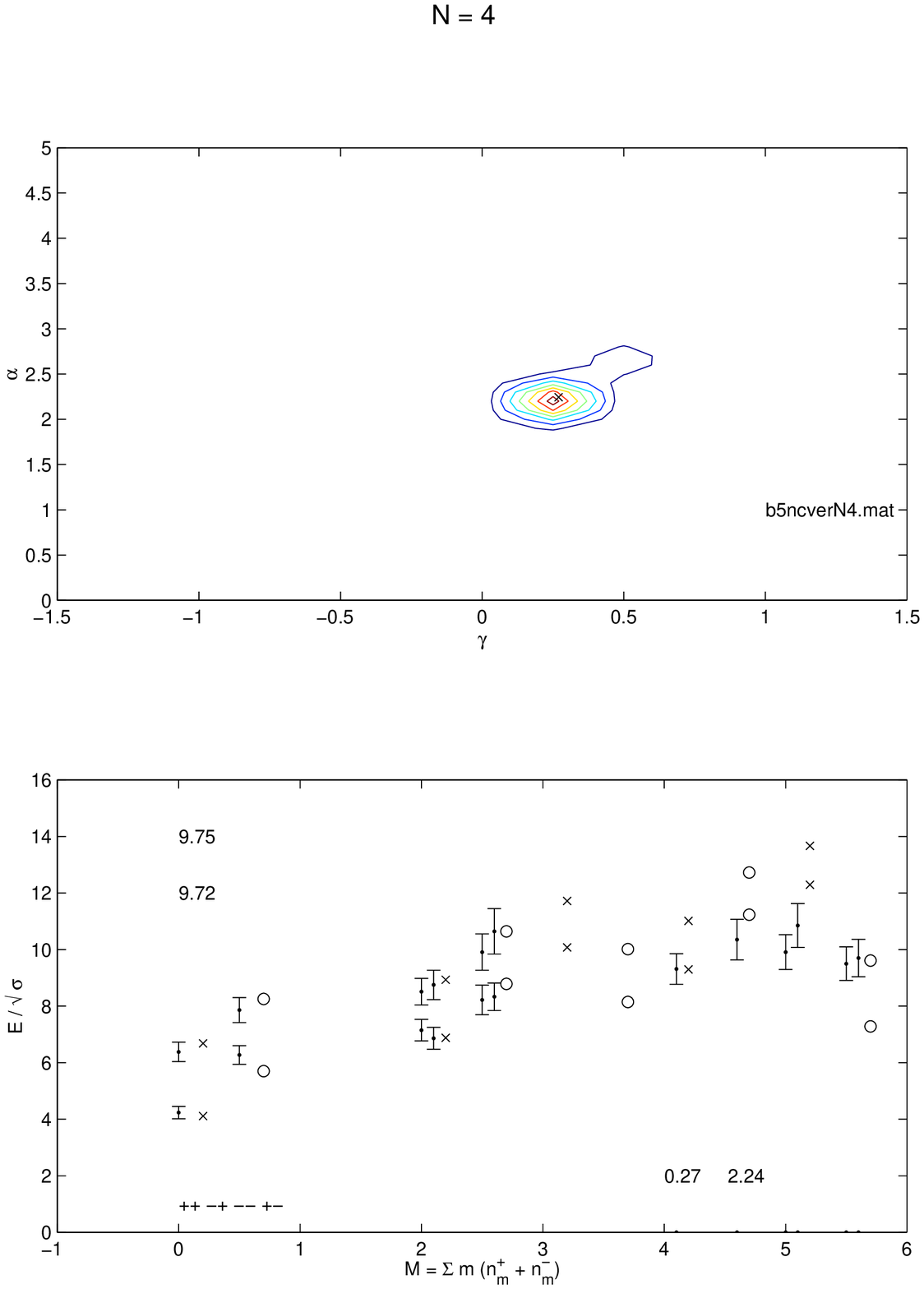}
\caption{The vertex mixing posterior and spectrum for N=4.}
\label{fig:b5ncverN4}
\end{figure}

\begin{figure}[!t]
\centering
\includegraphics[width=\textwidth]{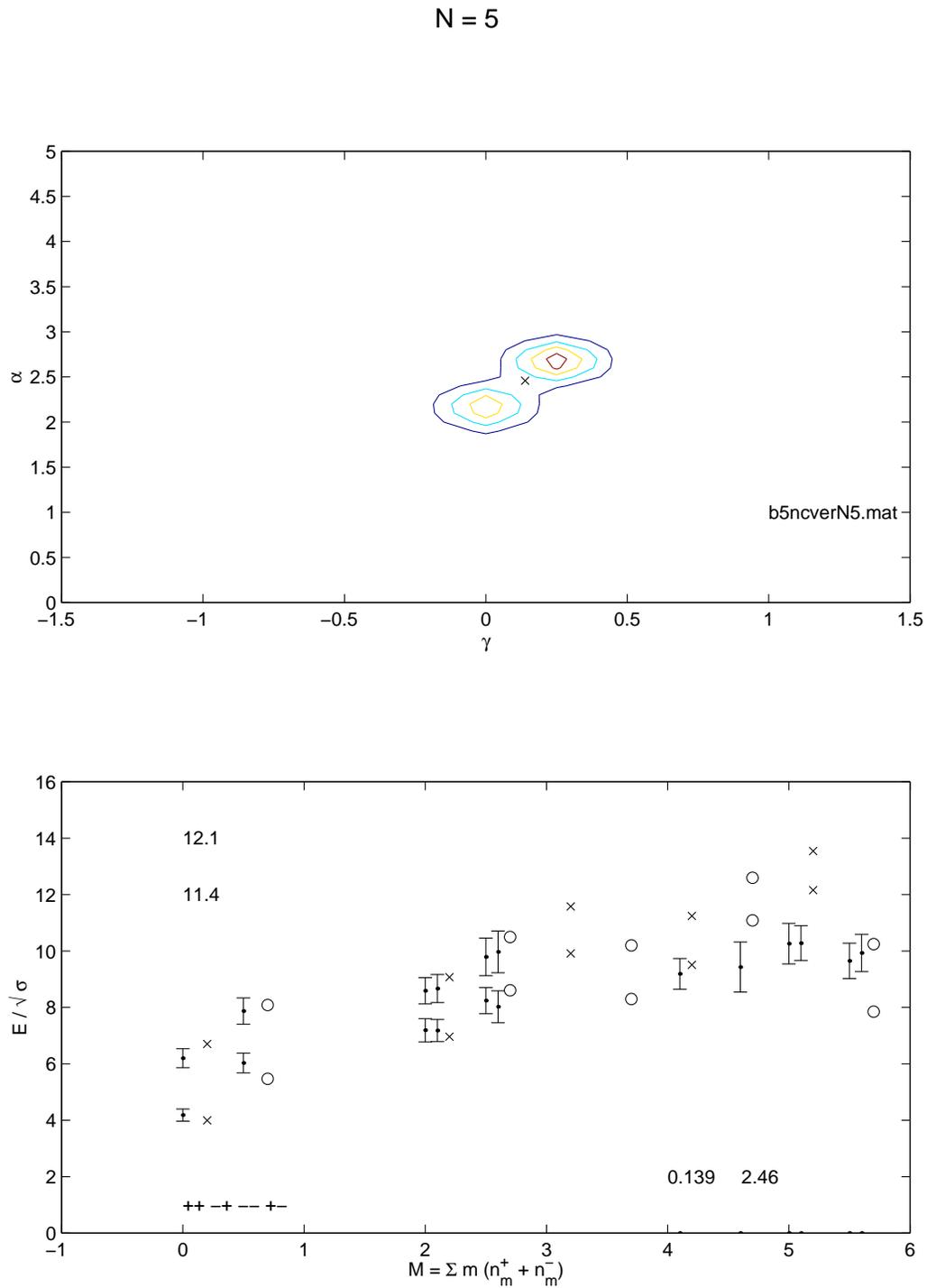}
\caption{The vertex mixing posterior and spectrum for N=5.}
\label{fig:b5ncverN5}
\end{figure}

\chapter{Analysis of the Parameters and the Merit of the Models} \label{chap:N2andchi2}

Following the observations of Teper~\cite{Teper:1997tq}, the value of physical quantities derived from lattice calculations should demonstrate a dependence linear in $(1/N^2)$.  The most obvious example is the masses themselves -- the fits shown with $N=\infty$ are actually fit to the lattice spectrum as extrapolated to $N=\infty$ in~\cite{Teper:1997tq}.  Having lattice data at several values of N is what allows the extrapolation to be made.  As we have collected values for the parameters at several N, we can do the same extrapolation to investigate how well the parameters follow the $(1/N^2)$ dependence.  We take the values at N=3, 4, and 5 when performing the fit for two reasons:  for consistency across the models and so that the values at N=2 are predictive as well as those at $N=\infty$.  In Table~\ref{table:ParametersB0}, the last row for each model is the parameter values as extrapolated by $(1/N^2)|_{N=\infty}$.

\begin{figure}[!t]
\centering
\subfigure[for $\beta = 0$]{
	\label{fig:gammavsN:b0}
	\includegraphics[width=\textwidth]{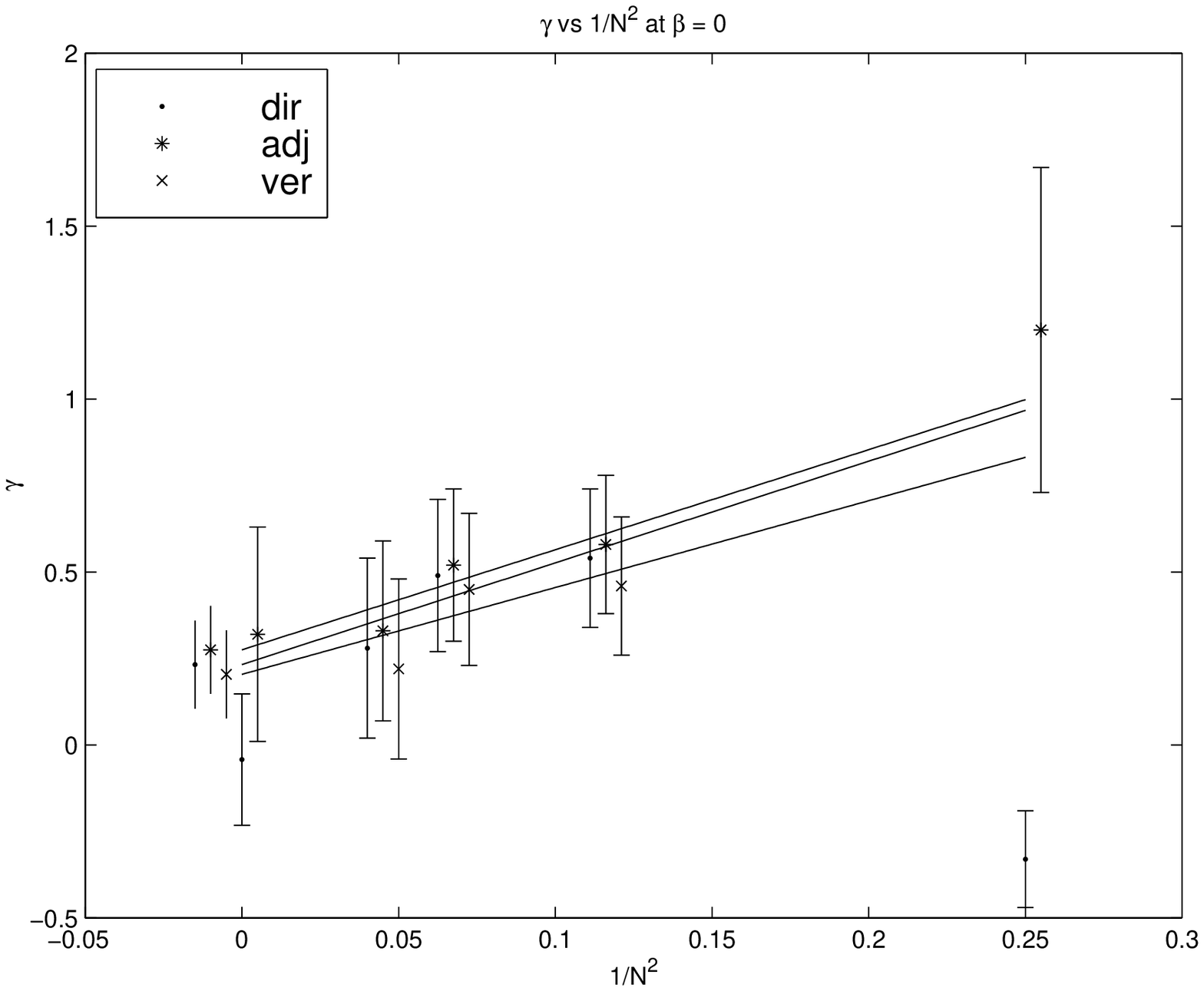}}
\subfigure[for $\beta = .05$]{
	\label{fig:gammavsN:b5}
	\includegraphics[width=\textwidth]{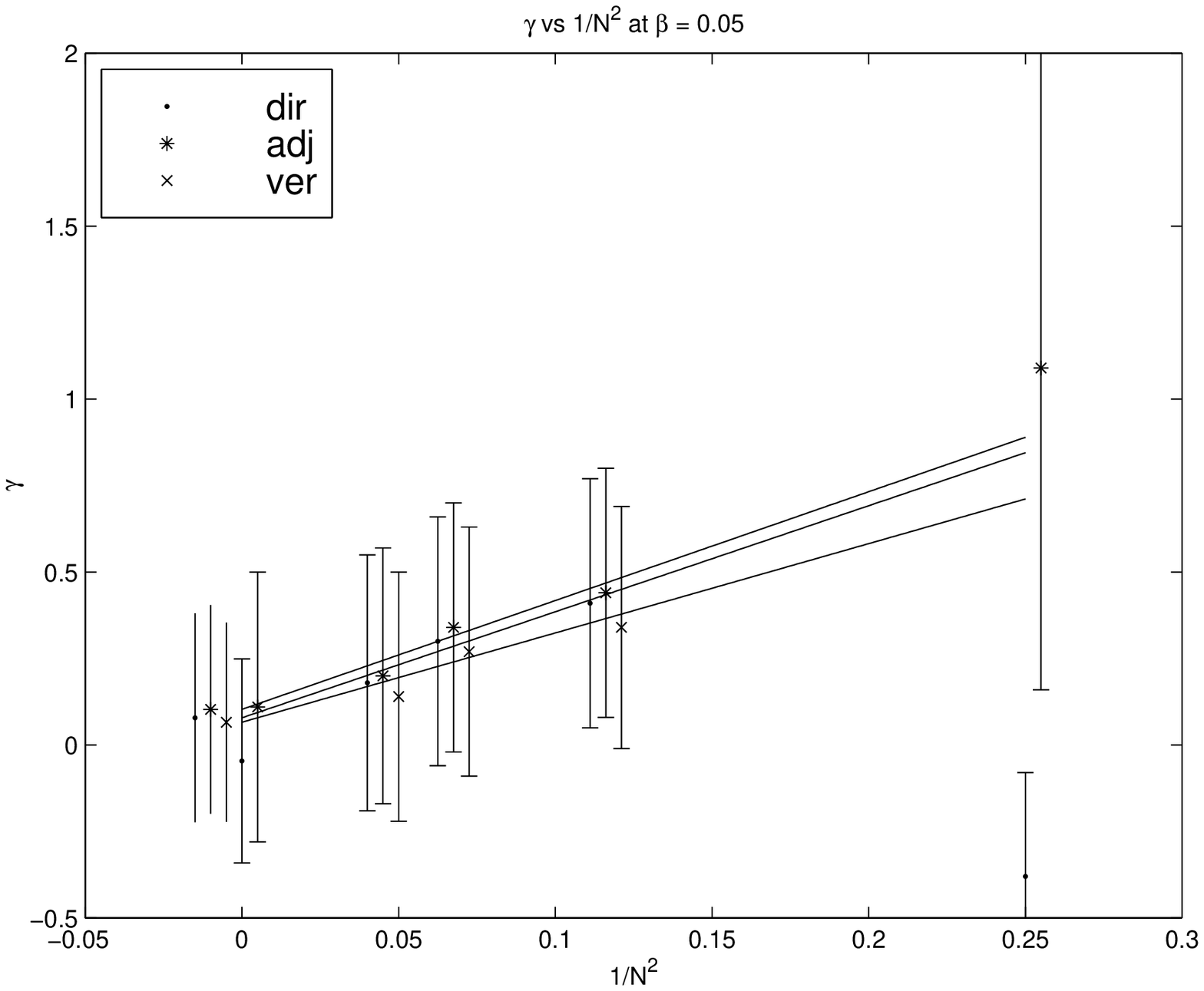}}
\caption{$\gamma$ vs $(1/N^2)$}
\label{fig:gammavsN}
\end{figure}

To perform the extrapolation, we find the best m and c in a least squares sense when minimizing
\beq \sum_{N=3}^5 \frac{(m/N^2 + c - \gamma_N)^2}{\sigma_{\gamma_N}^2} , \eeq
which can be evaluated directly from the data~\cite{Bretthorst:1988,Sivia:1996}.  Figure~\ref{fig:gammavsN:b0} shows the results for $\gamma$ at $\beta = 0$.  Across the range of N, a $(1/N^2)$ dependence is quite believable.  The adjoint model in particular shows a good dependence from N=2 down to $N=\infty$.  The direct model does not agree well at the endpoints, but the estimates of $\gamma_{N=\infty}$ from all three models agree within errors at $\gamma \sim .25$.  For $\beta=.05$, Figure~\ref{fig:gammavsN:b5}, much of the same behavior is observed.  The value of $\gamma_\infty$ drops slightly to be $\sim .1(3)$, which is consistent with zero within errors.  As we are seeing $(1/N^2)$ dependence, we should believe that we are measuring a physical parameter.

\begin{figure}[!t]
\centering
\subfigure[for $\beta = 0$]{
	\label{fig:alphavsN:b0}
	\includegraphics[width=\textwidth]{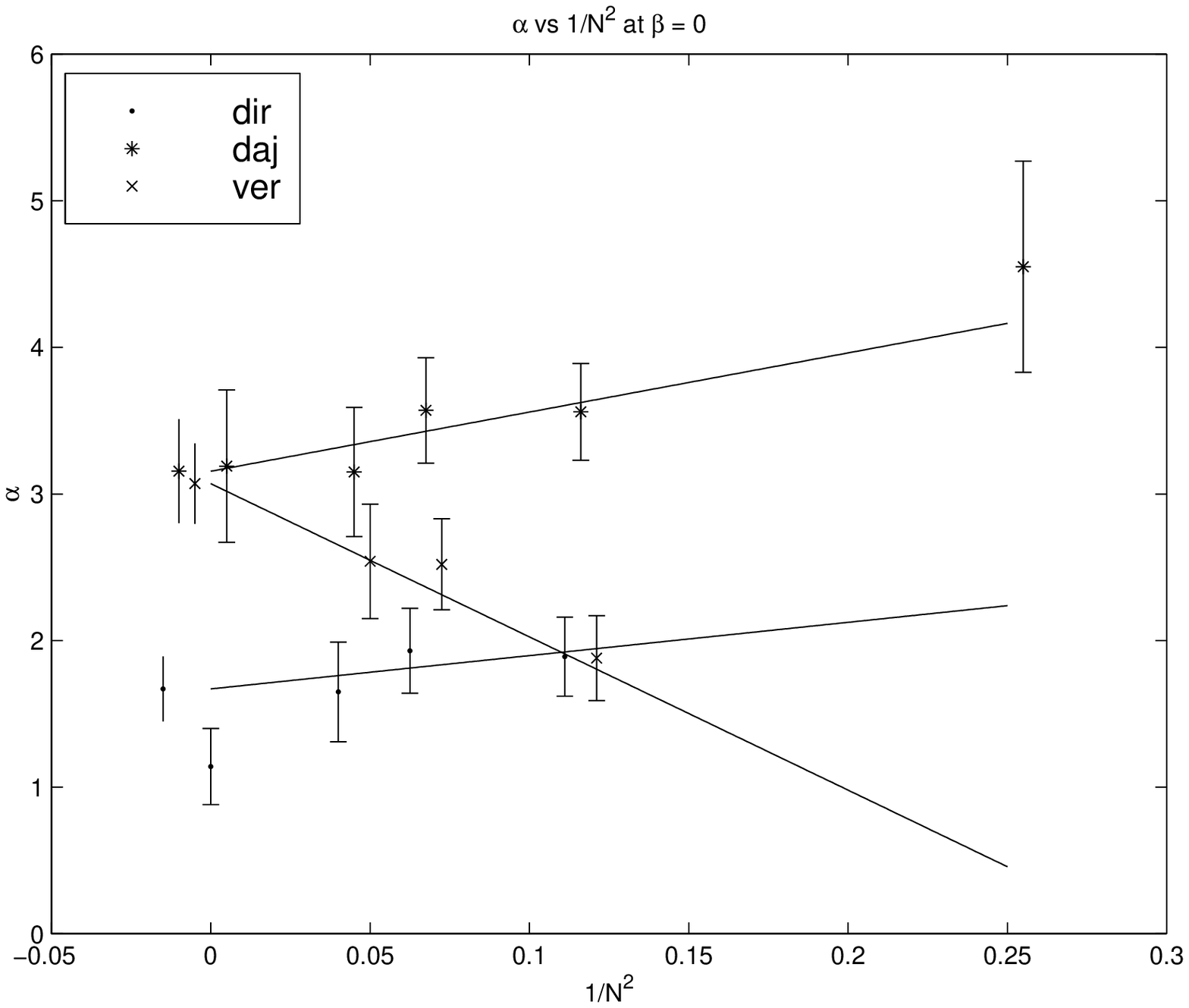}}
\subfigure[for $\beta = .05$]{
	\label{fig:alphavsN:b5}
	\includegraphics[width=\textwidth]{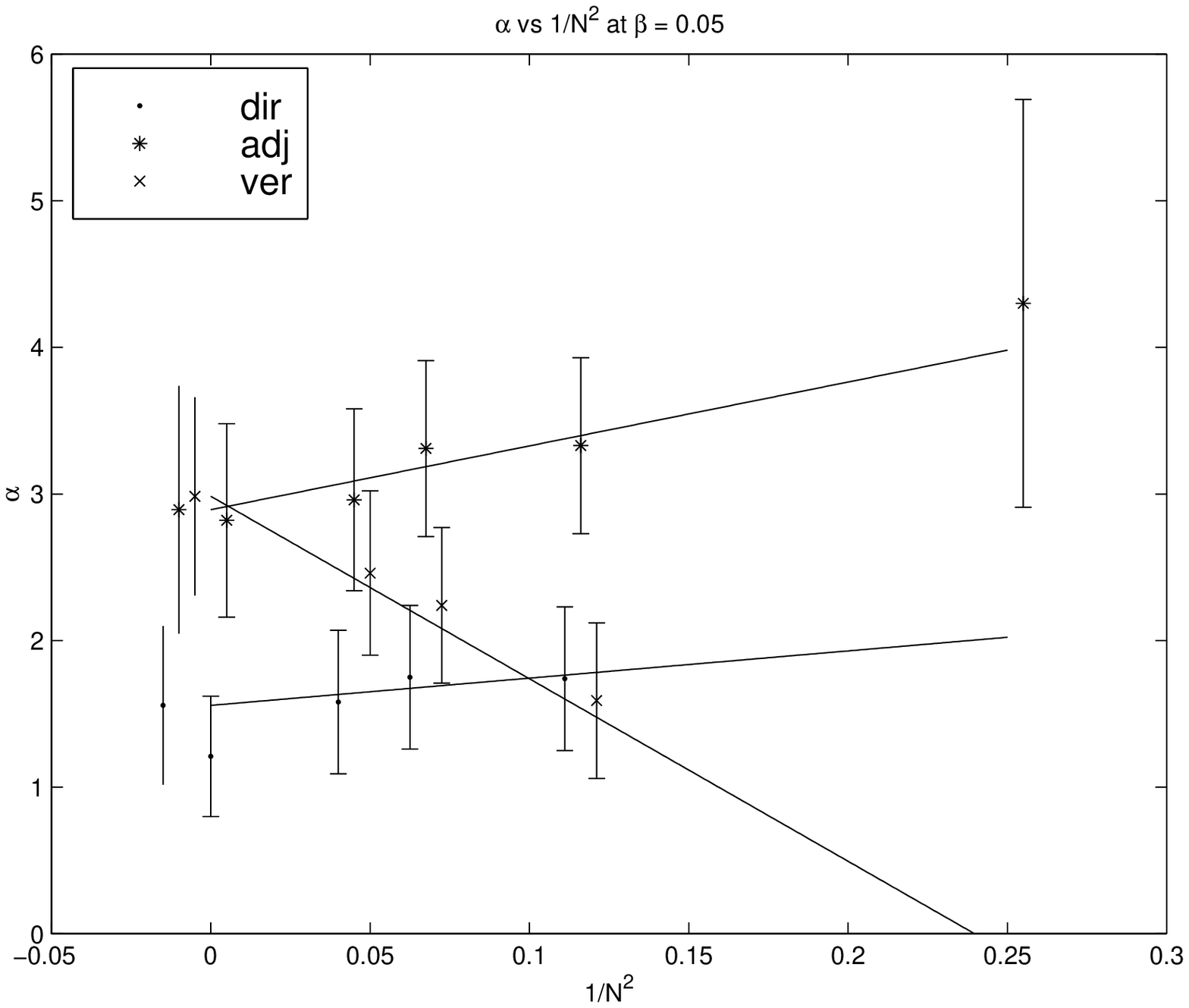}}
\caption{$\alpha$ vs $(1/N^2)$}
\label{fig:alphavsN}
\end{figure}

For the mixing parameter $\alpha$, we see more variety among the models.  At $\beta=0$, third column of Table~\ref{table:ParametersB0} and Figure~\ref{fig:alphavsN:b0}, we see $(1/N^2)$ dependence, but the two indirect mixing models have slopes of opposite sign.  However, the two models reach a common value at $N=\infty$ of $\gamma_0\sim 3$.  That these two alternatives reach agreement at the $N\rightarrow\infty$ limit suggests that this form of mixing mechanism is physically viable.  If these more exotic states exist in the pure gauge theory, we should construct a more generalized model {\it{a la}}~\cite{Schwinger:1964}, with a Hamiltonian
\beq H = \left[\begin{array}{cccc} H_+ & 0 & \alpha_1 & \alpha_3 \\ 0 & H_- & \alpha_2 & \alpha_4 \\ \alpha_1 & \alpha_2 & H_V & \alpha_5 \\ \alpha_3 & \alpha_4 & \alpha_5 & H_A \end{array}\right] , \eeq
but that work has yet to be done.  The direct mixing mechanism gives a reasonable $(1/N^2)$ fit, with an intercept lower than for the indirect mixing mechanisms.  As indirect mixing is second order in $\alpha$, and the direct mixing is first order, it seems plausible that $\alpha_\mathrm{direct}\approx \sqrt{\alpha_\mathrm{indirect}}$ at $N=\infty$.  When $\beta=.05$, the magnitude of the parameters decreases slightly, but the same behavior is observed (see Figure~\ref{fig:alphavsN:b5}).

\clearpage

\begin{table}[!t]
\centering
\begin{tabular}{|r|rl|rl|c|} \hline
N & $\gamma$ & $(\sigma_\gamma)$ & $\alpha$ & $(\sigma_\alpha)$ & $(\sigma_{\gamma \alpha})$ \\\hline
direct & & & & & \\
2 & -.33 & (14) & & & \\
3 & 0.54 & (20) & 1.89 & (27) & (23) \\
4 & 0.49 & (22) & 1.93 & (29) & (24) \\
5 & 0.28 & (26) & 1.65 & (34) & (29) \\
$\infty$ & -.042 & (190) & 1.14 & (26) & (20) \\
$(1/N^2)|_\infty$ & 0.23 & (13) & 1.67 & (22) & \\\hline
adjoint & & & & & \\
2 & 1.20 & (47) & 4.55 & (72) & (58) \\
3 & 0.58 & (20) & 3.56 & (33) & (26) \\
4 & 0.52 & (22) & 3.57 & (36) & (28) \\
5 & 0.33 & (26) & 3.15 & (44) & (34) \\
$\infty$ & 0.32 & (31) & 3.19 & (52) & (40) \\
$(1/N^2)|_\infty$ & 0.28 & (13) & 3.15 & (36) & \\\hline
vertex & & & & & \\
3 & 0.46 & (20) & 1.88 & (29) & (24) \\
4 & 0.45 & (22) & 2.52 & (31) & (25) \\
5 & 0.22 & (26) & 2.54 & (39) & (31) \\
$(1/N^2)|_\infty$ & 0.20 & (13) & 3.07 & (27) & \\\hline

\end{tabular}
\caption{Parameter values for $\beta$ = 0.}
\label{table:ParametersB0}
\end{table}

\begin{table}[!b]
\centering
\begin{tabular}{|r|rl|rl|c|} \hline
N & $\gamma$ & $(\sigma_\gamma)$ & $\alpha$ & $(\sigma_\alpha)$ & $(\sigma_{\gamma \alpha})$ \\\hline
direct & & & & & \\
2 & -.38 & (30) & & & \\
3 & 0.41 & (36) & 1.74 & (49) & (40) \\
4 & 0.30 & (36) & 1.75 & (49) & (40) \\
5 & 0.18 & (37) & 1.58 & (49) & (40) \\
$\infty$ & -.046 & (295) & 1.21 & (41) & (29) \\
$(1/N^2)|_\infty$ & 0.079 & (302) & 1.56 & (54) & \\\hline
adjoint & & & & & \\
2 & 1.09 & (93) & 4.30 & (1.39) & (1.14) \\
3 & 0.44 & (36) & 3.33 & (60) & (45) \\
4 & 0.34 & (36) & 3.31 & (60) & (45) \\
5 & 0.20 & (37) & 2.96 & (62) & (46) \\
$\infty$ & 0.11 & (39) & 2.82 & (66) & (50) \\
$(1/N^2)|_\infty$ & 0.10 & (30) & 2.89 & (84) & \\\hline
vertex & & & & & \\
3 & 0.34 & (35) & 1.59 & (53) & (41) \\
4 & 0.27 & (36) & 2.24 & (53) & (42) \\
5 & 0.14 & (36) & 2.46 & (56) & (43) \\
$(1/N^2)|_\infty$ & 0.066 & (302) & 2.98 & (68) & \\\hline

\end{tabular}
\caption{Parameter values for $\beta$ = .05.}
\label{table:ParametersB5}
\end{table}

\clearpage

How do we evaluate the merit of our models?  While rules of thumb abound, the rules of inferential statistics~\cite{Bretthorst:1988,Sivia:1996} highlight the difficulties.  Writing the posterior for the probability of model M given a set of data D with errors $\sigma$ as
\beq p(M|D,\sigma) = \frac{p(D,\sigma|M)p(M)}{p(D,\sigma)} , \eeq
the critical factor $p(D,\sigma)$ (named the ``evidence'') is incalculable.  What is the value of ``the probability that {\it{this particular}} data set was drawn from the myriad possibilities of the universe''?  When certain simplifying assumptions are met, the likelihood should follow the statistical noise, $\chi^2 \sim N$, but with few data points and/or nonnormal (systematic) errors, the magnitude of a ``believable'' $\chi^2$ becomes hard to calculate.  Of course, the right answer to the question ``Is this a good model?'' is the question ``Compared to what?''  Naturally, without alternatives we should believe the only model we have!  This ambiguity disappears when we compare the {\it{relative}} probabilities of two (or more) models
\beq \frac{p(M_1|D,\sigma)}{p(M_2|D,\sigma)} = \frac{p(D,\sigma|M_1)}{p(D,\sigma|M_2)} \times \frac{p(M_1)}{p(M_2)}, \eeq
as the evidence cancels between the numerator and the denominator.  The second factor is the ratio of priors, quantifying any prejudice on the models we might wish to incorporate and usually set equal to unity.  So, to compare two models we should compare the (exponential) of their $\chi^2$.  In detail, there are also factors arising from the priors (which cancel in our case, being equivalent) as well as the breadth of the posterior as embodied in $\sigma_\gamma$ and $\sigma_\alpha$.
\beq \frac{p(M_1|D,\sigma)}{p(M_2|D,\sigma)} = \frac{\sigma_{\gamma 1} \sigma_{\alpha 1}}{\sigma_{\gamma 2} \sigma_{\alpha 2}} \exp{\left[-\half (\chi^2_1 - \chi^2_2)\right] } \eeq
If we want the factor by which one model is more likely than another, we can take the exponential of the difference in $\chi^2$.  For our purposes, we can look directly at $\chi^2$, our merit function, Table~\ref{table:chi2}.

\begin{table}[t]
\centering
\begin{tabular}{|r|c|c|} \hline
N & $\beta = 0$ & $\beta = .05$ \\\hline
direct & & \\
2 & 49.1 & 3.39 \\
3 & 55.9 & {\bf{6.79}} \\
4 & 51.5 & 9.9 \\
5 & 43.6 & 9.56 \\
$\infty$ & 24.2 & 8.42 \\\hline
adjoint & & \\
2 & 8.43 & .777 \\
3 & 78.2 & 8.87 \\
4 & 79.1 & 14.1 \\
5 & 72.6 & 13.4 \\
$\infty$ & 47.7 & 12 \\\hline
vertex & & \\
3 & 85.3 & 8.04 \\
4 & 53.1 & 9.72 \\
5 & 55.5 & 11.4 \\\hline
\end{tabular}
\caption{$\chi^2$ for the Isgur-Paton model}
\label{table:chi2}
\end{table}

The first column is for $\beta = 0$.  The jump from N=2 to N=3 for adjoint and direct mixing results from the addition of more states to the fit.  The reduction in $\chi^2$ for N=$\infty$ is because of the larger errorbars on the extrapolated lattice data.  Looking at the physically interesting case of N=3, the direct mixing comes in with $\chi^2 = 55.9$, while the adjoint has $\chi^2 = 78.2$ and vertex mixing has $\chi^2 = 85.3$.  Thus we must conclude that the data support direct mixing over indirect mixing by a relative probability of $\sim exp(-55.9 +80) \approx 3\times 10^{10}$.  Looking at the N=$\infty$ fits, while the magnitude of the $\chi^2$ changes, the relative difference $\sim -24.2 + 50$ is about the same.  Comparing the indirect mechanisms, we note that for N=3, 4, and 5 the adjoint's $\chi^2$ is between 70 and 80, while the vertex method's $\chi^2$ drops to $\sim 50$ for N=4 and 5.  At N=4, the vertex and the direct mechanisms have $\chi^2$ different by only 2.

After we have adjusted the weightings by $\beta = .05$, we get $\chi^2$ at a much more familiar magnitude.  Most of the fits have 12 data points -- some have 11.  With two parameters, that leaves $\sim 10$ degrees of freedom. We can now apply our rule of thumb $\chi^2 \sim \mathrm{d.o.f.}$ to the question ``Do we believe this model can explain the data at a 10\% level?''  Looking again at Table~\ref{table:chi2}, and ignoring N=2, the lowest $\chi^2$ is for direct mixing at N=3, equal to 6.79.  As the effect of finite $\beta$ is to de-weight the $0^{++}$, we should conclude that the Isgur-Paton model can explain both the C = + and C = -- sectors of the lattice data to a reasonable level of agreement.  For higher N, the indirect mixings do not fit as well as the direct mechanism, and the vertex mechanism is slightly better than the adjoint.  Even at the 10\% level, the indirect mechanisms cannot quite be said to be ``believable'', though they do capture certain qualitative features of the spectrum.  Trying to capture the complicated dynamics in a simple nonrelativistic semi-classical Hamiltonian might be an impossible task.

We might try to improve the model by making it relativistically correct by including the chromo-magnetic contributions.  Another avenue is to explore what happens as $r\rightarrow 0$ and the string picture breaks down.  Shedding light here is a recent study of the dual Ginzburg-Landau flux ring~\cite{Koma:1999sm}.  The dGL picture might be applicable to the QCD vacuum, if the conditions of Abelian dominance and monopole condensation hold.  Defining their effective string tension as the total rest energy of a solution of the differential equations divided by its circumference, those authors find
\beq \sigma_\mathrm{eff} \rightarrow 0 \; \mathrm{as} \; r \rightarrow 0 . \eeq
The traditional Isgur-Paton model has the constant $\sigma \equiv 1 \: \forall r$, thus these two pictures have radically different viewpoints on the properties of a flux tube at small radius.

\chapter{Analysis II:  Including $\sigma \rightarrow \sigma_\mathrm{eff} $ } \label{chap:analysis2}

One problem for the Isgur-Paton model is understanding what happens as $r \rightarrow 0$. The string picture of a flux tube is only appropriate on scales where the radius is much larger than the thickness of the flux tube $\sim {1\over \sqrt{\sigma}}$.  With the radius $\sim {1\over \sqrt{\sigma}}$, the flux tube would look more like an annulus in two spatial dimensions or a torus in three dimensions, and the $M/r$ potential for the phonon contribution seems unreasonable.  Indeed, the fluctuations might not even be able to be described by a phonon potential.  A detailed theory of membrane dynamics might address the problem, but such is beyond the scope of this work.

\begin{figure}[h]
\centering
\includegraphics[width=\textwidth]{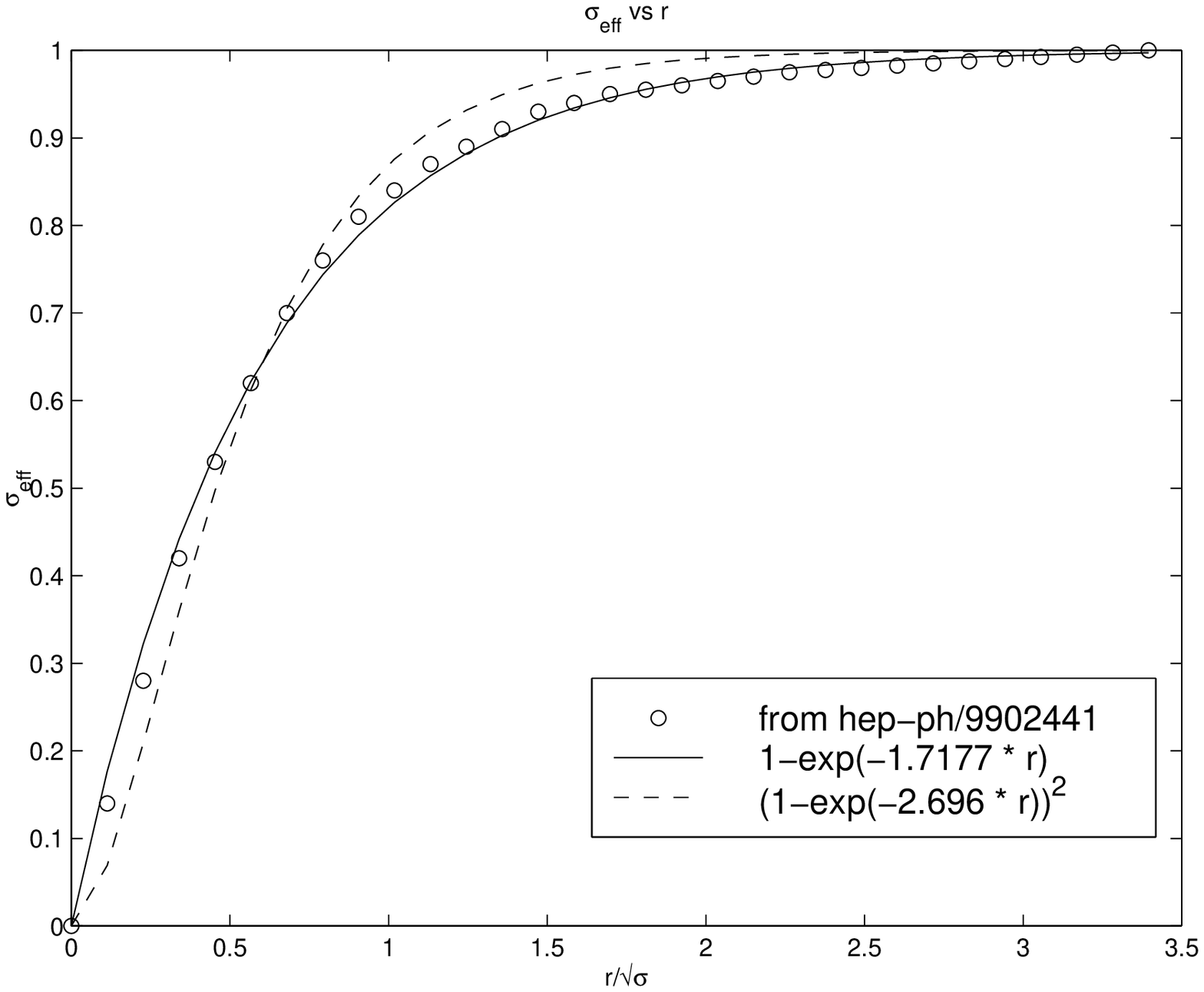}
\caption{$\sigma_\mathrm{eff}$ vs r}
\label{fig:sigeff}
\end{figure}

What if as $r \rightarrow 0$, the string itself ``dissipated''?  A study of the dual Ginzburg-Landau theory of type-II superconductors~\cite{Koma:1999sm} determined a function $\sigma_\mathrm{eff}(r)$, Figure~\ref{fig:sigeff}, defined to be the total energy of a flux ring solution of radius r divided by the circumference.  The total energy was calculated from the microscopic differential equations for the fields.  The non-abelian gauge theory's flux tube is often thought to arise from a dual Meissner effect for a type-II superconducting picture of the QCD vacuum~\cite{Mandelstam:1973,'tHooft:1981ht}.  As $\sigma$ is defined from the interquark potential as $r \rightarrow \infty$, its behavior as $r \rightarrow 0$ is uncertain, and we might apply $\sigma \rightarrow \sigma_\mathrm{eff}(r)$ to the Isgur-Paton model.  Transcribing the function from Figure~\ref{fig:sigeff} and converting scales, we can fit the function with the form
\beq \sigma_\mathrm{eff} = 1 - e^{-f r} \label{eqn:sigeff} \eeq
for f=1.7177, best in a least squares sense.  Knowing the functional form of $\sigma_\mathrm{eff}(r)$, we can investigate its effect on the Isgur-Paton model.

\begin{figure}[t]
\centering
\includegraphics[width=.5\textwidth]{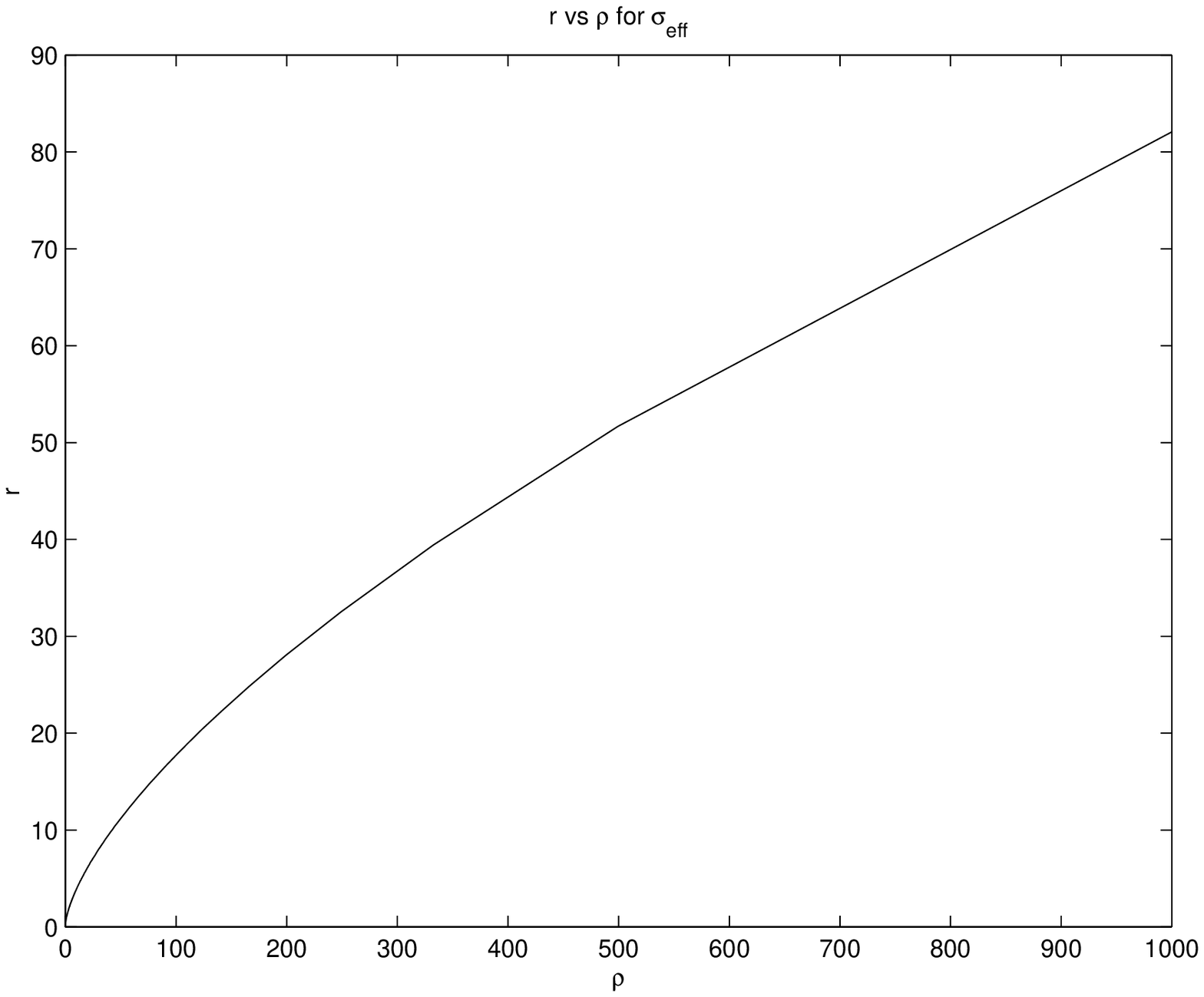}
\caption{r vs $\rho$}
\label{fig:rvsrho}
\end{figure}

When we quantized the model earlier, the difficult term was the kinetic energy, so that is where we will begin.  Abusing notation slightly,
\bea T &=& \frac{P_r^2}{2 \mu} \\
 &\rightarrow& \frac{-1}{4 \pi} \left( \frac{\frac{\pa}{\pa r}}{\sqrt{\sigma r}}\right)^2 , \eea
where $\sigma$ is now included in the r dependence of the kinetic energy.  If $\sigma$ were still a constant, we would get (as before)
\beq T \rightarrow \frac{-1}{4 \pi \sigma_\mathrm{const}} {9\over 4} \frac{\pa^2}{\pa \rho^2} , \eeq
since $d\tilde{\rho} = r^{\half} dr$, $\tilde{\rho} = {2\over 3} r^{3\over 2}$.  Now we must include $\sigma_\mathrm{eff}$ in our differential equation relating r and $\rho$.  For some canonical variable $\rho$,
\beq \frac{\frac{d\rho}{dr}\frac{d}{d\rho}}{\sqrt{\sigma_\mathrm{eff} r}} = \frac{d}{d\rho} , \eeq
so
\beq d\rho = dr \sigma_\mathrm{eff}^\half r^\half \eeq
with boundary conditions
\beq \rho(0)=0 , \; \rho(1) = 1 . \eeq
Substituting our function for $\sigma_\mathrm{eff}$, we find
\beq \rho = \int dr r^\half \left( 1 - e^{-f r} \right)^\half . \eeq
This integral cannot be put into closed form, yet we need the antiderivative explicitly to substitute into the potential $V(r(\rho))$.  While it would be nice to include explicitly the function from Equation~\ref{eqn:sigeff}, any similar function which captures the salient features should suffice to test its influence on the spectrum.  We need a function which goes to zero as $r \rightarrow 0$ and which approaches unity asymptotically, and which is integrable with respect to the measure $r^\half dr$.  After looking at several functional forms, the closest one meeting these criteria is
\beq \sigma_\mathrm{eff} = \left( 1 - e^{-f' r} \right)^2 , \eeq
where f' is now given by 2.696.  Performing the integral (and dropping the prime)
\beq \int dr r^\half \left( 1 - e^{-f r} \right) = \frac{\sqrt{r}}{f} \left(e^{-f r} + {2\over 3} f r \right) - \frac{\sqrt{\pi}}{2 f^{3\over 2}} \mathrm{erf}\sqrt{f r} , \eeq
thus
\beq \tilde{\rho} = {2\over 3} r^{3\over 2} + \frac{\sqrt{r}}{f}e^{-f r} - \frac{\pi}{2 f^{3\over 2}} \mathrm{erf}\sqrt{f r} . \eeq
Applying the condition $\rho(r=1) = 1$,
\beq \tilde{\rho}(1) = c = 0.495544 , \eeq
so
\beq \rho = {1\over c}\tilde{\rho} = \frac{\sqrt{r}}{f c} \left( e^{-f r} + {2\over 3} f r \right) - \frac{\sqrt{\pi}}{2 c f^{3\over 2}} \mathrm{erf}\sqrt{f r} . \label{eqn:rvrho} \eeq
We cannot find the functional inverse explicitly, but as ultimately we are working with a discrete set $\{\rho_j\}$ related to $\{t_j\}$, we can solve the equation numerically to give $r_j$,
\beq \rho_j = \frac{n+1}{j} \Rightarrow r_j , \eeq
which we can put into the potential
\beq V = V(r(\rho_j)) . \eeq
The solution of Equation~(\ref{eqn:rvrho}) is shown in Figure~\ref{fig:rvsrho}.
Returning to the kinetic energy,
\beq T \rightarrow \frac{-1}{4 \pi} {1\over c^2} \frac{\pa^2}{\pa \rho^2} , \eeq
so our Hamiltonian goes to
\bea T + V = E &\Rightarrow& \\
\frac{-\pa^2}{\pa \rho^2} + 4 \pi c^2 V(r(\rho)) &=& 4 \pi c^2 E = \lambda , \eea
which we transform as before to the matrix equation 
\beq S_b + S_d = \lambda \eeq
and solve.

So, how does this model perform?  Looking at Table~\ref{table:sigchi2}, we see the values of $\chi^2$ for various N at $\beta =0$.  For the direct mechanism, $\chi^2$ decreases from 117 at N=3 down to 15.3 for N=$\infty$.  This N dependence of $\chi^2$ seems a little odd.  For small N, $\chi^2$ is too large to believe the model including the effective string tension, but as $N\rightarrow \infty$, $\chi^2$ drops to a value which is nearly believable, even with the tight errors.  As the details of the conjugate splitting might depend on N, we are seeing evidence that the Isgur-Paton model is most applicable to the $N\rightarrow \infty$ limit.

Looking now at the indirect mechanisms, we see the same type of behavior or $\chi^2$.  For the adjoint mechanism, $\chi^2$ ranges from 135 for N=3 down to 26.5 for N=$\infty$.  Apparently, having $\sigma_\mathrm{eff} \rightarrow 0$ works reasonably well for $N\rightarrow \infty$, but for small N including such behavior is detrimental to the model.  For vertex mixing, $\chi^2$ simply remains unbelievably large, $\gtrsim 100$.  As this mechanism has an extra flux tube for its diameter, imposing $\sigma_\mathrm{eff} \rightarrow 0$ is a major complication to the model and is not supported by the evidence.

\begin{table}
\centering
\begin{tabular}{|l|ccccc|} \hline
N & 2 & 3 & 4 & 5 & $\infty$ \\\hline
direct & & 117 & 79.3 & 36.5 & {\bf{15.3}} \\\hline
adjoint & 13.2 & 135 & 101 & 81.1 & 26.5 \\\hline
vertex & & 257 & 130 & 96.4 & \\\hline
\end{tabular}
\caption{$\chi^2$ for $\sigma_\mathrm{eff}$}
\label{table:sigchi2} 
\end{table}

Comparing the values of $\chi^2$ at $\beta =0$ for both $\sigma \equiv 1$ and $\sigma_\mathrm{eff} \rightarrow 0$, we can say that the lattice data best supports the direct mixing mechanism in the limit as $N\rightarrow \infty$.  Were we to believe in using the incomplete gamma function~\cite{Press:1992} to assess this model's goodness-of-fit, we would find 
\bea Q &=& {{\Gamma({\nu\over 2},{\chi^2\over 2})}\over {\Gamma({\nu\over 2})}}\\
 &=& .88 \eea

\begin{table}[t]
\centering
\begin{tabular}{|r|c|c|} \hline
N & $\gamma$ & $\alpha$ \\\hline
direct & & \\
3 & .716 & 1.92  \\
4 & .602 & 1.89  \\
5 &-.007 & .973  \\
$\infty$ & -.069 & 1.06 \\
$(1/N^2)|_\infty$ &      &  \\\hline
adjoint & & \\
2 &-.104  & .897  \\
3 & .706  & 3.35  \\
4 & .594  & 3.26  \\
5 &-.292  & 1.22  \\
$\infty$ &-.324 & 1.29 \\
$(1/N^2)|_\infty$ &      & \\\hline
vertex & & \\
3 & .512  & 1.52  \\
4 & .472  & 2.12  \\
5 &-.223  & 1.28  \\
$(1/N^2)|_\infty$ &      & \\\hline
\end{tabular}
\caption{Parameter values for $\sigma_\mathrm{eff}$ for $\beta$ = 0.}
\label{table:sigParametersB0}
\end{table}

Turning now to the parameter values for the direct and adjoint mechanisms, Table~\ref{table:sigParametersB0}, we find some interesting behavior.  The $(1/N^2)$ dependence is not readily apparent for $\alpha$ and is dubious for $\gamma$.  Overall, the results seem to split into two sectors across mechanisms, one for N=3 and 4, the other for N=5 and $\infty$.  The smaller N return a $\gamma$ which is positive $\sim$ .6 or .7, while for higher N a negative $\gamma$ is supported $\sim$ -.07 or -.3.  As $\gamma$ shifts discontinuously, $\alpha$ also makes a jump.  In the direct scenario, $\alpha$ jumps from $\sim$ 1.9 to $\sim$ 1.0 as $\gamma$ goes from positive to negative.  For adjoint mixing, $\alpha$ goes from $\sim$ 3.3 down to $\sim$ 1.2.  No longer seeing our $(1/N^2)$ dependence, we need to explore the posteriors in more detail.

\newpage

\begin{figure}[!t]
\centering
\includegraphics[width=\textwidth]{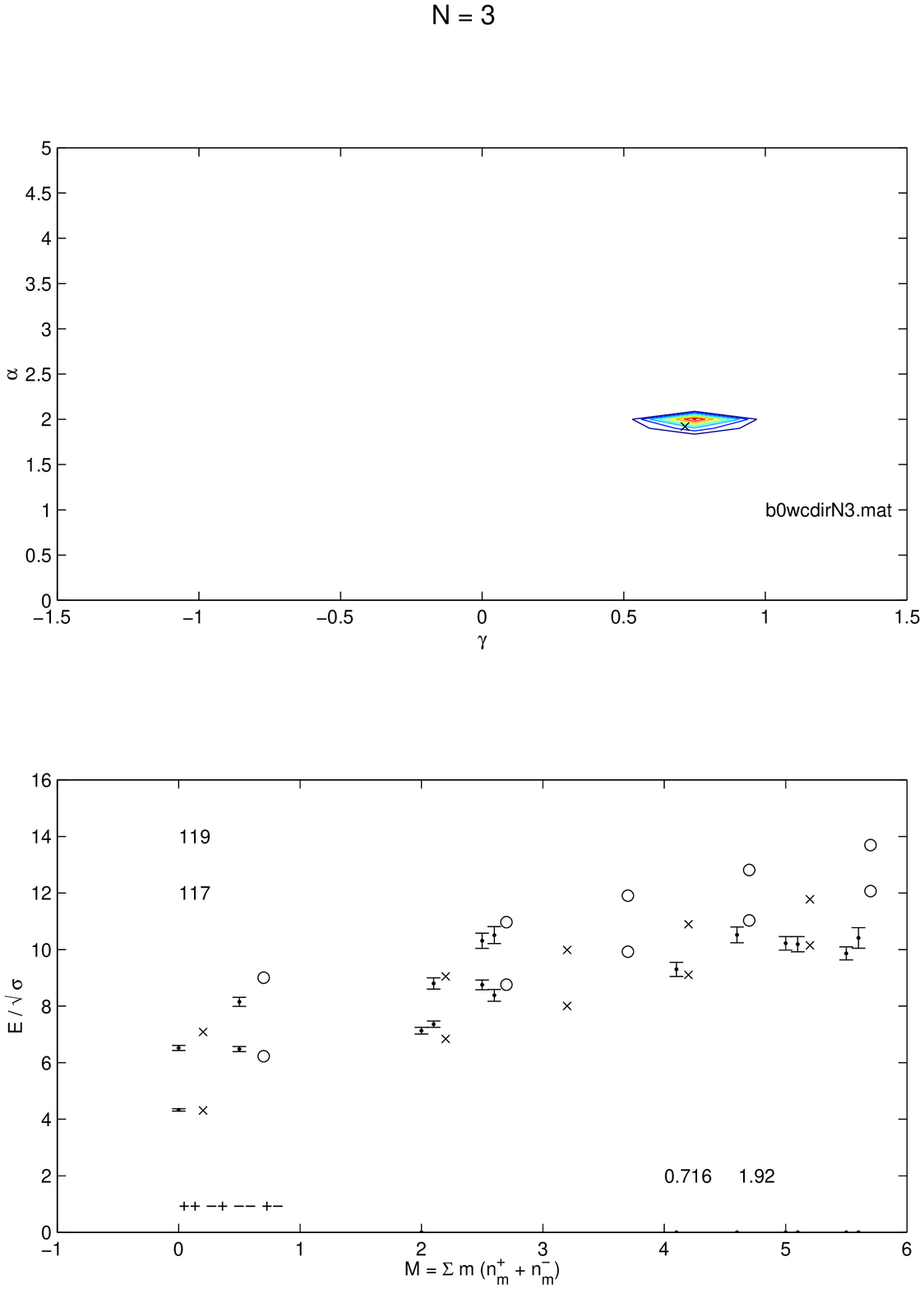}
\caption{The direct mixing posterior and spectrum for N=3.}
\label{fig:b0wcdirN3}
\end{figure}

\begin{figure}[!t]
\centering
\includegraphics[width=\textwidth]{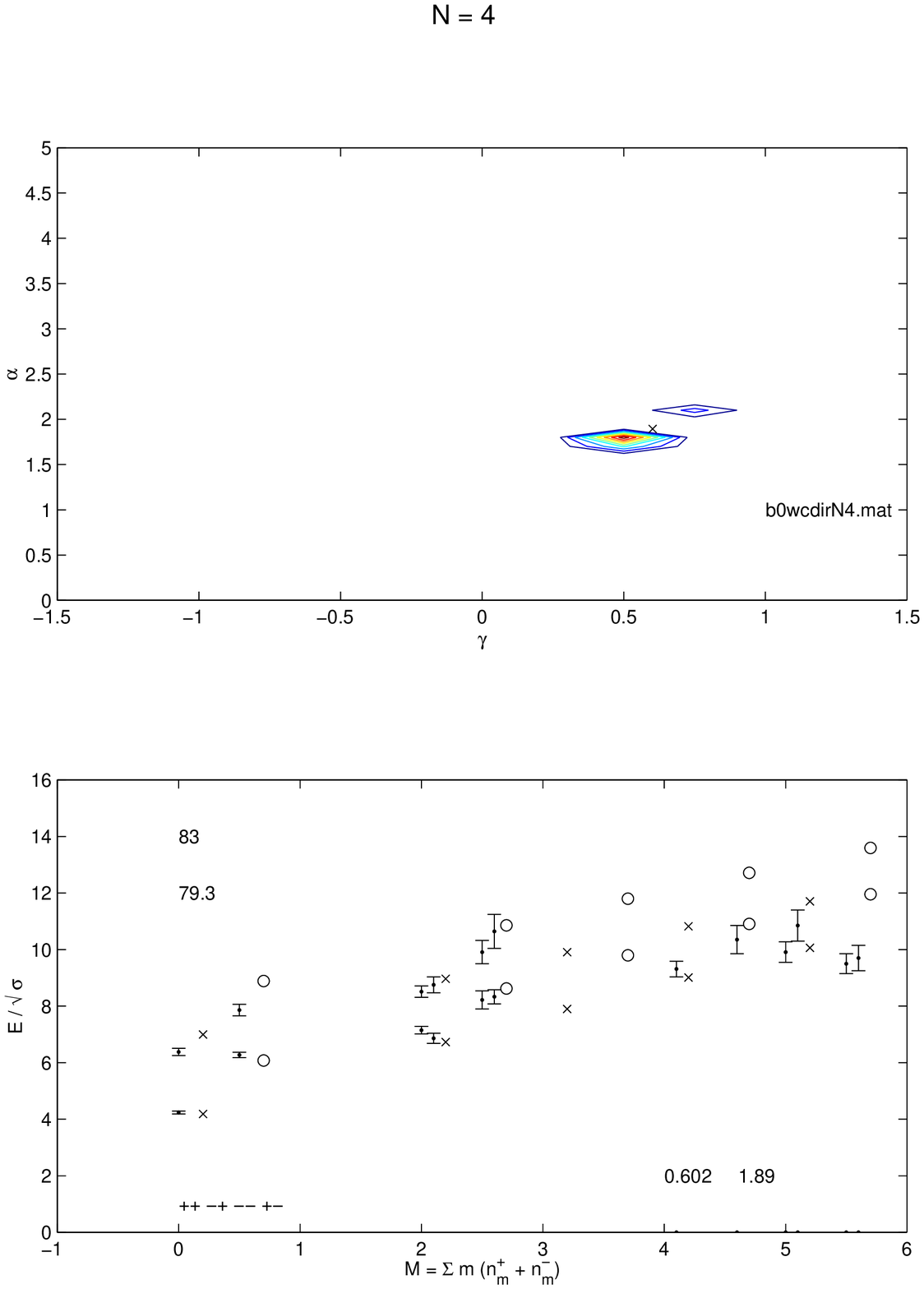}
\caption{The direct mixing posterior and spectrum for N=4.}
\label{fig:b0wcdirN4}
\end{figure}

\begin{figure}[!t]
\centering
\includegraphics[width=\textwidth]{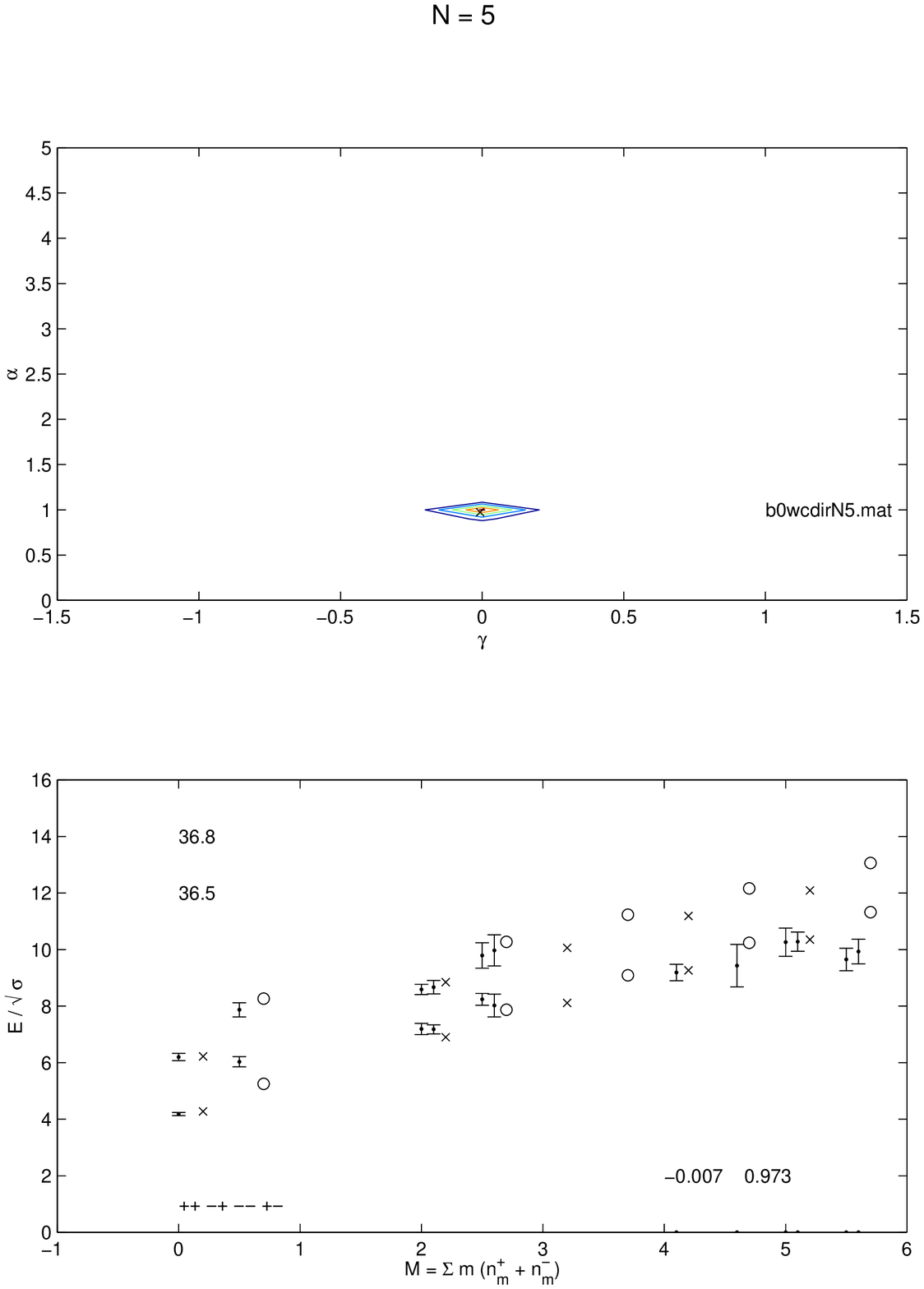}
\caption{The direct mixing posterior and spectrum for N=5.}
\label{fig:b0wcdirN5}
\end{figure}

\begin{figure}[!t]
\centering
\includegraphics[width=\textwidth]{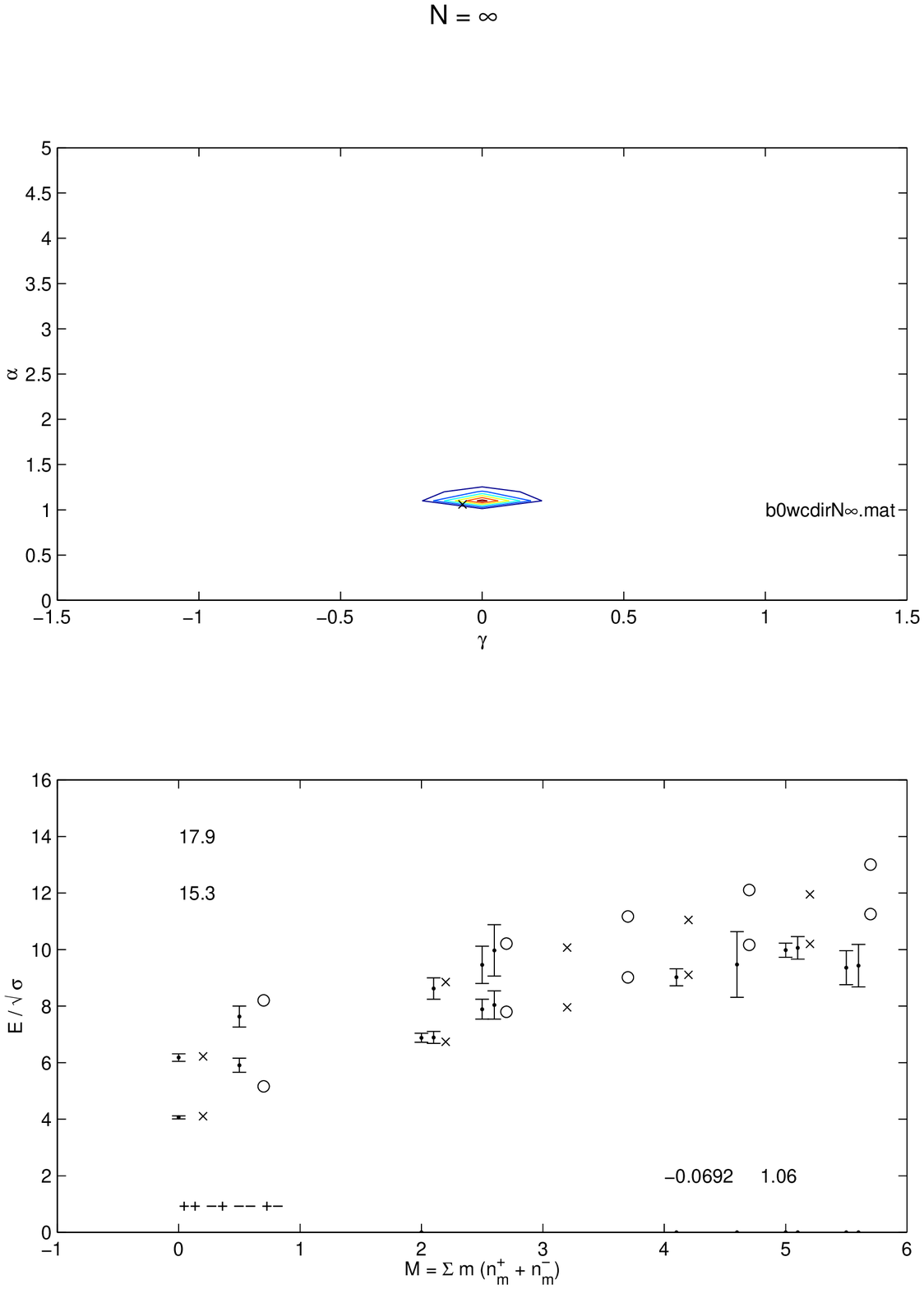}
\caption{The direct mixing posterior and spectrum for N=$\infty$.}
\label{fig:b0wcdirN6}
\end{figure}

Looking at the posterior for direct mixing first,
Figures~\ref{fig:b0wcdirN3} through \ref{fig:b0wcdirN6}, we see that the peaks for N=3 and 4 are likely to be related, while starting at N=5 the peaks shifts dramatically.  If we turn our attention to comparing the spectra, we can see what is driving the different peaks.  The prime difficulty is getting the J=0 sector to fall into place.  For N=4, Figure~\ref{fig:b0wcdirN4}, we see that the model is choosing to fit the conjugate splitting between the $0^{++}$ and the $0^{--}$ at the expense of the magnitude of the radial excitations.  (Such is also true for N=3.)  When N=5, the model shifts to favoring matching the radial excitations at the expense of the low lying $0^{--}$.  Now the excited states are within $\sim 1 \sigma$, while the $0^{--}$ is about $2\sim3 \sigma$ away.  This behavior explains the shift in the peak of the posterior.  What we would like is a model which can incorporate both the conjugate splitting as well as the radial excitations.  The J=2 states seem to be little affected by the shift of the peak -- so far they are compatible with either emphasis.

\begin{figure}[!t]
\centering
\includegraphics[width=\textwidth]{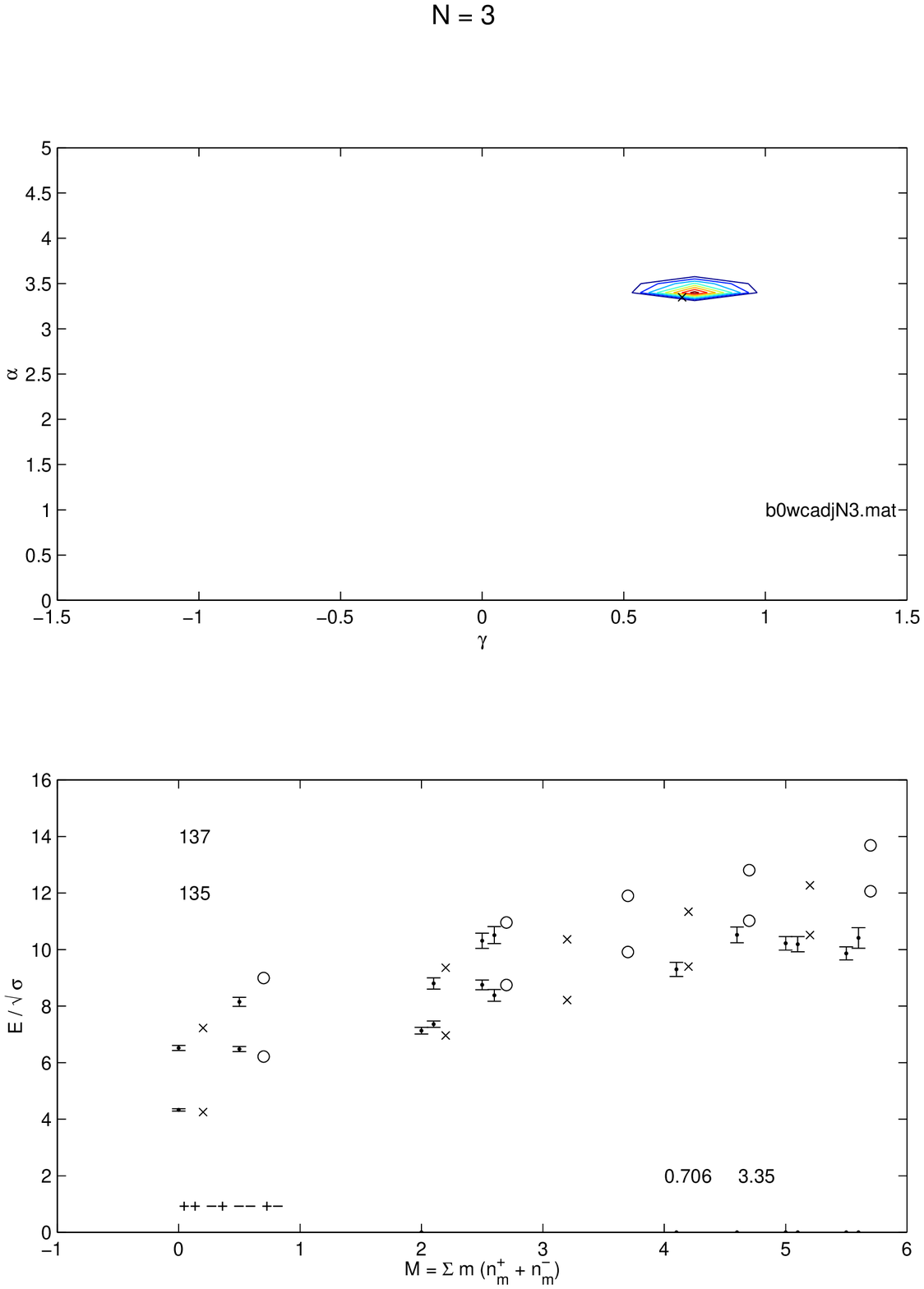}
\caption{The adjoint mixing posterior and spectrum for N=3.}
\label{fig:b0wcadjN3}
\end{figure}

\begin{figure}[!t]
\centering
\includegraphics[width=\textwidth]{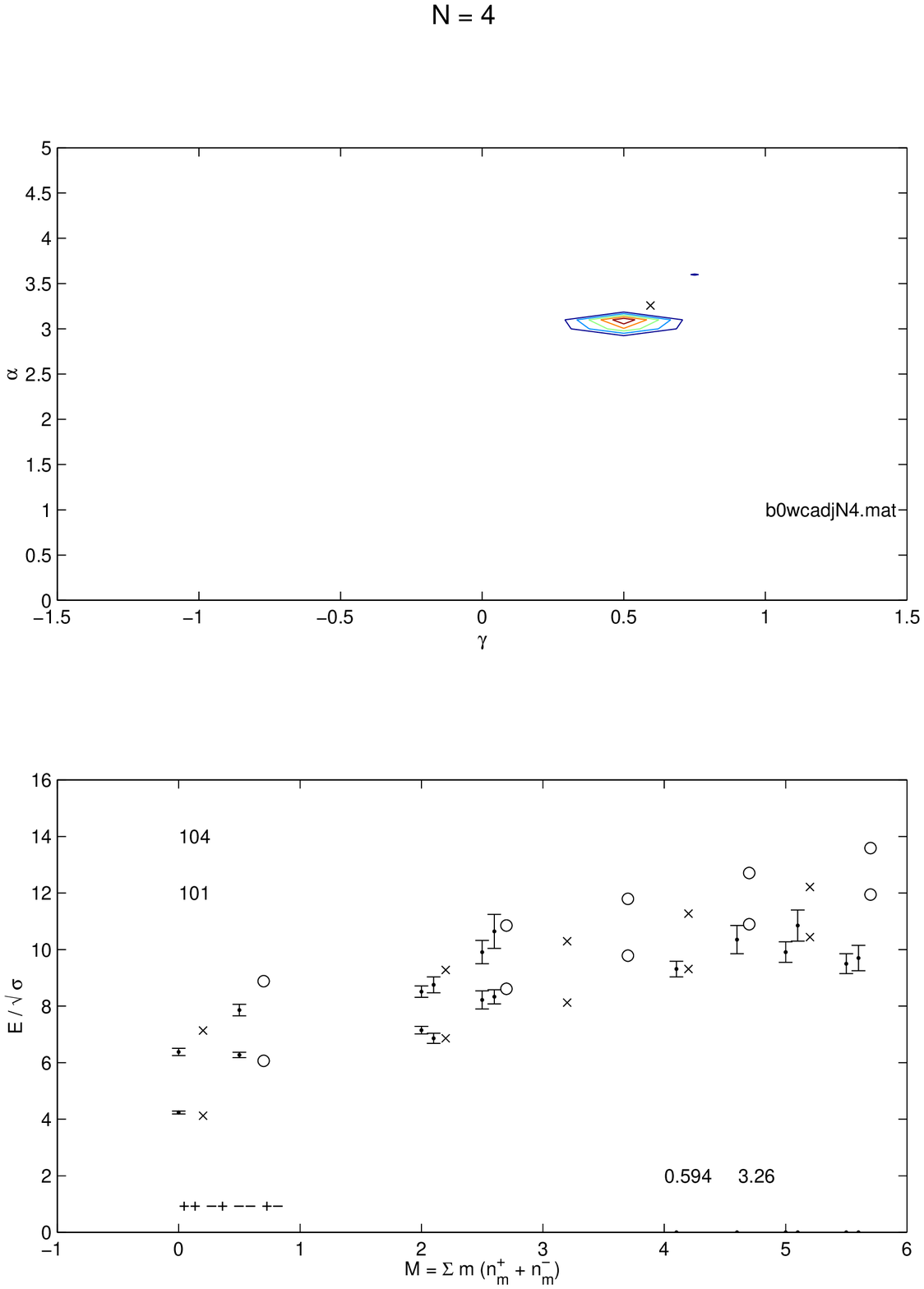}
\caption{The adjoint mixing posterior and spectrum for N=4.}
\label{fig:b0wcadjN4}
\end{figure}

\begin{figure}[!t]
\centering
\includegraphics[width=\textwidth]{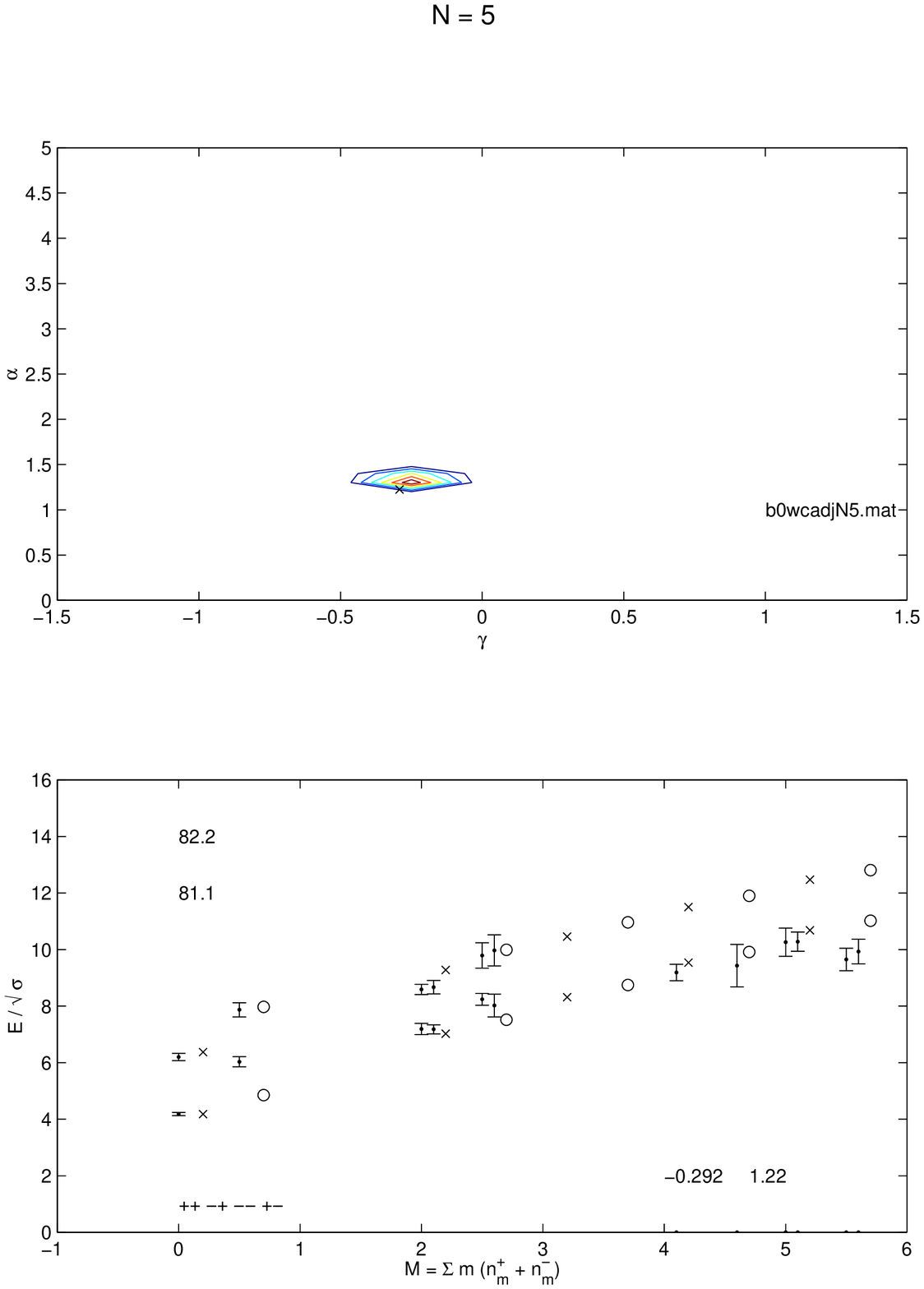}
\caption{The adjoint mixing posterior and spectrum for N=5.}
\label{fig:b0wcadjN5}
\end{figure}

\begin{figure}[!t]
\centering
\includegraphics[width=\textwidth]{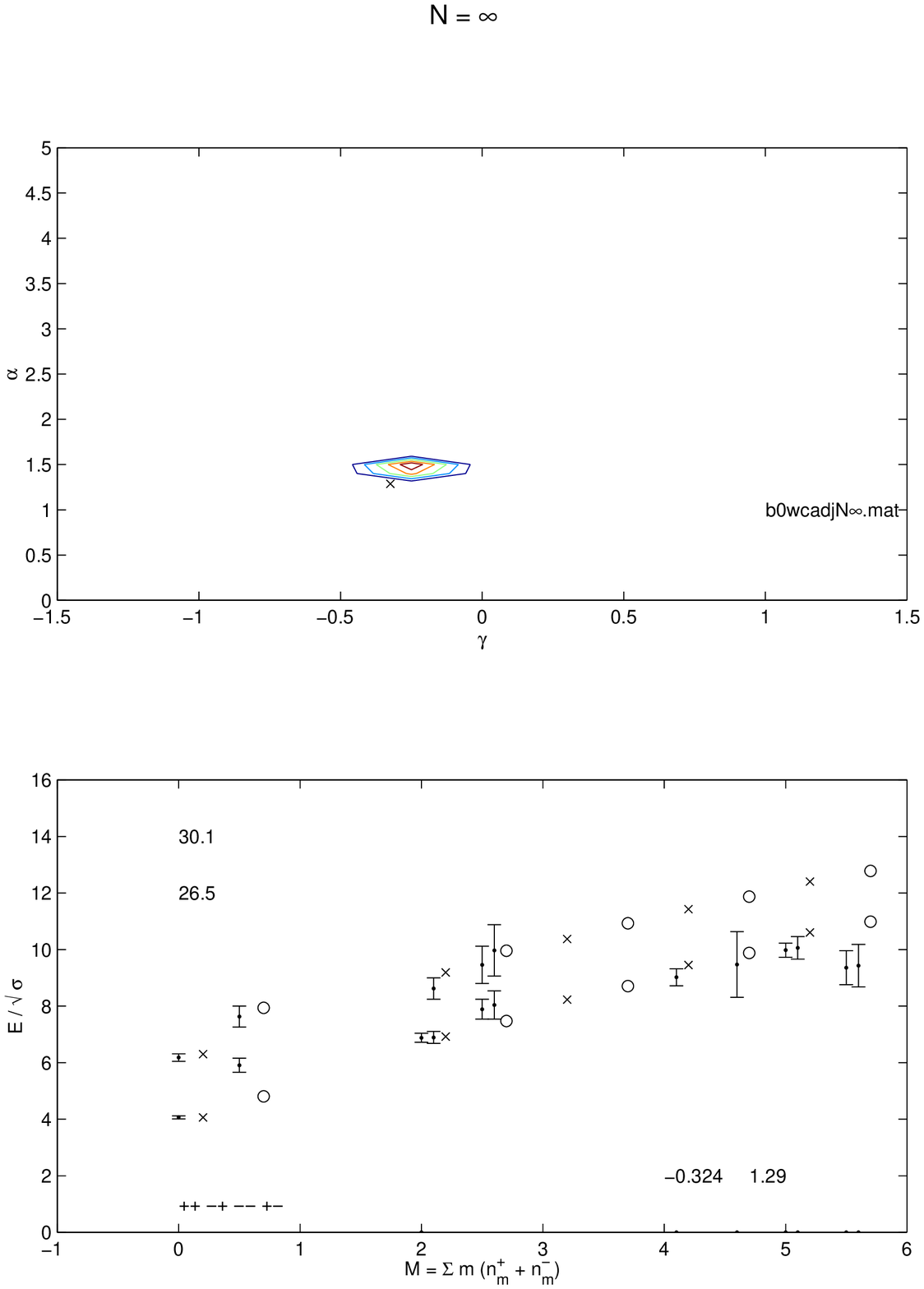}
\caption{The adjoint mixing posterior and spectrum for N=$\infty$.}
\label{fig:b0wcadjN6}
\end{figure}

\begin{figure}[!t]
\centering
\includegraphics[width=\textwidth]{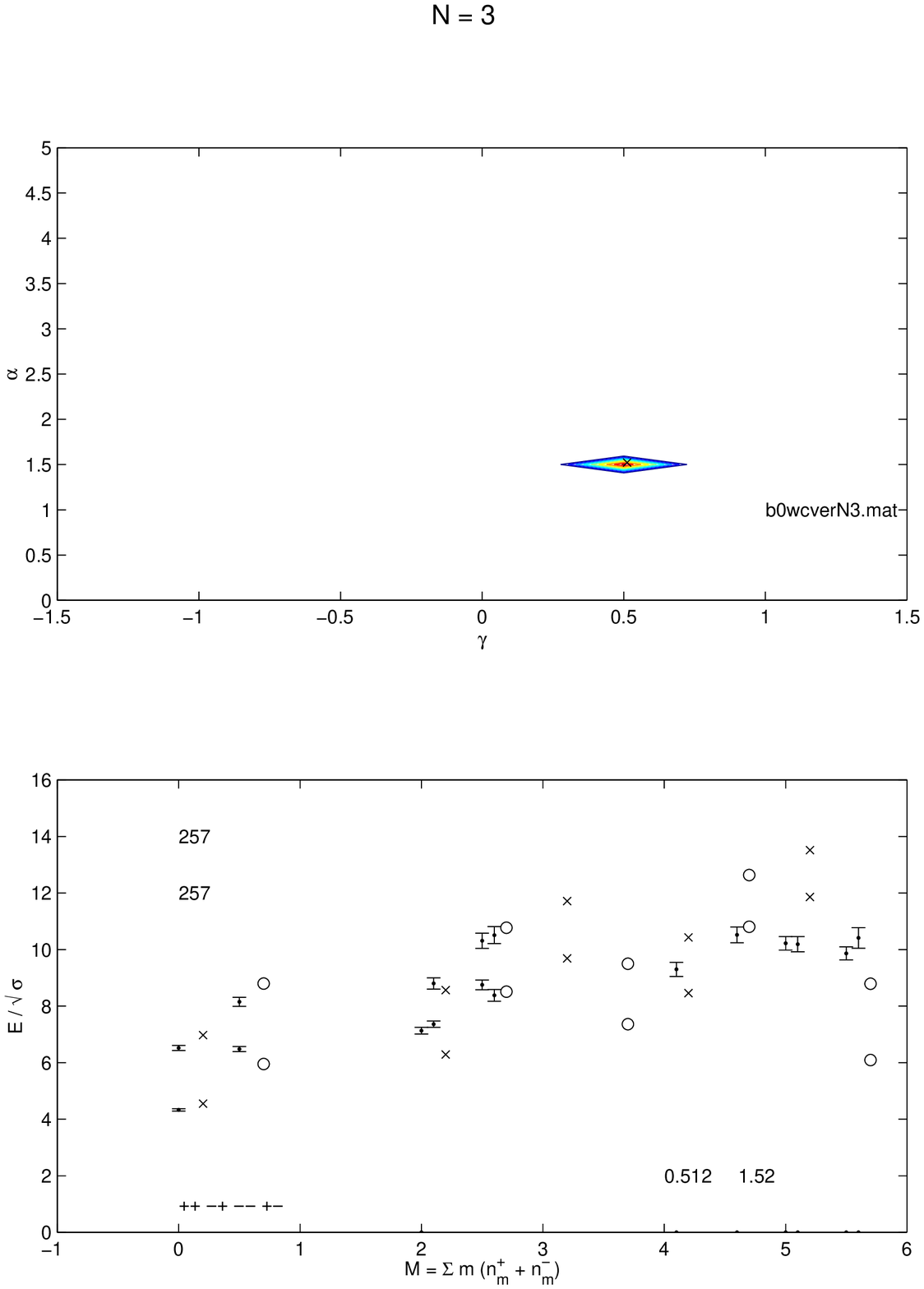}
\caption{The vertex mixing posterior and spectrum for N=3.}
\label{fig:b0wcverN3}
\end{figure}

\begin{figure}[!t]
\centering
\includegraphics[width=\textwidth]{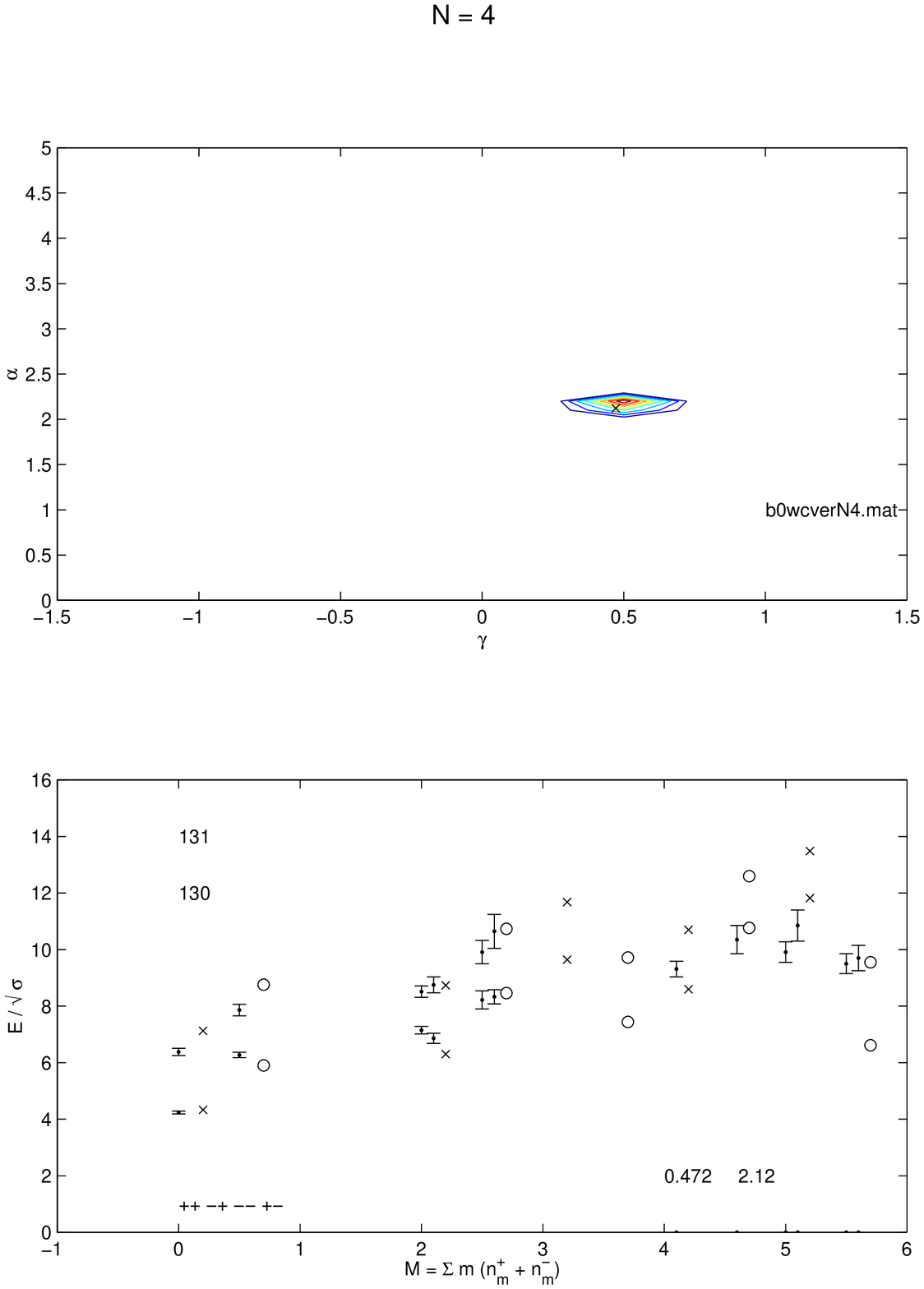}
\caption{The vertex mixing posterior and spectrum for N=4.}
\label{fig:b0wcverN4}
\end{figure}

\begin{figure}[!t]
\centering
\includegraphics[width=\textwidth]{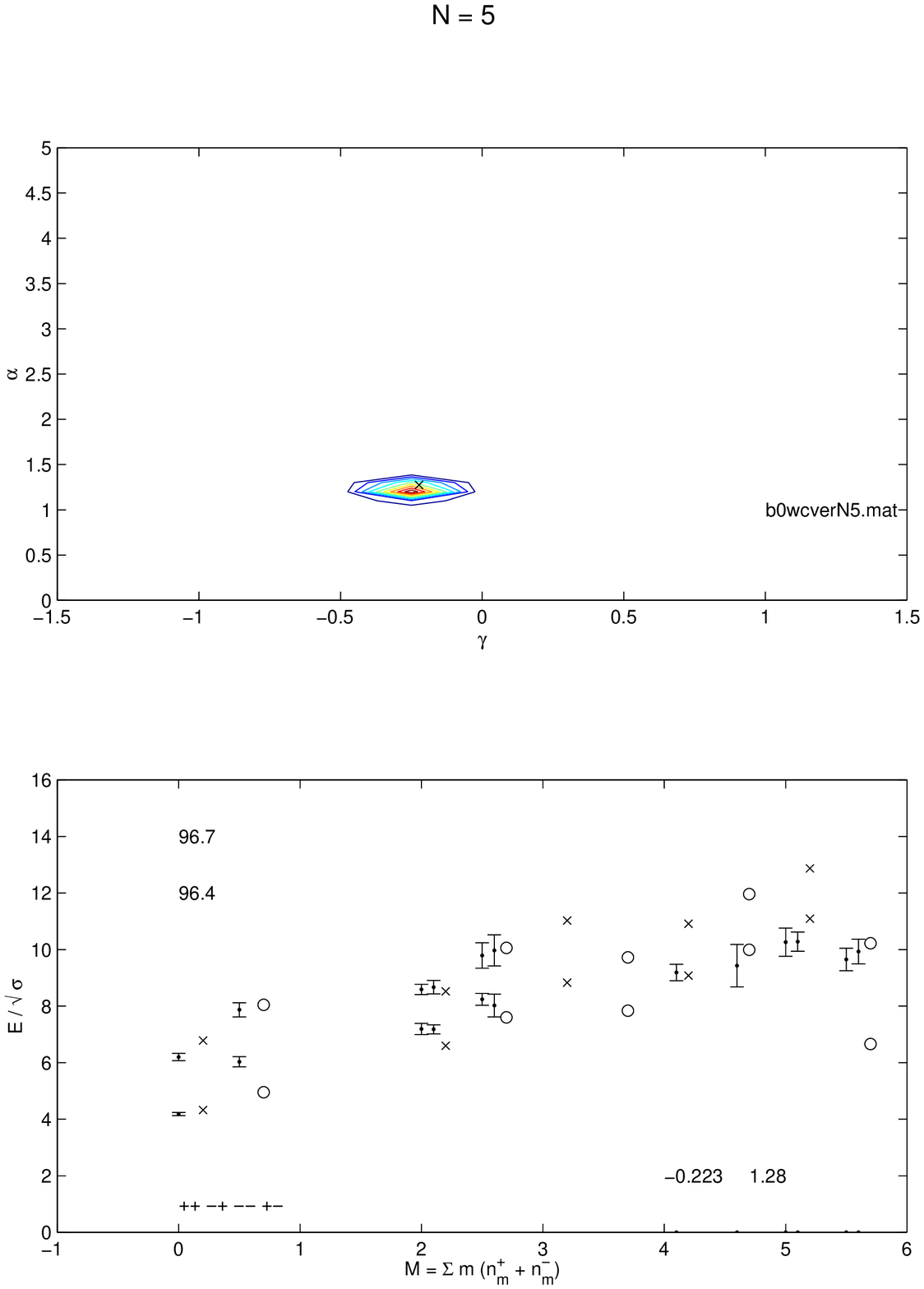}
\caption{The vertex mixing posterior and spectrum for N=5.}
\label{fig:b0wcverN5}
\end{figure}

Turning to the indirect mechanisms, we see much the same behavior.  For N=3 and 4, we see one peak corresponding to fitting the conjugate splitting, while for N=5 and $\infty$ we see the other peak corresponding to fitting the radial excitations.  Fitting the radial excitations gives a lower $\gamma$ than fitting the splitting, but the coefficient of elasticity is still positive, as the signal for zero elasticity is $\gamma \sim -1$, from which the peak is still some distance away.

Looking in detail at the spectrum resulting from the overall best fit (to the statistical errors), the $0^{--}$ notwithstanding, we see that the higher states are agreeing quite well with the hypothesis of spin misassignment.  The M=4 states match the J=0 states with mixed quantum numbers quite well.  When M=5, the $\pm+$ state agrees with the J=$1^{\pm+}$, but the $\pm-$ state is predicted to be too heavy, yet the lattice data would agree will with the M=3.  With the free predictions of the model giving such a signal, we should take seriously the issue of the spin assignments of the lattice operators.

\begin{table}[h]
\centering
\begin{tabular}{|lc|cc|cc|cc|cc|cc|} \hline
N & direct & $0^{++}$ & $0^{--}$ & $2^{\pm+}$ & $2^{\pm-}$ &
$3^{\pm+}$ & $3^{\pm-}$ & $4^{\pm+}$ & $4^{\pm-}$ & $1^{\pm+}$ &
$1^{\pm-}$ \\\hline

3 & $m_{J^{PC}}$ &
4.308  &  6.228  &  6.838  &  8.758  &  8.005  &  9.924  &  9.108  &
11.03  &  10.15  &  12.07 \\
  & $m_{J^{PC}}^\star$ & 
7.083  &  9.002  &  9.047  &  10.97  &  9.986  &  11.91  &   10.9  &
12.82  &  11.77  &  13.69 \\\hline

4 & & 
4.183  &  6.076  &  6.727  &   8.62  &  7.901  &  9.794  &  9.012  &
10.9  &  10.06  &  11.95 \\ & & 
6.994  &  8.887  &  8.965  &  10.86  &  9.907  &   11.8  &  10.82  &
12.71  &   11.7  &   13.6 \\\hline

5 & & 
4.276  &  5.249   &   6.9  &  7.873  &  8.114  &  9.087  &  9.263  &
10.24  &  10.35  &  11.32 \\ & & 
6.223  &  8.265  &  8.847  &  10.27  &  10.06  &  11.23  &  11.19  &
12.16  &  12.09  &  13.06 \\\hline

$\infty$ & & 
4.105  &  5.164  &  6.737  &  7.796  &  7.955  &  9.014  &  9.108  &
10.17  &   10.2  &  11.26 \\ & & 
6.223  &  8.201  &  8.855  &  10.21  &  10.07  &  11.17  &  11.04  &
12.1  &  11.95  &  13.01  \\\hline

\end{tabular}
\caption{The spectrum for direct mixing with $\sigma_{eff} \rightarrow
0$ at various N.  For each N, the
first mass is the lowest state in that channel, and the second mass is
the radial excitation.}
\label{table:b0wcdir}
\end{table}

\begin{table}[h]
\centering
\begin{tabular}{|lc|cc|cc|cc|cc|cc|} \hline
N & adjoint & $0^{++}$ & $0^{--}$ & $2^{\pm+}$ & $2^{\pm-}$ &
$3^{\pm+}$ & $3^{\pm-}$ & $4^{\pm+}$ & $4^{\pm-}$ & $1^{\pm+}$ &
$1^{\pm-}$ \\\hline

3 & $m_{J^{PC}}$ &
4.25  &  6.215  &  6.963  &  8.746  &  8.214  &  9.913  &  9.396  &
11.02  &  10.51  &  12.06 \\
  & $m_{J^{PC}}^\star$ & 
7.227  &  8.992  &  9.357  &  10.96  &  10.36   &  11.9  &  11.34  &
12.81  &  12.27  &  13.69 \\\hline

4 & & 
4.124  &  6.066  &   6.86  &  8.611  &  8.121  &  9.786  &  9.313  &
10.9  &  10.44  &  11.95 \\ & & 
7.134  &  8.879  &  9.281  &  10.85  &  10.29  &  11.79  &  11.27  &
12.71  &  12.21  &  13.59  \\\hline

5 & & 
4.178  &  4.855  &  7.021  &  7.516  &  8.317  &  8.748  &  9.534  &
9.916  &  10.68  &  11.02 \\ & & 
6.373  &  7.971  &   9.28  &  9.991  &  10.46  &  10.96  &   11.5  &
11.9  &  12.47  &  12.81  \\\hline

$\infty$ & & 
4.06   &  4.81  &  6.925  &  7.475   & 8.229   & 8.709  &  9.455   &
9.879   & 10.61  &  10.98 \\ & & 
6.298  &  7.937  &  9.191  &  9.959  &  10.37  &  10.93  &  11.43  &
11.87  &   12.4  &  12.78 \\\hline

\end{tabular}
\caption{The spectrum for adjoint mixing with $\sigma_{eff} \rightarrow
0$ at various N.  For each N, the
first mass is the lowest state in that channel, and the second mass is
the radial excitation.}
\label{table:b0wcadj}
\end{table}

\begin{table}[h]
\centering
\begin{tabular}{|lc|cc|cc|cc|cc|cc|} \hline
N & vertex & $0^{++}$ & $0^{--}$ & $2^{\pm+}$ & $2^{\pm-}$ &
$3^{\pm+}$ & $3^{\pm-}$ & $4^{\pm+}$ & $4^{\pm-}$ & $1^{\pm+}$ &
$1^{\pm-}$ \\\hline

3 & $m_{J^{PC}}$ &
4.55  &  5.955  &  6.287  &  8.511  &  9.691  &   7.36  &  8.459  &
10.81  &  11.86  &  6.091 \\
  & $m_{J^{PC}}^\star$ & 
6.973  &  8.796  &  8.565 &   10.77 &   11.72  &    9.5 &   10.43 &
12.63  &  13.52  &  8.793 \\\hline

4 & & 
4.33  &  5.901  &  6.301  &  8.463  &  9.645  &  7.439 &   8.592 &
10.76  &  11.82  &  6.612 \\ & & 
7.123  &  8.755  &  8.738   & 10.73  &  11.68  &  9.719  &   10.7  &
12.6  &  13.48  &  9.546  \\\hline

4 & & 
4.328  &  4.951  &    6.6 &   7.603  &  8.831  &  7.839  &  9.075 &
9.994  &  11.09  &  6.659 \\ & & 
6.782  &  8.042  &   8.52  &  10.06  &  11.03  &  9.725  &  10.91 &
11.96  &  12.87  &  10.22 \\\hline

\end{tabular}
\caption{The spectrum for vertex mixing with $\sigma_{eff} \rightarrow
0$ at various N.  For each N, the
first mass is the lowest state in that channel, and the second mass is
the radial excitation.}
\label{table:b0wcver}
\end{table}

\chapter{Mixing With the K-string} \label{chap:kstring}

\section{Adding another mixing parameter}

\begin{figure}[b]
\centering
\subfigure[$\sigma_1$]{
	\label{fig:kstring:1}
	\includegraphics[width=.3\textwidth]{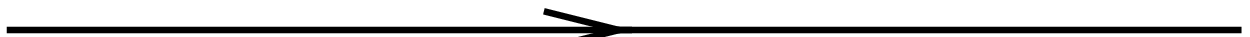}}
\hspace{1in}
\subfigure[$\sigma_k$]{
	\label{fig:kstring:2}
	\includegraphics[width=.3\textwidth]{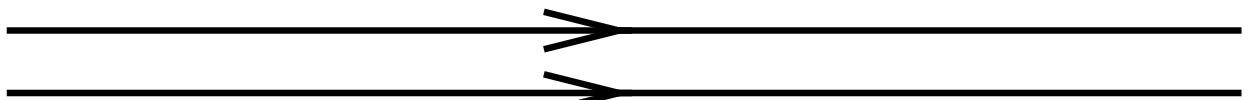}}
\caption{The k-string.}
\label{fig:kstring}
\end{figure}

Recent investigations~\cite{Bali:2000un,Lucini:2000qp} have lent support to the existence of a so-called k-string amongst the spectrum of pure gauge theory.  The number and string tension of these new strings depends on N.  For N=4 or 5, there exists a string with k=2.  Following predictions from brane theory (see~\cite{Lucini:2000qp}), the string tension $\sigma_k$ for a k-string goes as
\bea \frac{\sigma_k}{\sigma_1} &=& \frac{\sin{k\pi\over N}}{\sin{\pi\over N}} \\ &=& k + O(1/N^2) . \label{eqn:H_k} \eea
The k=2 string may be schematically represented as the proximate union of two fundamental strings, Figure~\ref{fig:kstring}.
This bound state is lighter than two free flux tubes of equivalent length.  As $N\rightarrow \infty$, the string tension approaches $2\sigma_1$.  For N=4, this ratio equals $\sqrt{2}$, while for the physically interesting case of N=3, it equals unity.

The important difference with the adjoint model is that this k-string carries an arrow of chromoelectric flux.  Hence, eigenstates with negative charge conjugation may be built.  This model possesses (at least) two towers of states:  those arising from the traditional Isgur-Paton flux tube and states heavier in proportion by the ratio of the k-string tension to the fundamental sting tension.  We are led to a basis consisting of four components: the L and R from the fundamental string, and another set of L and R from the k=2 string.  Eigenstates are built analogously between the two sectors, and so we expect the sub-Hamiltonians $H_{IP}$ and $H_k$ to be identical in form.  With a four component basis, we write
\beq H  = \left[ \begin{array}{cc} H_{IP} & M \\ M & H_k \end{array}\right] , \eeq
where $H_{IP}$ and $H_k$ are each two component direct mixing Hamiltonians with mixing $\alpha_1$ (and elasticity $\gamma$).  $H_k$ is equivalent to $H_{IP}$, Equation~\ref{eqn:Hip}, with the replacement of $\sigma$ by $\sigma_k$
\beq H_{k} = H_{IP}(\sigma \rightarrow \sigma_k) . \eeq
The mixing matrix M is ours to determine (sensibly).  Proposing that an L-orientated is much more likely to fluctuate into an L k-string rather than an R k-string (and vice versa), we write
\beq M = \left[ \begin{array}{cc} \alpha_2 & 0 \\ 0 & \alpha_2 \end{array}\right] \eeq
for some new mixing parameter $a_2$.  While we have now increased our parameter count by one, further analysis will yield quite an interesting result.

\begin{figure}[!t]
\centering
\includegraphics[width=\textwidth]{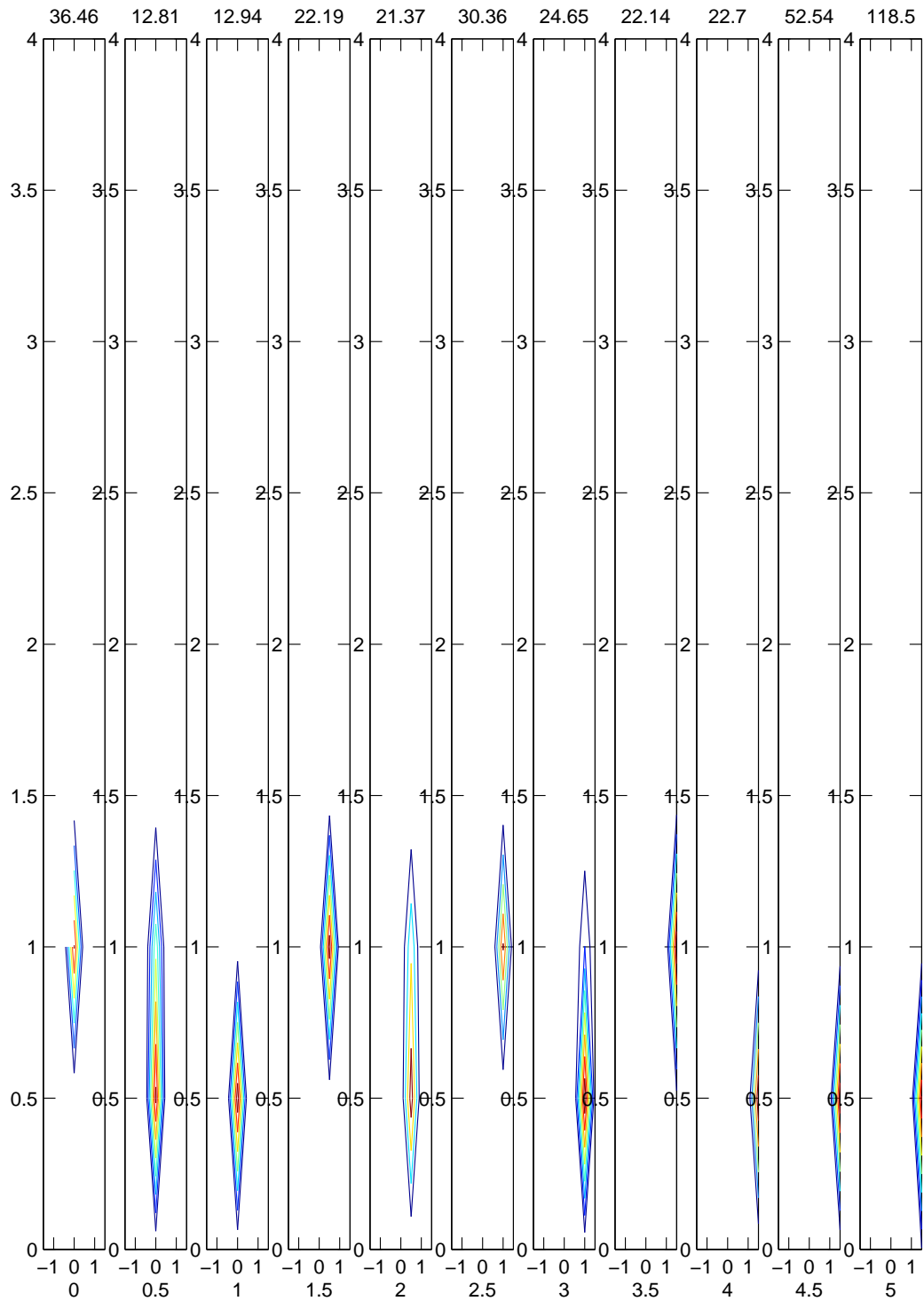}
\caption{The posterior for N=5 with adjusted weightings.}
\label{fig:b5a2newkN5post}
\end{figure}

Following earlier chapters, our merit function $\chi^2$ is defined and subject to minimization, only this time the parameter space is three-dimensional, making visualization of minimum $\chi^2$ surfaces slightly harder and the search for the global minimum more difficult.  Note that two more states, the triply excited states at $M=0$, have been added to the fits, making these $\chi^2$ not directly comparable to earlier ones.  These states are in a range of mass compatible with $J=4$, but with congruent quantum numbers ++ and --, these states most likely have had their spins properly identified as 0.  Pressing on, we map the posterior of a sensible range of parameter space and start the simplex from its best point, arriving at the spectra presented in Figures~\ref{fig:b0a2newkN3} through \ref{fig:b5a2newkN5} and Tables~\ref{table:b0a2newk}~and~\ref{table:b5a2newk}.  (An example of the posterior is shown in Figure~\ref{fig:b5a2newkN5post}.)

While the theoretical structure of the model carries for N down to 3 (or even 2), the calculations do not yield as pleasing a fit for such low N.  At N=3, the ratio of the string tensions is unity, in which case we wonder if such a k-string really exists.  The bases would be degenerate except for the identification of k-string states, and this is precisely how the calculation is performed at $N=3$.  However, the structure of the lattice spectra continues smoothly from high N down to N=2~\cite{Teper:1997tq}, and so a model which performs well at higher N ought to remain applicable at lower N.  Perhaps for N=3 there does exist a string heavier than the fundamental (such as the adjoint) but carrying an arrow of chromoelectric flux which no one has yet conjectured...  Such speculations are beyond this work.

For N=4 or 5, the results are much more pleasing, with the independence of the mixing parameters called into question by their incredible similarity, see Table~\ref{table:kstringmix2}.  Looking at the best fit parameters, we see that for these N the two mixing parameters are virtually equal!  As the numerical calculation treats these numbers as independent, to see them converge might well be pointing at worthwhile physics to explore.  At these N, this convergence of parameters lends some credibility to the model.  We might even reconsider the k-string mixing model with the two mixing parameters replaced by a single parameter, thus reducing our parameter count again and allowing us to calculate some error bounds on our parameters.

\begin{table}[!b]
\centering
\begin{tabular}{|r|ccc|c|} \hline
N & $\gamma$ & $\alpha_1$ & $\alpha_2$ & $\chi^2$ \\\hline
$\beta=0$ & & & & \\
3 & 1.49 & .925 & 1.94 & 81.8        \\
4 & .262 & .899 & .818 & 34.5        \\
5 & .056 & .770 & .786 & 27.9        \\
$\infty$ & -.085 & .676  & .961 & 14.3        \\\hline

$\beta=0.05$ & & & & \\
3 & 1.45 & .848 & 1.92 & 7.44        \\
4 & .213 & .803 & .790 & 6.80        \\
5 & .042 & .736 & .728 & 7.36        \\
$\infty$ & -.086 & .666 & .957 & 5.59         \\\hline

\end{tabular}
\caption{Parameter values and $\chi^2$ for k-string mixing with 2 mixing parameters.}
\label{table:kstringmix2}
\end{table}

\clearpage

\begin{table}[!h]
\centering
\begin{tabular}{|lc|cc|cc|cc|cc|cc|} \hline
N & k-string & $0^{++}$ & $0^{--}$ & $2^{\pm+}$ & $2^{\pm-}$ &
$3^{\pm+}$ & $3^{\pm-}$ & $4^{\pm+}$ & $4^{\pm-}$ & $1^{\pm+}$ &
$1^{\pm-}$ \\\hline

3 & $m_{J^{PC}}$ & 4.371 & 6.221 & 6.799 & 8.649 & 7.916 & 9.766 & 8.972 & 10.82 & 9.970 & 11.82 \\
  & $m_{J^{PC}}^\star$ & 6.908 & 8.758 & 8.829 & 10.68 & 9.745 & 11.60 & 10.63 & 12.48 & 11.49 & 13.34 \\\hline

4 & & 4.290 & 6.088 & 6.950 & 8.749 & 8.175 & 9.974 & 9.332 & 11.13 &
10.42 & 12.22 \\
  & & 6.154 & 7.952 & 8.945 & 10.74 & 10.14 & 11.94 & 11.17 & 12.97 &
12.11 & 13.91 \\\hline

5 & & 4.218 & 5.757 & 6.911 & 8.450 & 8.149 & 9.688 & 9.318 & 10.86 &
10.42 & 11.96 \\
  & & 6.098 & 7.637 & 9.010 & 10.55 & 10.19 & 11.73 & 11.20 & 12.74 &
12.14 & 13.68 \\\hline

$\infty$ & & 4.081 & 5.432 & 6.817 & 8.168 & 8.071 & 9.422 & 9.252 &
10.60 & 10.37 & 11.72 \\
  & & 6.279 & 7.853 & 9.155 & 10.29 & 10.26 & 11.39 & 11.28 & 12.42 &
12.24 & 13.38 \\\hline
\end{tabular}
\caption{The spectrum for k-string mixing at various N.  For each N, the
first mass is the lowest state in that channel, and the second mass is
the radial excitation.}
\label{table:b0a2newk}
\end{table}
\begin{table}[!h]
\centering
\begin{tabular}{|lc|cc|cc|cc|cc|cc|} \hline
N & k-string & $0^{++}$ & $0^{--}$ & $2^{\pm+}$ & $2^{\pm-}$ &
$3^{\pm+}$ & $3^{\pm-}$ & $4^{\pm+}$ & $4^{\pm-}$ & $1^{\pm+}$ &
$1^{\pm-}$ \\\hline

3 & $m_{J^{PC}}$ & 4.413 & 6.109 & 6.847 & 8.542 & 7.967 & 9.662 & 9.025 & 10.72 & 10.03 & 11.72 \\
  & $m_{J^{PC}}^\star$ & 6.963 & 8.659 & 8.885 & 10.58 & 9.803 & 11.50 & 10.69 & 12.39 & 11.55 & 13.24 \\\hline

4 & & 4.334 & 5.939 & 7.001 & 8.606 & 8.229 & 9.834 & 9.389 & 10.99 &
10.48 & 12.09 \\
  & & 6.134 & 7.740 & 8.950 & 10.56 & 10.17 & 11.78 & 11.22 & 12.83 &
12.17 & 13.78 \\\hline

5 & & 4.271 & 5.743 & 6.961 & 8.433 & 8.197 & 9.669 & 9.364 & 10.84 &
10.47 & 11.94 \\
  & & 6.068 & 7.540 & 9.020 & 10.49 & 10.23 & 11.70 & 11.25 & 12.72 &
12.19 & 13.66 \\\hline

$\infty$ & & 4.093 & 5.425 & 6.828 & 8.161 & 8.082 & 9.414 & 9.263 &
10.59 & 10.38 & 11.71 \\
  & & 6.242 & 7.575 & 9.112 & 10.45 & 10.21 & 11.54 & 11.20 & 12.53 &
12.13 & 13.46 \\\hline
\end{tabular}
\caption{The spectrum for k-string mixing with adjusted weightings at various N.  For each N, the
first mass is the lowest state in that channel, and the second mass is
the radial excitation.}
\label{table:b5a2newk}
\end{table}

\clearpage

\begin{figure}[!t]
\centering
\includegraphics[width=\textwidth]{b0a2newkN3.epsi}
\caption{The spectrum for N=3.}
\label{fig:b0a2newkN3}
\end{figure}

\begin{figure}[!t]
\centering
\includegraphics[width=\textwidth]{b0a2newkN4.epsi}
\caption{The spectrum for N=4.}
\label{fig:b0a2newkN4}
\end{figure}

\begin{figure}[!t]
\centering
\includegraphics[width=\textwidth]{b0a2newkN5.epsi}
\caption{The spectrum for N=5.}
\label{fig:b0a2newkN5}
\end{figure}

\begin{figure}[!t]
\centering
\includegraphics[width=\textwidth]{b0a2newkN6.epsi}
\caption{The spectrum for N=$\infty$.}
\label{fig:b0a2newkN6}
\end{figure}

\begin{figure}[!t]
\centering
\includegraphics[width=\textwidth]{b5a2newkN3.epsi}
\caption{The spectrum for N=3 with adjusted weightings.}
\label{fig:b5a2newkN3}
\end{figure}

\begin{figure}[!t]
\centering
\includegraphics[width=\textwidth]{b5a2newkN4.epsi}
\caption{The spectrum for N=4 with adjusted weightings.}
\label{fig:b5a2newkN4}
\end{figure}

\begin{figure}[!t]
\centering
\includegraphics[width=\textwidth]{b5a2newkN5.epsi}
\caption{The spectrum for N=5 with adjusted weightings.}
\label{fig:b5a2newkN5}
\end{figure}

\begin{figure}[!t]
\centering
\includegraphics[width=\textwidth]{b5a2newkN6.epsi}
\caption{The spectrum for N=$\infty$ with adjusted weightings.}
\label{fig:b5a2newkN6}
\end{figure}

\clearpage

\section{Combining the mixing parameters}

Returning our attention to Equation~\ref{eqn:H_k}, we now suppose that the mixing matrix M does {\it{not}} contain a new parameter $a_2$ but instead uses our old parameter $a_1$ to perform the mixing between sectors.  That such an identification of parameters may be performed is not immediately obvious:  $a_1$ stems from the direct mixing between the L and R orientations of the same type of string, while $a_2$ represents mixing between the two types of string.  However, the results immediately above strongly suggest that, at least for $N \geq 4$, such identification may be valid.  As always, the proof is in the pudding.

We repeat the calculations (again), but this time only for $N=4,5,{\mathrm{and\;}} \infty$.  In Table~\ref{table:kstringmix1} we present the resulting best fit parameters and their associated $\chi^2$.  The details of the spectra are given in Figures~\ref{fig:b0newk2N4} through \ref{fig:b5newk2N5} and Tables~\ref{table:kstringmassB0} and \ref{table:kstringmassB5}.  The most interesting feature of this model is that it works, and quite well at that.  The values of $\chi^2$ are comparable or even better than the earlier models.  However, we must not put too much faith in the k-string model as a successful explanation of the lattice spectrum, as the continuation down to N=2 which manifests in the lattice data cannot be reproduced.  Nonetheless, the k-string model must have some validity:  it produces (for adjusted weightings) a $\chi^2$ / d.o.f. of 7 / (8 - 2) $\approx$ 1!

Ultimately, the complete spectrum of pure gauge theory is unlikely to be adequately described by such simple effective Hamiltonians as these.  However, by distilling the multitudinous gauge degrees of freedom down to a few clearly defined parameters, we hope to capture the essential aspects necessary to describe the gluonic physics relevant to the real world.

\begin{table}[!h]
\centering
\begin{tabular}{|r|rl|rl|c|} \hline
N & $\gamma$ & $(\sigma_\gamma)$ & $\alpha$ & $(\sigma_\alpha)$ & $\chi^2$ \\\hline
$\beta=0$ & & & & & \\

4 & .276 & (6) & .878 & (26) & 35.4 \\
5 & .054 & (38) & .774 & (39) & 27.9 \\
$\infty$ & -.080 & (45) & .779 & (49) & 17.2 \\\hline

$\beta=.05$ & & & & & \\

4 & .215 &  (1) & .799 &  (1) & 6.81 \\
5 & .043 & (76) & .735 & (92) & 7.36 \\
$\infty$ & -.112 & (79) & .720 & (100) & 6.31 \\\hline

\end{tabular}
\caption{Parameter values and $\chi^2$ for k-string mixing with 1 mixing parameter.}
\label{table:kstringmix1}
\end{table}

\clearpage

\begin{figure}[!t]
\centering
\includegraphics[width=\textwidth]{b0newk2N4.epsi}
\caption{The spectrum for N=4.}
\label{fig:b0newk2N4}
\end{figure}

\begin{figure}[!t]
\centering
\includegraphics[width=\textwidth]{b0newk2N5.epsi}
\caption{The spectrum for N=5.}
\label{fig:b0newk2N5}
\end{figure}

\begin{figure}[!t]
\centering
\includegraphics[width=\textwidth]{b0newk2N6.epsi}
\caption{The spectrum for N=$\infty$.}
\label{fig:b0newk2N6}
\end{figure}

\begin{figure}[!t]
\centering
\includegraphics[width=\textwidth]{b5newk2N4.epsi}
\caption{The spectrum for N=4 with adjusted weightings.}
\label{fig:b5newk2N4}
\end{figure}

\begin{figure}[!t]
\centering
\includegraphics[width=\textwidth]{b5newk2N5.epsi}
\caption{The spectrum for N=5 with adjusted weightings.}
\label{fig:b5newk2N5}
\end{figure}

\begin{figure}[!t]
\centering
\includegraphics[width=\textwidth]{b5newk2N6.epsi}
\caption{The spectrum for N=$\infty$ with adjusted weightings.}
\label{fig:b5newk2N6}
\end{figure}

\clearpage

\begin{table}[!b]
\begin{center}
\begin{tabular}{|l|c|c|c|}\hline

state & SU(4) & SU(5) & SU($\infty$) \\ \hline
$0^{++}$         & 4.283& 4.219 & 4.105 \\
$0^{++\ast}$     & 6.24 & 6.08  & 6.10 \\
$0^{++\ast\ast}$ & 7.40 & 7.43  & 7.39 \\ 
$2^{\pm+}$       & 6.95 & 6.91  & 6.81 \\
$2^{\pm+\ast}$   & 8.99 & 9.00  & 9.09 \\
$4^{-+}$         & 9.33 & 9.32  & 9.23 \\
$4^{-+\ast}$     & 11.18& 11.20 & 11.16 \\
$1^{\pm+}$       & 10.42& 10.42 & 10.34 \\
$1^{\pm+\ast}$   & 12.11& 12.14 & 12.01 \\ \hline

$0^{--}$         & 6.04 & 5.77  & 5.66 \\
$0^{--\ast}$     & 8.00 & 7.63  & 7.66 \\
$0^{--\ast\ast}$ & 9.15 & 8.98  & 8.95 \\ 
$2^{\pm-}$       & 8.70 & 8.46  & 8.37 \\
$2^{\pm-\ast}$   & 10.75& 10.55 & 10.65 \\
$4^{+-}$         & 11.09& 10.87 & 10.79 \\
$4^{+-\ast}$     & 12.93& 12.75 & 12.72 \\
$1^{\pm-}$       & 12.18& 11.97 & 11.89 \\
$1^{\pm-\ast}$   & 13.87& 13.69 & 13.65 \\ \hline

\end{tabular}
\caption{The spectrum for k-string mixing with 1 parameter.}
\label{table:kstringmassB0}
\end{center}
\end{table}

\begin{table}[!t]
\begin{center}
\begin{tabular}{|l|c|c|c|}\hline

state & SU(4) & SU(5) & SU($\infty$) \\ \hline
$0^{++}$         & 4.332& 4.270& 4.148 \\
$0^{++\ast}$     & 6.15 & 6.07 & 6.04 \\
$0^{++\ast\ast}$ & 7.45 & 7.46 & 7.41 \\ 
$2^{\pm+}$       & 7.00 & 6.96 & 6.86 \\
$2^{\pm+\ast}$   & 8.96 & 9.02 & 9.11 \\
$4^{-+}$         & 9.39 & 9.36 & 9.27 \\
$4^{-+\ast}$     & 11.23& 11.25& 11.21 \\
$1^{\pm+}$       & 10.48& 10.46& 10.38 \\
$1^{\pm+\ast}$   & 12.17& 12.12& 12.13 \\ \hline

$0^{--}$         & 5.93 & 5.74 & 5.59 \\
$0^{--\ast}$     & 7.74 & 7.54 & 7.48 \\
$0^{--\ast\ast}$ & 9.05 & 8.93 & 8.85 \\ 
$2^{\pm-}$       & 8.60 & 8.43 & 8.30 \\
$2^{\pm-\ast}$   & 10.56& 10.49 & 10.55 \\
$4^{+-}$         & 10.99& 10.83 & 10.71 \\
$4^{+-\ast}$     & 12.82& 12.72 & 12.64 \\
$1^{\pm-}$       & 12.08& 11.93 & 11.82 \\
$1^{\pm-\ast}$   & 13.77& 13.66 & 13.57 \\ \hline

\end{tabular}
\caption{The spectrum for k-string mixing with 1 parameter with adjusted weightings.}
\label{table:kstringmassB5}
\end{center}
\end{table}

\chapter{A Novel Construction of Lattice Operators} \label{chap:lattice}

Working with the Isgur-Paton model has brought to light the importance of correctly identifying the spin of states measured by lattice operators.  For a review of lattice gauge theory, see~\cite{Creutz:1983,Montvay:1994,Teper:1999}.  While the lattice is an excellent method for determining the gross properties of a state, such as its mass or branching ratios, it can give only limited information regarding a state's structure.  Yet that structure has a strong impact on the state's quantum numbers.  The modern technology of smeared lattice operators~\cite{Teper:1987wt} further complicates extracting structure information from correlation functions.  As we have already seen, differentiating between models of these gluonic states requires understanding the spin structure of the entire spectrum.

Our problem demonstrates that what is a solution to one problem can become another problem in itself.  In the continuum, field theory is plagued by ultraviolet divergences~\cite{Ramond:1990}.  Introducing the lattice discretization of space-time successfully regularizes these divergences, but at a cost.  By discretizing space-time, we break the continuous rotational symmetry down to that of the lattice, so that rotations of arbitrary degree are reduced to a discrete set of rotations.  For conventional square lattices, the rotation group is broken to the discrete group O(2), rotations by multiples of $\pi/2$.

Let $R(\theta)$ be the operator for a rotation of a loop by an angle $\theta$. Then, its action on a loop may be written
\beq \psi_\theta = R(\theta) \psi_{\theta=0} . \eeq
To represent a state of arbitrary J in the continuum, we then write
\beq \Psi_J = \int_0^{2\pi} d\theta e^{i J \theta} \psi_\theta)\; |0>, \eeq
where $\psi_{\theta = 0}$ represents the state along some fixed axes, which is then multiplied by the phase $i J \theta$ when rotated to angle $\theta$, and $|0>$ is the vacuum.  On the lattice, this becomes
\beq \Psi_J = \sum_{n=0}^3 e^{i J \theta_n} \psi_{R(\theta_n)} \; |0> = \sum_{n=0}^3 e^{i J n \pi/2} \psi_{R_n} \; |0> . \eeq
Accordingly, the lattice can only distinguish rotations of $\pi/2$.  This rotational ambiguity shows up as an ambiguity in the spin represented by the state.  For example, taking J=4,
\bea \Psi_{J=4} &=& \sum_{n=0}^3 e^{i n 2 \pi} \psi_{R(\theta_n)} \; |0> \\
 &=& \sum_{n=0}^3 \psi_{R(\theta_n)} \; |0> \\
 &=& \Psi_{J=0} , \eea
thus an operator which couples to J=0 will also couple to J=4 on a cubic lattice.  So far, lattice methods only extract the two or three lowest states in a given symmetry channel, and the convention of assigning the lowest compatible spin seems applicable, but as we continue to explore the heavier gluonic states, we need to critically reevaluate our methods.  

Earlier work~\cite{Johnson:1998ev} explored a technique to distinguishing the J=4 and the J=0 in the $PC=++$ sector on D=2+1 lattices, but its result was inconclusive.  Recently progress has been made in distinguishing the J=4 from the J=0 on D=3+1 cubic lattices~\cite{Liu:2001}.  By methodically and painstakenly sorting out the appropriate symmetry contributions for quite twisted (and smeared) conventional lattice operators, those authors claim to separate out the J=4 channel from the J=0.  However, their technique is quite specific to the spins 0 and 4 and the 3+1 cubic lattice.  What we are after is a technique that may be applied to arbitrary spins in arbitrary dimensions (and indeed even on arbitrary lattices).  \footnote{For the initial idea to try this new method, one must thank Mike Teper.}

To overcome our difficulties with arbitrary spin resolution on a cubic lattice, we need operators which we can place onto the lattice at arbitrary angles with a minimum of distortion.  These operators may be placed on the lattice at relative angles other than $\pi/2$, thereby picking up different phase contributions than operators constrained to align to the axes of the lattice.  These phase contributions will not be identical for spins which are equal $\mathit{modulo} 4$, and thus the J=4 will be distinguishable from the J=0.  Such new operators must still be smooth on physical length scales, requiring the development of the arbitrary smeared link between any two spatial lattice sites.  

\begin{figure}[!b]
\centering
\subfigure[Operators coupling to J=2]{
	\label{fig:conops:ops2}
	\includegraphics[width=.3\textwidth]{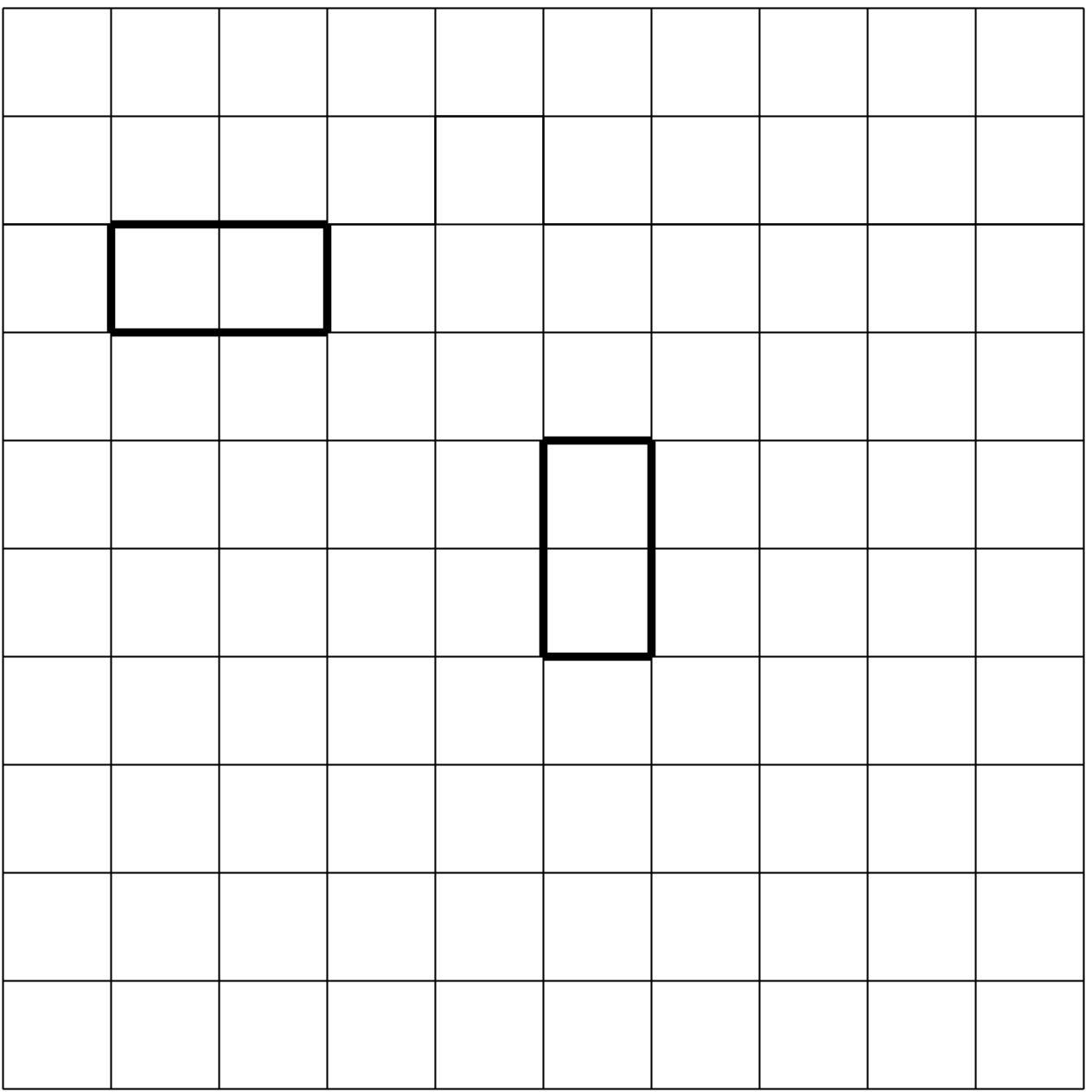}}
\hspace{1in}
\subfigure[Operators coupling to J=1]{
	\label{fig:conops:ops1}
	\includegraphics[width=.3\textwidth]{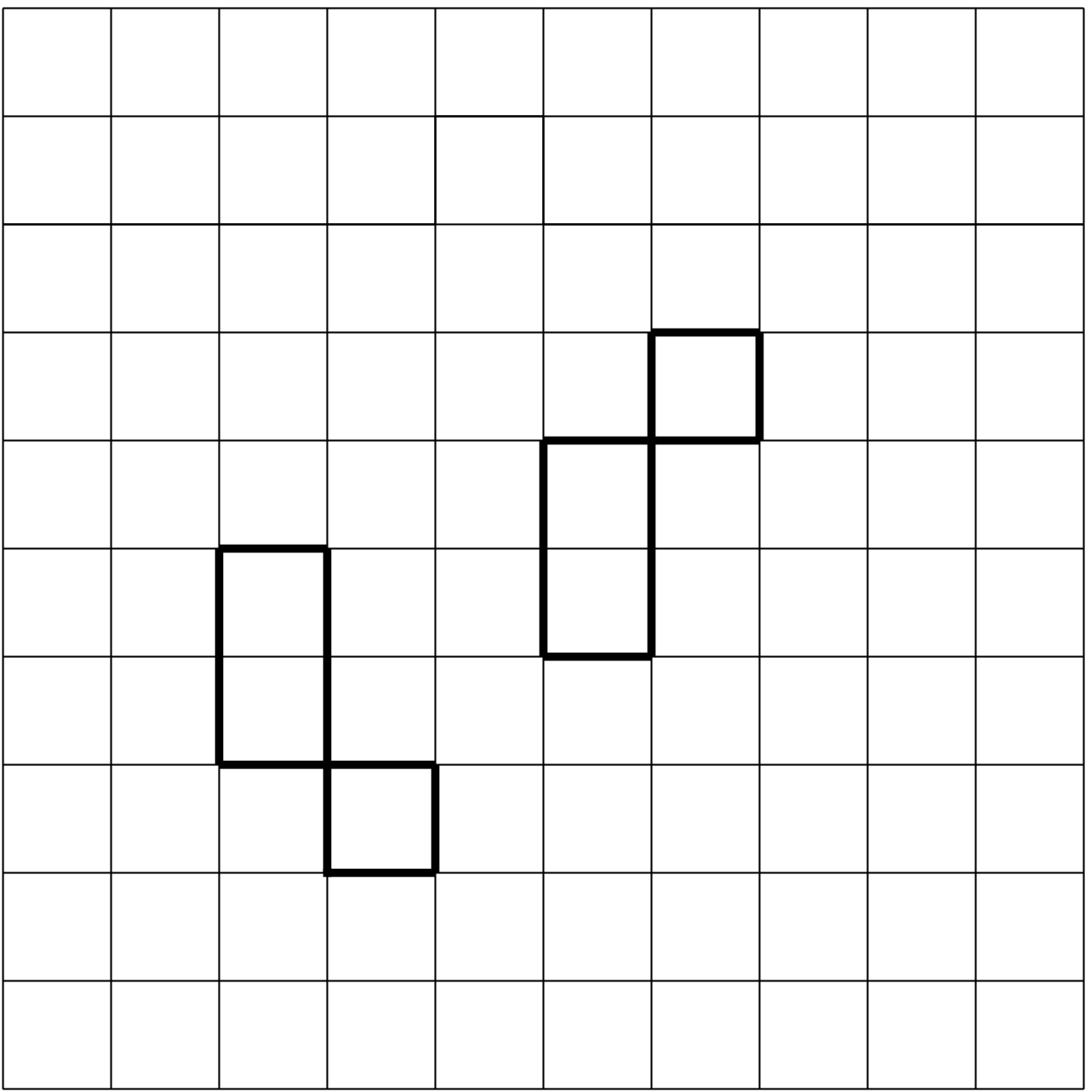}}
\caption{Conventional lattice operators.}
\label{fig:conops}
\end{figure}

Conventional operators are constructed from closed contours of gluonic links.  The requirement of gauge invariance is what necessitates using closed contours; otherwise under a gauge transformation $U(x) \rightarrow V(x)U_{\hat{\mu}}(x)V^\dagger(x+\hat{\mu})$ the operator acquires unmatched contributions from its endpoints which break gauge invariance.  Thus, the simplest gluonic operator is the real part of the trace of the basic plaquette, which has the quantum numbers of the vacuum.  To find the $C=-$ analogue of the operator we take the imaginary part of the trace.  As the plaquette is rotationally invariant, to construct an operator of nonzero spin, we must use operators with more interesting shapes.  States with J=2 can be made from operators which are longer than they are broad, Figure~\ref{fig:conops:ops2}.  To couple to J=1, operators with very little intrinsic symmetry must be used, eg Figure~\ref{fig:conops:ops1}.  To construct an operator of negative parity, basic shapes of negative parity are combined in an appropriate linear combination.
As we go to the continuum limit, $a\rightarrow 0$, the size of our basic operator becomes infinitesimal, while the size of the physical glueball remains finite.  To overcome this difficulty, basic links are replaced by their smeared counterparts~\cite{Teper:1987wt}.  Smearing increases both the length and breadth of the links used.  As the level of iterative smearing increases, the resultant smeared link is a complicated combination of the raw links of the lattice which is ``smooth'' on physical length scales.  These techniques, while improving the overlap with physical states, prevent an easy visualization of the structure of the state.

\begin{figure}[t]
\centering
\includegraphics[scale=.5]{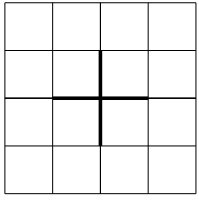}
\caption{Four paths of length 2.}
\label{fig:M2}
\end{figure}

\begin{figure}[!h]
\centering
\subfigure[Twelve paths of unity]{
	\label{fig:M4:a}
	\includegraphics[scale=.5]{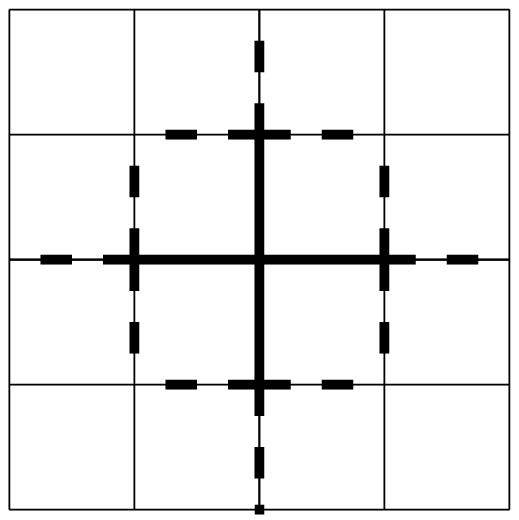}}
\hspace{1in}
\subfigure[Four plaquettes]{
	\label{fig:M4:b}
	\includegraphics[scale=.5]{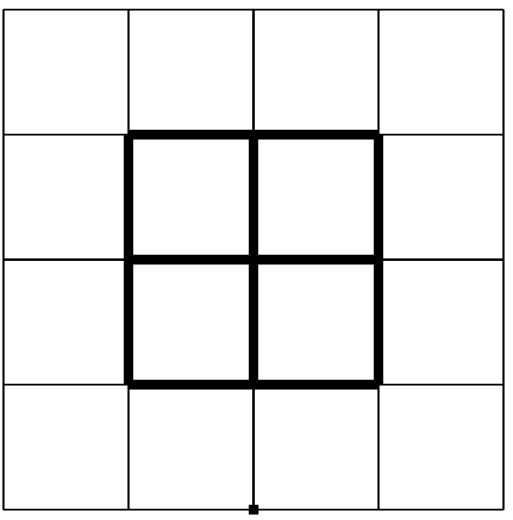}}
\caption{Paths of length 4.}
\label{fig:M4}
\end{figure}

Adopting a broader perspective, we would like to construct operators using an arbitrary selection of lattice points.  What we need is a ``smeared link'' between {\it{any}} two points in a timeslice, not just along an axis.  Labelling the link starting at i between adjacent lattice sites i and j by $U_{ij}$, we construct the matrix M for each timeslice by
\bea M_{ij} = \left\{ \begin{array}{cc} U_{ij} & \mathrm{if}\; U_{ij}\; \mathrm{exists} \\ 0 & \mathrm{otherwise} \end{array} \right\}  \eea
M is an $n\times n$ sparse matrix with a high bandwidth.  
As $U_{ji} = U_{ij}^\dagger$, $M_{ji} = M_{ij}^\dagger$, and M is Hermitian.
Using the well known expansion, we define
\bea K &=& {1\over {1 - \alpha M}} \\ &=& 1 + \alpha M + \alpha^2 M^2 + \dots  \label{eqn:K} \eea
with parameter $\alpha$.  When M is multiplied by itself, nonzero entries are the sum of the products of the link variables between the sites consisting of a number of links equal to the number of times M appears.  The full matrix K represents a matrix of propagators between every site in our timeslice, i.e. $K_{ij} = K_{ji}^\dagger$ is the propagator from site i to site j.  Within its radius of convergence, the expansion for K can be used to determine which paths are contributing to K.  The weighting parameter $\alpha$ controls the extent to which longer paths are contributing.  For very small $\alpha$,
\beq K \approx 1 + \alpha M + \alpha M^2 . \eeq
Looking at the main diagonal of K, paths of length 2 are contributing, thus
\beq K_{ii} \approx 1 + 4 \alpha^2 , \eeq
as there are 4 paths from the center site i to its nearest neighbors and back again, Figure~\ref{fig:M2}.  These paths contribute unity each as the product along a link and back again is simply $U \times U^\dagger$.
For slightly greater $\alpha$, along the diagonal
\beq K_{ii} \approx 1 + \alpha^2 [M^2]_{ii} + \alpha^4 [M^4]_{ii} . \eeq
Now that we have paths of length 4, there are 12 paths which contribute unity each to $[M^4]_{ii}$, three off of each link coming from i,  Figure~\ref{fig:M4:a}.  And we also have the 4 plaquettes in the plane which have site i as a vertex, Figure~\ref{fig:M4:b}.  
To illustrate our method, we can estimate the expectation value of the plaquette by
\beq <[]> \approx {1\over 4n} \left( Tr\,K - 1 - 4 \alpha^2 - 12
\alpha^4 \right) \label{eqn:kplaq} \eeq

As $\alpha$ increases, paths of increasing length contribute to K.  We must beware, however, of the divergence of K as $\alpha$ approaches some critical value $\alpha_c$ which makes the denominator of Equation (\ref{eqn:K}) equal zero.  To find this critical value, we can expand the trace of K as a sum over terms containing the reciprocal of the eigenvalues of M:
\bea Tr\,K &=& Tr\,(\mathbf{1} + \alpha M + \alpha^2 M^2 + ...) \\
&=& Tr\,\mathbf{1} + \alpha Tr\,M + \alpha^2 Tr\,M^2 + ... \\
&=& n + \alpha \sum_{i=1}^n \lambda_i + \alpha^2 \sum_{i=1}^n \lambda_i^2 + ... \\
&=& \sum_{i=1}^n (1 + \alpha \lambda_i + \alpha^2 \lambda_i^2 + ...) \\
&=& \sum_{i=1}^n {1\over {1 - \alpha \lambda_i}} \eea
which can equal $\infty$ if any $\lambda_i$ equals $\alpha$.  As we take $\alpha$ as ranging from 0 to $\alpha_c$, it is the largest eigenvalue of M which determines the divergence of $Tr\,K$.
A plot of $<Tr\,K>$ is shown in Figure~\ref{fig:traceK} -- we can easily see the divergence at $\alpha_c$ as well as the change in sign when $\alpha > \alpha_c$.  To be effective in constructing operators, we need $\alpha$ large enough that paths significantly longer than the shortest path between two sites i and j contribute to $K_{ij}$, while not so large that we are diverging.  The range of effective $\alpha$ in this calculation is found empirically.

\begin{figure}[!hb]
\centering
\includegraphics[scale=.5]{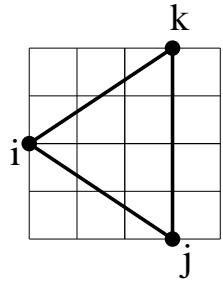}
\caption{Triangle based lattice operators.}
\label{fig:triops}
\end{figure}

\begin{figure}[!hb]
\centering
\includegraphics[scale=.3]{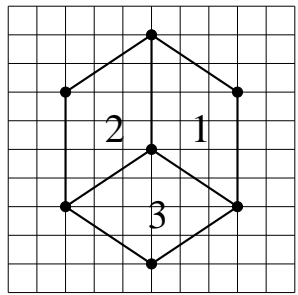}
\caption{Rhombus based lattice operators.}
\label{fig:myops}
\end{figure}

From our matrix K, we can construct an operator by taking a closed path of propagators, so that, e.g.
\beq \CO = K_{ij}K_{jk}K_{ki} \eeq
defines a triangular operator, Figure~\ref{fig:triops}.  While triangular operators are quite pretty, they possess too much symmetry, coupling effectively only to J=0.
In order to have less symmetry, we construct operators by joining two triangles together to produce a rhombus, Figure~\ref{fig:myops}.
Each rhombus can be put onto the lattice in one of three orientations, but as we are still working with a cubic lattice, these orientations will have slightly different actual sizes, and hence different vacuum expectation values.  (Actually, the two ``vertical'' orientations should be equivalent, while the ``horizontal'' orientation will be different.)  A solution, following~\cite{Johnson:1998ev}, is to normalize the operators so that 
\beq <\CO_{\theta_1}^2> = c_{\theta_j}^2 <\CO_{\theta_j}^2> \; \mathrm{for} \; j \neq 1 . \label{eq:normc} \eeq
Such normalization should account for any directional variance.  From these normalized operators we can construct our physical operators by multiplying each base operator by a phase $i J \theta$ corresponding to its desired spin and its angle of orientation.  We take the three orientations shown in Figure~\ref{fig:myops} as well as their $\pi/2$ counterparts as our physical operators, for a total of six base operators.  We need pairs of loops with a relative angle of $\pi/2$ to produce good J=2 operators as well as to affect the necessary cancellations detailed below.  Next we take the zero momentum sum of $\CO$ across each timeslice.  From these numbers we can calculate the correlation function
\beq C_{\CO}(t) = <\CO(t)|\CO(0)> \eeq
which decomposes into a sum of decaying exponentials
\beq C_{\CO}(t) = \sum_n a_n e^{- m_n t} .  \eeq
For large t, we can extract the lowest mass from
\beq C_{\CO}(t) \propto e^{-m_{\CO}(t)} + e^{-m_{\CO}(T-t)} \eeq
in Euclidean space-time, where the second term arises from our finite lattice with time extent T=L.  As we are interested in distinguishing the $4^{++}$ from the $0^{++}$, as well as from the vacuum which has the same quantum numbers, in practice we use vacuum subtracted correlation functions
\beq C_{\CO}(t) \rightarrow C_{\CO}(t) - < \CO >^2 . \eeq

Operator construction is as much an art as it is science.  Finding the right combination of shapes and smearing level which couple to the desired state is educated guesswork.  Our first operators are based on two-triangle rhombi, which may be rotated by $\pi/3$ while remaining roughly the same size, Figure~\ref{fig:myops}.  As these operators are at relative angles other than $\pi/2$, the J=0 and J=4 will now have different phases on each rhombus, and hence are distinguishable.

\begin{figure}[t]
\centering
\subfigure[Unwanted torelon contribution.]{
	\label{fig:torel:ij}
	\includegraphics[width=.3\textwidth, bb = 100 190 510 600]{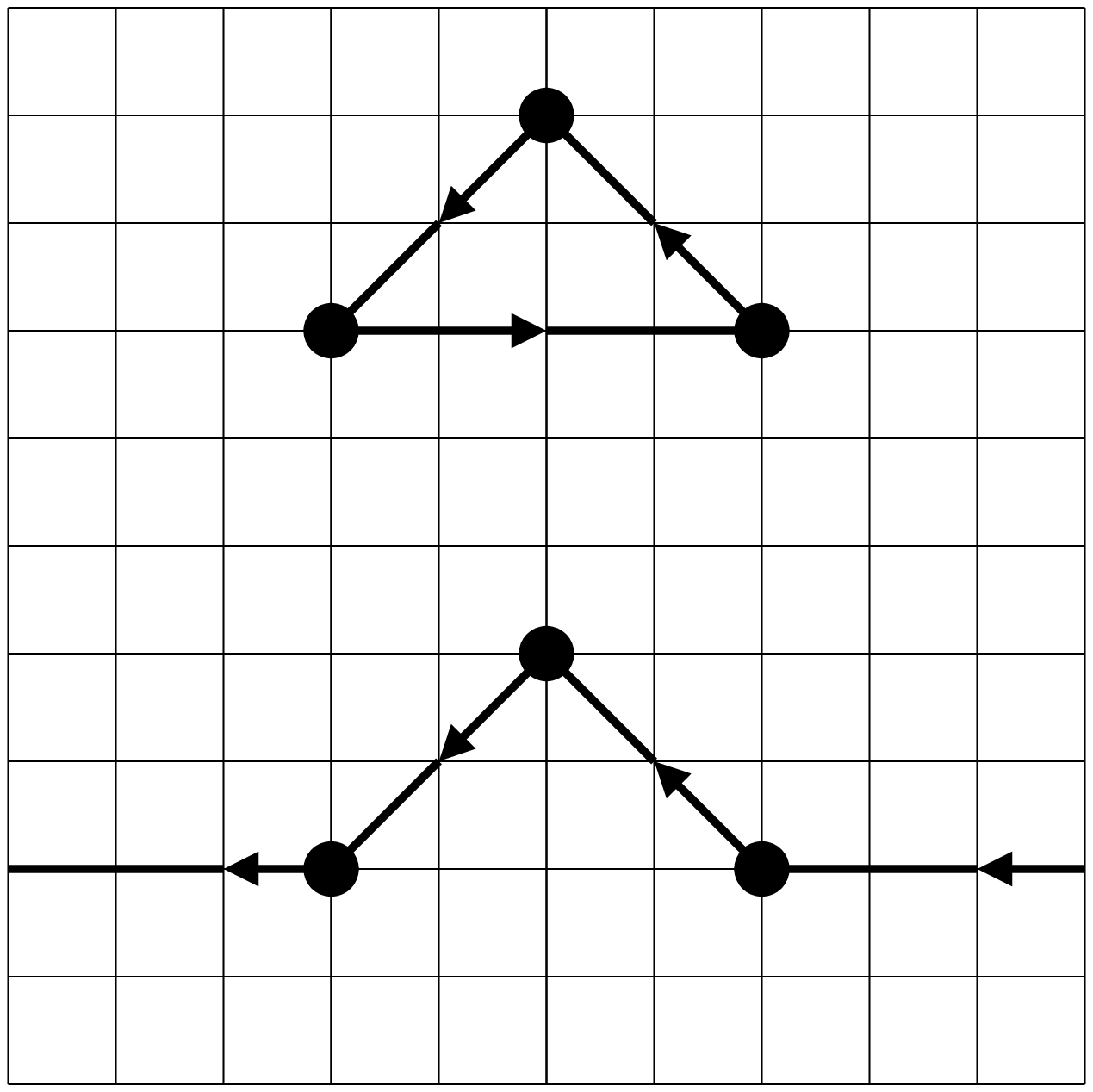}}
\hspace{1in}
\subfigure[Removing torelon contributions.]{
	\label{fig:torel:rem}
	\includegraphics[width=.3\textwidth]{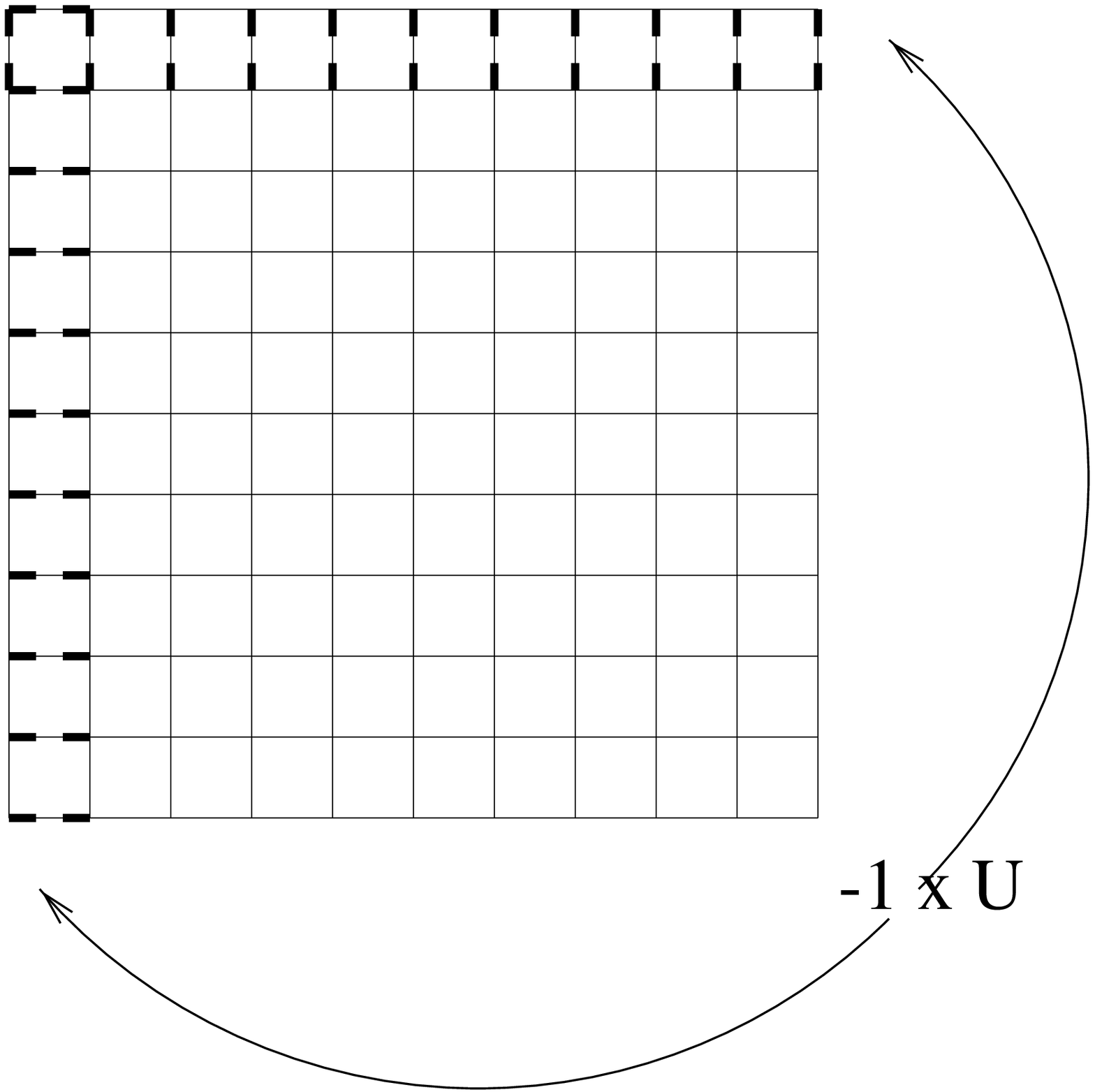}}
\caption{Torelon contributions on the lattice.}
\label{fig:torel}
\end{figure}

One problem encountered is when flux from torelons contributes to $K_{ij}$.  This flux becomes a problem at the higher $\alpha$ necessary to get a significant contribution to K.  As $\alpha$ increases, not only flux along the shortest path between sites i and j but also flux stretching the long way around the lattice will contribute, Figure~\ref{fig:torel:ij}.  As we are interested in operators based on flux along the shortest closed path, this extra flux needs to be removed.  Luckily, a technique for removing torelon flux contributions exists.  For each timeslice, the links perpendicular to a boundary layer in either direction are multiplied by -1 (see Figure~\ref{fig:torel:rem}) and the calculation of the operators is repeated.  When the values are combined with the earlier values, flux which crosses a boundary only once will cancel away, leaving the flux contributions that we desire.  In practice, this procedure is done three times:  once each for either direction alone, and once for both directions together.  Thus we increase our computational load by a factor of 4, not an insignificant amount.

Operators of various sizes may be constructed by these means.  We label the size of the operator by half the number of lattice units on a side of its constituent equilateral triangles, so that, eg, Figure~\ref{fig:triops} is a ``size 2'' triangle, and the full operator Figure~\ref{fig:myops} is also a ``size 2''.  We should note that these are not perfect equilateral triangles, owing to the minor distortion necessary to put the vertices onto sites of the lattice, and so the angles involved are not perfect multiples of $\pi/3$.  A simple center-of-mass calculation for the loops yields the true angles, but in practice the difference is miniscule, and for these calculations perfect $\pi/3$ rotations are assumed.

Let's look in detail at why we need to use two sets of rhombi rotated by $\pi/2$.  Suppose we have loops $l_j$ for $j = 1...N$ identical (in the continuum) upto rotations of $\theta_j$, where $\sum_j \theta_j = 2\pi ({\mathrm{mod}}\theta_1)$.  (The mod$\theta_1$ simply accounts for whatever offset the first loop is at.)  For example, the operator in Figure~\ref{fig:myops} has $\theta_j \in \{\pi/6, 5\pi/6, 3\pi/2\}$.  Consider as a trial operator for definite spin J
\beq L_J = \sum_j e^{iJ\theta_j}l_j . \eeq
Does this mean that its projection onto the vacuum really is of spin J, ie
\beq \langle \Omega|L_J|\tilde{J} \rangle \propto \delta_{J\tilde{J}} \; ? \eeq
Consider
\beq \langle L_J^\dagger L_J \rangle = \sum_{jj'} e^{i(\theta_j - \theta_{j'})J} \langle l_{j'} l_j \rangle . \eeq
Inserting a complete set of states $\sum_{\tilde{J}} |\tilde{J}\rangle \langle \tilde{J}|$ into the expectation value on the r.h.s,
\bea \langle l_{j'} l_j \rangle = \sum_{\tilde{J}} \langle \Omega| l_{j'}|\tilde{J}\rangle \langle \tilde{J}|l_j|\Omega \rangle \eea
by rotating $l_{j'}$ over to $l_j$,
\bea \langle l_{j'} l_j \rangle &=& \sum_{\tilde{J}} e^{i(\theta_{j'} - \theta_j)\tilde{J}} |\langle \Omega|l_j|\tilde{J} \rangle |^2 \\
 &=& \sum_{\tilde{J}} e^{i(\theta_{j'} - \theta_j)\tilde{J}} C_{\tilde{J}} \eea
Thus,
\bea \langle L_J^\dagger L_J \rangle &=& \sum_{\tilde{J}} C_{\tilde{J}} \sum_{jj'} e^{i(\theta_{j'} - \theta_j)(J - \tilde{J})} \\
 &=& \sum_{\tilde{J}} C_{\tilde{J}} F(J,\tilde{J}) , \eea
where
\beq F(J \tilde{J}) = \sum_j e^{i \theta_j (J - \tilde{J})} \sum_{j'} e^{-i \theta_{j'} (J - \tilde{J})} . \eeq
Thus our state with spin J will have no overlap with states of spin $\tilde{J}$ as long as either factor of $F(J\tilde{J})$ is zero.

For example, let's consider using only the three loops shown in Figure~\ref{fig:myops} as our operator, with J=4 and $\theta_j \in \{\pi/6, 5\pi/6, 3\pi/2\}$.  Evaluating the first factor of $F(J\tilde{J})$ we have:
\bea 
\tilde{J}=0 \; &:& e^{4i {\pi\over 6}} + e^{4i {5\pi\over 6}} + e^{4i {3\pi\over 2}} = 0\; \; {\mathrm{(cool...)}} \\
\tilde{J}=2 \; &:& e^{2i {\pi\over 6}} + e^{2i {5\pi\over 6}} + e^{2i {3\pi\over 2}} = 0\; \; {\mathrm{(cool...)}} \\
\tilde{J}=3 \; &:& e^{1i {\pi\over 6}} + e^{1i {5\pi\over 6}} + e^{1i {3\pi\over 2}} = 0\; \; {\mathrm{(cool...)}} \\
\tilde{J}=1 \; &:& e^{3i {\pi\over 6}} + e^{3i {5\pi\over 6}} + e^{3i {3\pi\over 2}} = 3i\; \; {\mathrm{(ouch!)}} \eea
And so this operator is going to have some overlap with the J=1 state.  Now consider adding three more loops to the operator with angles $\theta_j \in \{\pi/2, 7\pi/6, 11\pi/6\}$.  Looking only at its overlap with $\tilde{J}=1$,
\beq
\tilde{J}=1 \; : e^{i 3 {\pi\over 2}} + e^{i 3 {7\pi\over 6}} + e^{i 3 {11\pi\over 6}} = -3i \eeq
which cancels the earlier contribution of 3i from $\tilde{J}=1$.  Thus, by using six loops at the angles given above, we can construct an operator with J=4 and no overlap with states having $J \leq 3$.

\chapter{Performing the Calculation} \label{chap:lattcalc}

For this calculation we work with SU(2) on a D=2+1 Euclidean lattice with L=16 and $\beta=6$.  As we are in pure gauge theory, the only field present is the gauge field defined on the links between lattice sites:
\beq U^{\hat{\mu}}(x_i) = \exp{\int_{x_i}^{x_i + \hat{\mu}} A(x) dx} , \eeq
where $x_i$ gives the site where the link begins and $\hat{\mu}$ is a unit vector in the direction of the link.  Working with the gauge group SU(2), the link variables are SU(2) matrices, which require four real numbers to represent.  The gauge field is updated using the Kennedy-Pendleton heatbath algorithm~\cite{Kennedy:1985nu} supplemented by over-relaxation sweeps~\cite{Creutz:1987xi} and the occasional global gauge transformation.  Calculations are performed on a stored set of 1000 configurations separated by 40 heatbath sweeps each.  Since we want to calculate the rest energy, we take the zero momentum average of our operators in each timeslice,
\beq \CO \rightarrow {1\over {L^2}} \sum_{x_i} \CO(x_i) . \eeq
One Euclidean direction is taken to be the time direction, $x_3$ say, and sites with a common value of $x_3$ comprise a single timeslice.  There are a total of $L^3$ lattice sites, and it is convenient to define $n \equiv L^2$ to be the number of sites in a given timeslice.

\section{Building M}

First, a set of configurations is generated using standard techniques.  After an initial period of 1000 sweeps thermalizing the lattice, configurations are saved every 40 sweeps.  (Actually, they were saved after every 10 sweeps, but to reduce the computational time to something reasonable, only one quarter of that data was used.)  Measurements are then performed on each configuration as follows.

\begin{figure}[!h]
\centering
\includegraphics[scale=.5]{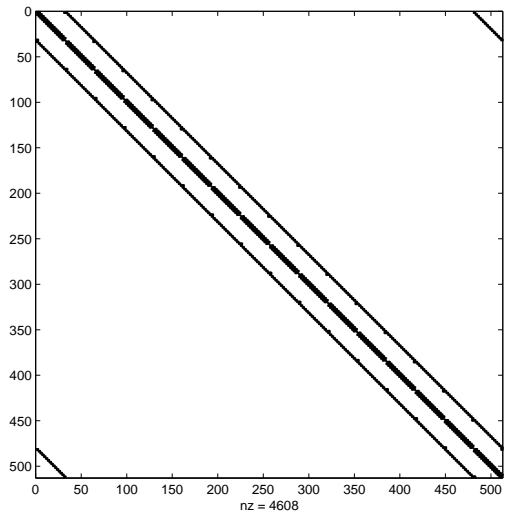}
\caption{The band structure of M.}
\label{fig:M}
\end{figure}

For each timeslice, sites are labelled by an index i = 1...n.  Then, the $n \times n$ matrix M is formed from the gauge fields U such that the entry for the link from site i to site j is
\beq M_{ij} = U_{ij} = U_{ji}^\dagger . \eeq
Note that $M = M^\dagger = {\mathbf{1}}$, so that M is Hermitian.
Because the gauge fields U have a nearest-neighbor relationship, only certain entries of M are non-zero.  For each row i of M, the only non-zero entries are in the four columns j corresponding to the four neighbors of i.  This construction leads to a very sparse matrix M having a band diagonal structure.  For L=16, the diagonals of M are located at offsets (where 0 means the main diagonal):
\beq {\mathrm{diags}} \; = \pm 1,2,3,29,30,31,32,33,479,480,481 , \eeq
See Figure~\ref{fig:M}.  

We wish to calculate
\beq K = {1\over {1 - \alpha M}} , \eeq
and so M is multiplied by $-\alpha$ and added to the $n \times n$ identity matrix to form $K^{-1}$.  Next comes the most computationally expensive part of the calculation:  inverting $K^{-1}$.  Several algorithms were tested for speed and accuracy:  the one used in the calculation was a modified version of the SuperLU program, version 1.1~\cite{Li:1997}.  SuperLU is a good tool to use when inverting any general sparse matrix, however an algorithm specific to the structure at hand will prove useful in future calculations.  

\section{Hermitian Inverse by LU Decomposition} \label{sect:hermLU}

While the following algorithm was not implemented in the current work because of time restrictions, any future work on this technique will find it usefull.
The high bandwidth of $K^{-1}$ means there is much filling in of non-zero entries as the inverse is calculated.  The following algorithm seeks to minimize such fill-ins as well as storage requirements, necessary steps when building a production version of this calculation to handle lattices larger than L=16.  First, we write
\bea K^{-1} &=& \mathbf{1} - \alpha M \\
 &=& L \times U , \eea 
where $K^{-1}$ is Hermitian and L and U are complex.
Taking the $3 \times 3$ case as a graphical example, let's display all the degrees-of-freedom:
\beq \left[\begin{array}{ccc} R_{11} & R_{21}-I_{21} & R_{31}-I_{31} \\ 
R_{21}+I_{21} & R_{22} & R_{32}-I_{32} \\ R_{31}+I_{31} & R_{32}+I_{32} & R_{33} \end{array}\right] = \left[\begin{array}{ccc} L & 0 & 0 \\ L & L & 0 \\ L & L & L \end{array}\right] \times \left[\begin{array}{ccc} U & U & U \\ 0 & U & U \\ 0 & 0 & U \end{array}\right] \eeq
Counting degrees of freedom:
\beq {n(n+1)\over 2} {\mathrm{real}} + {n(n-1)\over 2} {\mathrm{imag}} \sim n(n+1) \; \oplus \; n(n+1) \eeq
The l.h.s has $n^2$ degrees of freedom, while the r.h.s. has $2 \times n(n+1)$.  If we impose the conditions $U=L^\dagger$ and $L_{ii}=L_{ii}^\star$, then we are back down to $n^2$ on the r.h.s.  As the entries on the diagonal are unity while all other entries have a magnitude less than unity, the property of diagonal dominance insures we can ignore pivoting.  The storage requirements for the algorithm can be met with one full, complex matrix of size $n \times n$.  First, store L in the lower part of that matrix via forward substitution (read Numerical Recipes \cite{Press:1992} for more details of forward and backward substitution):
\beq L_{jj} = \sqrt{K^{-1}_{jj} - \sum_{i < j}|L_{ji}|^2} \quad ; \quad
L_{ij} = {1\over L_{jj}}(K^{-1}_{ij} - \sum_{k=1}^{j-1}L_{ik}L_{jk}^\star) \quad \mathrm{for\;} i>j  \eeq
Then compute $L^{-1}$, using a transposed storage scheme into the upper portion of our storage matrix and neglecting to store the diagonal elements (which are trivially related to L's diagonal elements):
\beq L^{-1}_{jj} = {1\over L_{jj}} \quad ; \quad L^{-1}_{ij} = {-1\over L_{ii}} \sum_{k=1}^{j-1} L_{ik} L_{kj}^{-1} \quad \mathrm{for \;} i>j \eeq
Next, compute $\tilde{K}$, where the tilde indicates that we are close but not quite there yet, using back substitution and storing over $L^{-1}$:
\beq \tilde{K}_{nn} = {L^{-1}_{nn}\over L^\star_{nn}} \quad ; \quad \tilde{K}_{ij} = {1\over L_{ii}^\star} (L^{-1}_{ij} - \sum_{k=i+1}^n L^\star_{ik} \tilde{K}_{kj}) \quad \mathrm{for\;} i<j \eeq
If we write $K = \tilde{K} + \delta$, some iterative improvement gives us the error $\delta$
\bea A_{ij} &=& \sum_k K^{-1}_{ik} \tilde{K}_{kj} \quad ; \quad A \rightarrow A - \mathbf{1} \\
\delta_{jj} &=& {1\over A_{jj}} \quad ; \quad \delta_{ij} = {-1\over A_{ii}} \sum_{k=1}^{j-1} A_{ik} \delta_{kj} \eea
which may be subtracted off from $\tilde{K}$ to give our final K.  Schematically, the storage requirements go like:
\begin{displaymath}
\left[\begin{array}{cc} \ddots & \\ L & \ddots \end{array}\right] \: , \:
\left[\begin{array}{cc} \ddots & L^{-1}\\ L & \ddots \end{array}\right] \: , \:
\left[\begin{array}{cc} \ddots & \tilde{K} \\ L & \ddots \end{array}\right] \: , \:
\left[\begin{array}{cc} \ddots & \tilde{K} \\ A & \ddots \end{array}\right] \: , \:
\left[\begin{array}{cc} \ddots & \tilde{K} \\ \delta & \ddots \end{array}\right]
\end{displaymath}
until we fill the entire matrix with K.  The last stage of iterative improvement may be done efficiently by taking into account the sparsity of $K^{-1}$.

\section{Evaluating the operators}

Once we have our matrix K we can begin evaluating our operators.  We use our rhomboid operators at sizes 3,4,5, and 6.  Size 2 was also computed, but never showed much promise for being useful, and so is hereafter neglected.  For each site in a timeslice we compute the values for all six loops and all four sizes (and for several values of alpha).  Then we compute the normalizing constants $c_\theta$ for the loops via Equation~\ref{eq:normc}.  If everything were perfectly symmetric, these would all be unity, but the slight distortions caused by the cubic lattice will produce non-zero normalizing constants.  A measure of how similar our loops are is given by $c_\theta$.  We now have the choice of applying our constants $c_\theta$ either to all of the loops at hand or only to those two loops which are obviously going to be different than the other four, namely loop 3 in Figure~\ref{fig:myops} and its $\pi/2$ analogue.  If we apply our constants to the two loops only, we are expecting the $\pi/2$ analogues to produce cancellations (over the duration of the calculation) when computing the J=2 so that there is no overlap with the J=0.  If, however, we apply our constants to all the loops, we expect no fortuitous cancellations and instead rely on the actual phases applied to our loops to affect the cancellations.  In practice, there is not much difference between the two methods.  If anything, the more general method of normalizing all the loops does produce a slightly better result and will be the focus for the remainder of this work.  We can measure this effect by looking at the overlap of our J=2 operator with the J=0:  ideally, subtracting $\pi/2$ analogues directly, ie $\psi_{J=2} = \psi_{\theta} - \psi_{\theta+\pi/2}$, will produce an overlap equal to the statistical error of the calculation.  As we are looking for a method which can be applied to even more general situations than the current one, we normalize all the loops and believe the calculation of the overlap to be the true overlap of our operators (upto statistical noise).

We now have a set of operators $\CO$ with several indices:  size, loop orientation, and $\alpha$.  Actually, we have four sets of these operators in order to cancel the torelons.  An important detail is that the torelon cancellation must be done {\em before} the normalizing constants are applied.  In order for the torelon flux to cancel, the four sets of loops are averaged together to produce a torelon canceled set of loops, and then the normalizing constants applied.  Any other ordering of these steps will not work.
We wish to reduce all these numbers down to a few correlation functions from which we can estimate masses.  To produce an operator coupling to a particular spin J, we add our six loops together with the appropriate phases.  Note that the center of the first loop, number 1 in Figure~\ref{fig:myops}, is at angle $\pi/6$.  Including notation for the normalizing constants, we have explicitly:
\bea \CO_{J=0} &=& \psi_{\pi/6} + c_{5\pi/6} \psi_{5\pi/6} + c_{3\pi/2} \psi_{3\pi/2} + \nonumber \\
 & & c_{\pi/2} \psi_{\pi/2} + c_{7\pi/6} \psi_{7\pi/6} + c_{11\pi/6} \psi_{11\pi/6} \eea
\bea \CO_{J=2} &=& e^{2i \pi/6} \psi_{\pi/6} + e^{2i 5\pi/6} c_{5\pi/6} \psi_{5\pi/6} + e^{2i 3\pi/2} c_{3\pi/2} \psi_{3\pi/2} + \nonumber \\
 & & e^{2i \pi/2} c_{\pi/2} \psi_{\pi/2} + e^{2i 7\pi/6} c_{7\pi/6} \psi_{7\pi/6} + e^{2i 11\pi/3} c_{11\pi/6} \psi_{11\pi/6} \label{eqn:O_2} \eea
\bea \CO_{J=4} &=& e^{4i \pi/6} \psi_{\pi/6} + e^{4i 5\pi/6} c_{5\pi/6} \psi_{5\pi/6} + e^{4i 3\pi/2} c_{3\pi/2} \psi_{3\pi/2} + \nonumber \\
 & & e^{4i \pi/2} c_{\pi/2} \psi_{\pi/2} + e^{4i 7\pi/6} c_{7\pi/6} \psi_{7\pi/6} + e^{4i 11\pi/3} c_{11\pi/6} \psi_{11\pi/6} \eea
As the vacuum has the same quantum numbers as our ground state, $0^{++}$, we need to use vacuum-subtracted operators for the J=0 correlation function, and as we are unsure of what is happening at J=4, we use vacuum-subtracted operators there too, just for good measure.
\beq \tilde{\CO} = \CO - <\CO> \eeq

Now that we have our operators calculated for each timeslice, we are in position to compute our correlation functions:
\bea C_J(t) &=& <\tilde{\CO_J}^\star(t) | \tilde{\CO_J}(0)> \\
 &=& <(\CO_J^\star(t) - <\CO_J^\star>) | (\CO_J(0) - <\CO_J>)> \\
 &=& <\CO_J^\star(t) | \CO_J(0)> - <\CO_J^\star><\CO_J> \eea
These correlation function will be normalized such that $C_J(0) = 1$, ie
\beq C_J(t) \rightarrow C_J(t)/C_J(0) \eeq
Similarly, we can define and compute the normalized cross-correlation for states of different J, eg:
\beq C_{40}(t) = {<\tilde{\CO}^\star_4(t) | \tilde{\CO}_0>\over \sqrt{<\tilde{\CO}^\star_4(t) | \tilde{\CO}_4(0)><\tilde{\CO}^\star_0(t) | \tilde{\CO}_0(0)>}} \label{eq:overlap} \eeq
If we take $t=0$ in Equation~\ref{eq:overlap}, we have computed the actual overlap of our operators.  This number will be our measure of how well we are distinguishing the J=4 from the J=0.
From these correlation functions we can extract effective masses in units of the lattice spacing a:
\beq m_{eff}^J(t) = \ln{C_J(t)\over C_J(t+1)} \eeq

\section{Error analysis}

To estimate the errors on our secondary quantities, ie our correlation functions C and effective masses m, we perform a standard jackknife analysis~\cite{Montvay:1994}.  Consider a primary quantity x with N measured values ${x_i}, i=1...N$.  The sample average is simply $\bar{x} = {1\over N}\sum_{i=1}^N x_i$.  Now consider a secondary quantity $y=y(x)$.  Our best estimate of y is given by
\beq \bar{y} = y(\bar{x}) \neq \overline{y(x)} \eeq
The jackknife procedure estimates the error on y by evaluating the distribution of N jackknife estimates $y^J$ calculated with one sample missing from ${x}$ for each, ie:
\beq x^J_j = {1\over {N-1}}\sum_{i \neq j} x_i \eeq
Then, the jackknife estimates are $y^J_j = y(x^J_j)$ with mean
\beq \bar{y}^J = {1\over N} \sum_{j=1}^N y^J_j \eeq
and variance
\beq \sigma^2_{\bar{y}^J} = {{N-1}\over N} \sum_{j=1}^N (y^J_j - \bar{y}^J)^2 \eeq
For primary quantities this variance is equivalent to the simple variance of the measurements, while for secondary quantities (such as effective masses) it produces a more stable estimate of the error
\beq \bar{y} = \bar{y} \pm \sigma_{\bar{y}^J} \eeq
These are the errors reported in the following chapter.

\chapter{Results From the Lattice} \label{chap:lattres}

\section{Preliminaries}

\begin{figure}[t]
\centering
\includegraphics[scale=.5]{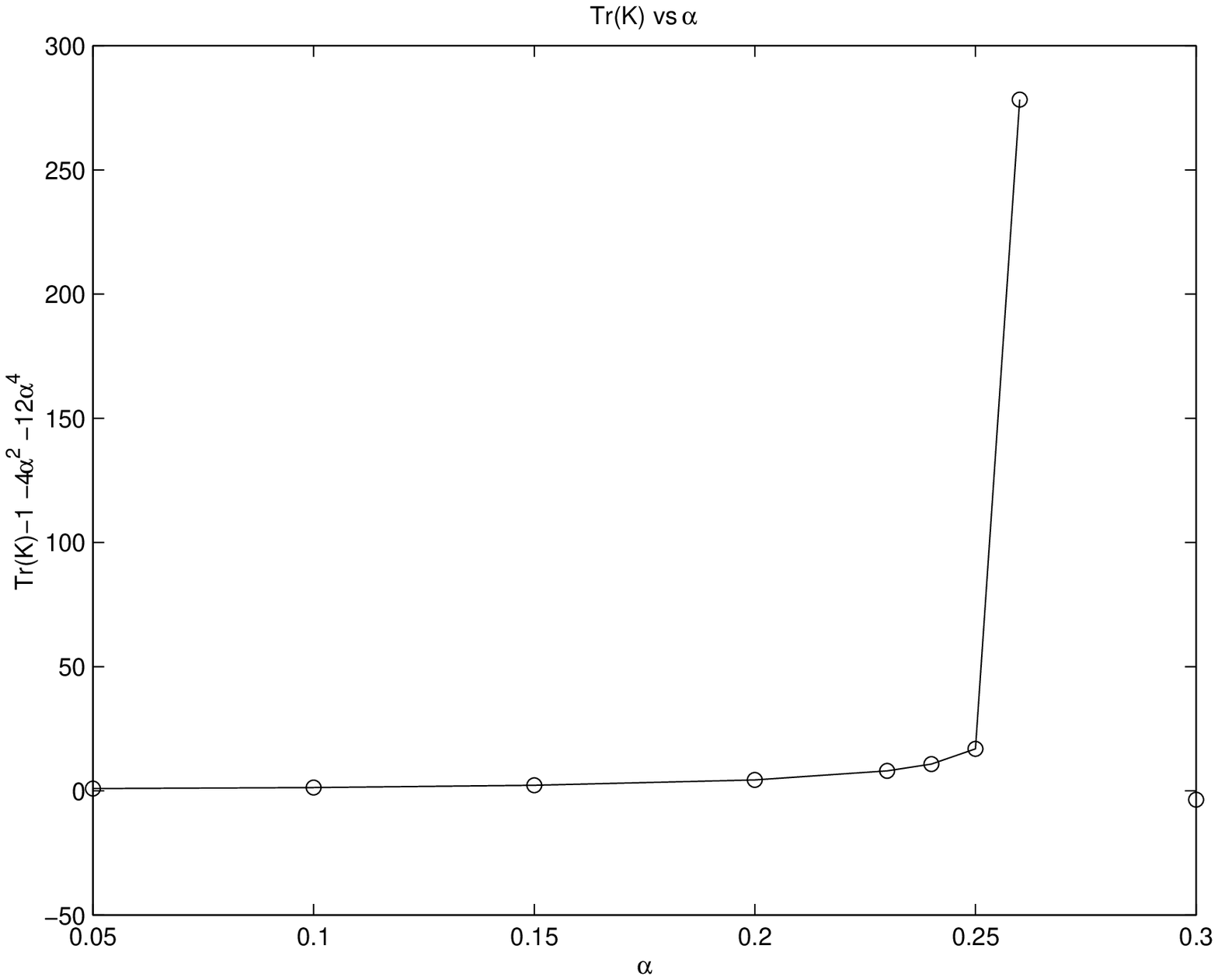}
\caption{$<Tr\,K>$ as a function of $\alpha$.}
\label{fig:traceK}
\end{figure}

First, let's look at the behavior of $<Tr\,K>$ as a function of $\alpha$.  As $<Tr\,K>$, for small enough $\alpha$, is a measure of the expectation value of a plaquette, upto some constants, its value gives us an idea of how our calculation is performing.  As our matrix M is in the denominator of our expression for K, Equation~\ref{eqn:K}, there exists a critical $\alpha_c$ such that K diverges.  We can estimate this $\alpha_c$ by evaluating the Frobenius norm of M.  In general, M will be a complex matrix, and its norm will be difficult to evaluate analytically.  However, we can derive our result by considering a simple limiting case:  that of a minimal two dimensional lattice at its ``cold'' start before thermalization.  This minimal lattice possesses 4 sites arranged as a square (or plaquette).  Using the symbols
\beq \eta = \left[\begin{array}{cc} 0 & 1 \\ 1 & 0 \end{array}\right] \; \mathrm{and} \; \zeta = \left[\begin{array}{cc} 0 & 0 \\ 0 & 0 \end{array}\right] \; , \eeq
M is given by
\beq M_{cold} = \left[\begin{array}{cccc} \zeta & \eta & \eta & \zeta \\ \eta & \zeta & \zeta & \eta \\ \eta & \zeta & \zeta & \eta  \\ \zeta & \eta & \eta & \zeta \end{array} \right] \; . \eeq
The Frobenius norm of this matrix is easily calculated to equal 4.  Thus, with an $\alpha_c = 1/4$ the norm of the denominator of Equation~\ref{eqn:K} will equal 0, and $K^{-1}$ will be non-invertible.  For a ``hot'' configuration, the unitary matrices making up M will be rotated away from their principle axes, and the norm of M will be slightly less than 4, so that $\alpha_c > .25$.
While these are heuristic arguments, we see precisely this behavior in actual calculation.  As $\alpha \rightarrow 0$, $<Tr\,K>$ approaches $<[]>$; as $\alpha \rightarrow .25$, $<Tr\,K>$ diverges.  Figure~\ref{fig:traceK} plots Equation~\ref{eqn:kplaq} as a function of $\alpha$.  We easily see $<Tr\,K>$ diverging as it approaches some $\alpha \sim .25$.  Actually, the final positive point is at $\alpha = .26$, indicating that $\alpha_c > .25$.  The last point, at $\alpha = .3$, has a negative value for $<Tr\,K>$, an unphysical result if $<Tr\,K>$ is to represent the expectation value of some sum of closed loops on the lattice.  Calculations performed at $\alpha > \alpha_c$ yield operators with a divergent vacuum expectation value and correlation functions which fall to 0 after only one lattice spacing.  While this regime is not useful to the present calculation, perhaps future work will find something worthwhile to explore.

Next, let's look at the normalizing constants $c_\theta$ which are applied to our individual loops to overcome any spatial rotational discrepancies our finite cubic lattice imposes.  In Table~\ref{table:nc} we display these constants for the six loops at the four values of $\alpha$ used in this calculation.  We notice that as $\alpha$ increases these constants approach unity.  Apparently, as longer and more spread-out paths contribute to the loop, the orientation of the loop on the lattice becomes less significant, which makes sense.  The size 6 loops show the most variation between orientations, to the order of $10^7$ at $\alpha=.1$!  Even at the more modest $\alpha=.2$ the variation is of the order $10^3$.  If we did not apply these normalization constants, we would not get a balanced contribution from the different orientations of loop, but perhaps we should be concerned about using such large normalization constants!  At least the behaviour of these constants is doing what we would expect:  the four equivalent orientations have constants consistently close to unity, the smallest loop needs the least correction, and the largest loop has its smallest (non-unity) constants close to $\alpha_c$ where the extreme spread of the loops should help cancel the difference in orientation.  Thus, we proceed to take these constants at face value and implement them in the calculation.

\begin{table}[!h]
\centering
\begin{tabular}{|c|cccc|} \hline

loop & & \multicolumn{2}{c}{size}  & \\
\# & 3 & 4 & 5 & 6 \\\hline

 &  &  \multicolumn{2}{c}{$\alpha=.1$}    & \\\hline
1 & 1.0000e+00 & 1.0000e+00 & 1.0000e+00 & 1.0000e+00 \\
2 & 1.0082e+00 & 9.6166e-01 & 1.0237e+00 & 1.0161e+00 \\
3 & 1.8708e+01 & 1.3566e+02 & 1.3982e+05 & 1.4130e+07 \\
4 & 1.0055e+00 & 9.5868e-01 & 1.0346e+00 & 1.0211e+00 \\
5 & 9.9742e-01 & 9.7717e-01 & 1.0386e+00 & 1.0196e+00 \\
6 & 1.8997e+01 & 1.4096e+02 & 1.4026e+05 & 1.4339e+07 \\\hline

 &  &  \multicolumn{2}{c}{$\alpha=.15$}    & \\\hline
1 & 1.0000e+00 & 1.0000e+00 & 1.0000e+00 & 1.0000e+00 \\
2 & 9.9819e-01 & 9.5151e-01 & 1.0282e+00 & 1.0201e+00 \\
3 & 4.6171e+00 & 1.7857e+01 & 2.5579e+03 & 8.5268e+04 \\
4 & 1.0044e+00 & 9.5565e-01 & 1.0428e+00 & 1.0210e+00 \\
5 & 9.9262e-01 & 9.7976e-01 & 1.0322e+00 & 1.0253e+00 \\
6 & 4.7088e+00 & 1.8702e+01 & 2.5747e+03 & 8.6693e+04 \\\hline

 &  &  \multicolumn{2}{c}{$\alpha=.2$}    & \\\hline
1 & 1.0000e+00 & 1.0000e+00 & 1.0000e+00 & 1.0000e+00 \\
2 & 9.8341e-01 & 1.0141e+00 & 1.0188e+00 & 1.0209e+00 \\
3 & 1.9445e+00 & 4.8168e+00 & 1.2366e+02 & 1.5746e+03 \\
4 & 9.9933e-01 & 9.9477e-01 & 1.0340e+00 & 1.0198e+00 \\
5 & 9.8311e-01 & 1.0596e+00 & 1.0502e+00 & 1.0290e+00 \\
6 & 1.9989e+00 & 4.9811e+00 & 1.2663e+02 & 1.6053e+03 \\\hline

 &  &  \multicolumn{2}{c}{$\alpha=.25$}    & \\\hline
1 & 1.0000e+00 & 1.0000e+00 & 1.0000e+00 & 1.0000e+00 \\
2 & 9.7862e-01 & 9.6075e-01 & 9.9969e-01 & 9.7553e-01 \\
3 & 1.0331e+00 & 1.0580e+00 & 3.3177e+00 & 1.5299e+01 \\
4 & 9.9450e-01 & 9.7818e-01 & 9.7776e-01 & 9.8693e-01 \\
5 & 9.5446e-01 & 9.7849e-01 & 1.0289e+00 & 1.0236e+00 \\
6 & 1.1033e+00 & 1.1611e+00 & 3.3920e+00 & 1.5382e+01 \\\hline

\end{tabular}
\caption{Loop normalization constants.}
\label{table:nc}
\end{table}

\section{J = 0 and the effect of torelons}

How well does this technique work?  With only 1000 measurements, we are dealing with really low statistics.  That these measurements are separated by 40 thermal sweeps each rather than the conventional 10 should help reduce the autocorrelations a bit, but we really should not expect too much an exact agreement with earlier lattice results~\cite{Teper:1997tq}.  At this $\beta$ and lattice size, 6 and 16 respectively, the mass of the $0^{++}$ has been determined to be 1.193(18) lattice units, using a standard notation for the error -- this is the number we are shooting for.  Let's look at our correlation functions and see how we do.

In Table~\ref{table:corr0p1} we have correlation functions before torelon subtraction.  For lower $\alpha$, the correlation function is overcome by noise around t=3, at least for the larger operators.  At higher $\alpha$ we have a good correlation function for the size 4 operator -- this operator seems to be our best for extracting the $0^{++}$ mass.  The corresponding masses, and their errors, and shown in Table~\ref{table:mass0p1}.

After the torelon cancellation has been performed, we get the correlation functions in Table~\ref{table:corr0p4} and the masses in Table~\ref{table:mass0p4}.  We see that the correlation functions are better behaved by one more lattice spacing, lending more credibility to these mass estimates.  In general, there is more consistency when the torelons are canceled.  Also, the error on the mass estimates has been reduced.  Looking at the second column of effective masses, we see that our best estimate of the $0^{++}$ goes from 1.09(10) down to 1.06(08).  We might even be seeing a mass plateau, as the next effective mass is 1.04(24), but saying that could be stretching our interpretation of the data by more than a little bit.  While slightly low numerically, these estimates do agree well, within errors, with the canonical value of 1.193(18).

\clearpage

\begin{table}[!t]
\centering
\begin{tabular}{|c|cccc|} \hline
t =  & 0 & 1 & 2 & 3 \\\hline

size &  &  \multicolumn{2}{c}{$\alpha=.1$}    & \\\hline
3 & 1 & 2.5339e-01 & 6.7977e-02 & 1.3958e-02 \\
4 & 1 & 2.0777e-01 & 4.8134e-02 &-3.2623e-03 \\
5 & 1 & 2.2440e-01 & 5.3068e-02 &-3.4754e-03 \\
6 & 1 & 2.5338e-01 & 7.1044e-02 & 9.0107e-03 \\\hline

 &  &  \multicolumn{2}{c}{$\alpha=.15$}    & \\\hline
3 & 1 & 2.5854e-01 & 7.0588e-02 & 1.4937e-02 \\
4 & 1 & 2.2336e-01 & 5.3347e-02 &-2.1334e-03 \\
5 & 1 & 2.3454e-01 & 5.5890e-02 &-2.3498e-03 \\
6 & 1 & 2.5892e-01 & 7.3078e-02 & 9.7205e-03 \\\hline

 &  &  \multicolumn{2}{c}{$\alpha=.2$}    & \\\hline
3 & 1 & 2.6267e-01 & 7.3012e-02 & 1.5629e-02 \\
4 & 1 & 2.4336e-01 & 6.2958e-02 & 8.0794e-04 \\
5 & 1 & 2.4717e-01 & 5.9903e-02 &-1.8301e-03 \\
6 & 1 & 2.6206e-01 & 7.3653e-02 & 9.2552e-03 \\\hline

 &  &  \multicolumn{2}{c}{$\alpha=.25$}    & \\\hline
3 & 1 & 2.2035e-01 & 5.9404e-02 & 8.9081e-03 \\
4 & 1 & 2.4352e-01 & 8.2043e-02 & 1.9869e-02 \\
5 & 1 & 2.3038e-01 & 6.4103e-02 & 4.1644e-03 \\
6 & 1 & 2.2014e-01 & 5.1320e-02 &-6.0157e-03 \\\hline

\end{tabular}
\caption{Correlation functions for $0^{++}$ before torelon removal.}
\label{table:corr0p1}
\end{table}

\begin{table}[!b]
\centering
\begin{tabular}{|c|cccc|} \hline
t =  &  & 1 & 2 & 3 \\\hline

size &  &  \multicolumn{2}{c}{$\alpha=.1$}    & \\\hline
3 & &  1.37(03) & 1.32(10) & 1.58(50)  \\
4 & &  1.57(03) & 1.46(12) & -  \\
5 & &  1.49(03) & 1.44(13) & -  \\
6 & &  1.37(03) & 1.27(11) & 2.06(88)  \\\hline

 &  &  \multicolumn{2}{c}{$\alpha=.15$}    & \\\hline
3 & &  1.35(03) & 1.30(10) & 1.55(45) \\
4 & &  1.50(03) & 1.43(12) & -    \\
5 & &  1.45(03) & 1.43(13) & -    \\
6 & &  1.35(03) & 1.27(10) & 2.02(79) \\\hline

 &  &  \multicolumn{2}{c}{$\alpha=.2$}    & \\\hline
3 & &  1.34(03) & 1.28(10) & 1.54(45) \\
4 & &  1.41(03) & 1.35(11) & -     \\
5 & &  1.40(03) & 1.42(13) & -     \\
6 & &  1.34(03) & 1.27(10) & 2.07(80)  \\\hline

 &  &  \multicolumn{2}{c}{$\alpha=.25$}    & \\\hline
3 & &  1.51(04) & 1.31(14) &  1.90(86)  \\
4 & &  1.41(04) & 1.09(10) &  1.42(38)  \\
5 & &  1.47(04) & 1.28(14) &  -     \\
6 & &  1.51(04) & 1.46(17) &  -     \\\hline

\end{tabular}
\caption{Masses for $0^{++}$ before torelon removal.}
\label{table:mass0p1}
\end{table}

\begin{table}[!t]
\centering
\begin{tabular}{|c|cccc|} \hline
t = & 0 & 1 & 2 & 3 \\\hline

size &  &  \multicolumn{2}{c}{$\alpha=.1$}    & \\\hline
3 & 1 & 2.5339e-01 & 6.7980e-02 & 1.3961e-02 \\
4 & 1 & 2.2010e-01 & 4.9200e-02 & 8.4372e-03 \\
5 & 1 & 2.2544e-01 & 5.4961e-02 &-2.2496e-03 \\
6 & 1 & 2.5348e-01 & 7.1141e-02 & 9.0913e-03 \\\hline

 &  &  \multicolumn{2}{c}{$\alpha=.15$}    & \\\hline
3 & 1 & 2.5858e-01 & 7.0627e-02 & 1.4970e-02 \\
4 & 1 & 2.3390e-01 & 5.8306e-02 & 1.1263e-02 \\
5 & 1 & 2.3729e-01 & 6.0708e-02 & 9.6583e-04 \\
6 & 1 & 2.5955e-01 & 7.3749e-02 & 1.0272e-02 \\\hline

 &  &  \multicolumn{2}{c}{$\alpha=.2$}    & \\\hline
3 & 1 & 2.6312e-01 & 7.3452e-02 & 1.6011e-02 \\
4 & 1 & 2.5080e-01 & 7.0270e-02 & 1.4823e-02 \\
5 & 1 & 2.5306e-01 & 7.0020e-02 & 6.0936e-03 \\
6 & 1 & 2.6515e-01 & 7.7361e-02 & 1.2315e-02 \\\hline

 &  &  \multicolumn{2}{c}{$\alpha=.25$}    & \\\hline
3 & 1 & 2.2624e-01 & 6.8414e-02 & 2.0320e-02 \\
4 & 1 & 2.4196e-01 & 8.3649e-02 & 2.9644e-02 \\
5 & 1 & 2.3288e-01 & 7.2597e-02 & 1.7940e-02 \\
6 & 1 & 2.2597e-01 & 6.5776e-02 & 1.1193e-02 \\\hline

\end{tabular}
\caption{Correlation functions for $0^{++}$ after torelon removal.}
\label{table:corr0p4}
\end{table}

\begin{table}[!b]
\centering
\begin{tabular}{|c|cccc|} \hline
t =  &  & 1 & 2 & 3 \\\hline

size &  &  \multicolumn{2}{c}{$\alpha=.1$}    & \\\hline
3 & &  1.37(03) & 1.32(10) & 1.58(50)  \\
4 & &  1.51(04) & 1.50(18) & 1.76(84)  \\
5 & &  1.49(03) & 1.41(12) & -         \\
6 & &  1.37(03) & 1.27(10) & 2.06(87)  \\\hline

 &  &  \multicolumn{2}{c}{$\alpha=.15$}    & \\\hline
3 & &  1.35(03) & 1.30(10) & 1.55(45) \\
4 & &  1.45(03) & 1.39(14) & 1.64(66) \\
5 & &  1.44(03) & 1.36(10) & -        \\
6 & &  1.35(03) & 1.26(10) & 1.97(75) \\\hline

 &  &  \multicolumn{2}{c}{$\alpha=.2$}    & \\\hline
3 & &  1.34(03) & 1.28(10) & 1.52(40) \\
4 & &  1.38(03) & 1.27(10) & 1.57(52) \\
5 & &  1.37(03) & 1.28(08) & -        \\
6 & &  1.33(03) & 1.23(09) & 1.84(59) \\\hline

 &  &  \multicolumn{2}{c}{$\alpha=.25$}    & \\\hline
3 & &  1.49(04) & 1.20(07) &  1.21(32)  \\
4 & &  1.42(04) & {\bf{1.06(08)}} &  1.04(24)  \\
5 & &  1.46(04) & 1.17(08) &  1.40(41)  \\
6 & &  1.49(04) & 1.23(08) &  1.77(65)  \\\hline

\end{tabular}
\caption{Masses for $0^{++}$ after torelon removal.}
\label{table:mass0p4}
\end{table}

\clearpage

\section{J = 2:  Coupling to higher spin states}

Now that we are confident in our techniques for producing spin 0 correlation functions and for removing the torelon contributions, we turn our attention to states with higher spin.  The next lightest state has J=2.  These operators have the phase contributions given by Equation~\ref{eqn:O_2}.  This operator is our first real test of our rotational phase method.  For brevity, we will only present the results after torelon cancellation has been applied.  The conventional estimate for the mass of the $2^{++}$, at these parameter values, from~\cite{Teper:1997tq}, is 1.80(8).

Table~\ref{table:corr2p4} displays our correlation functions, and Table~\ref{table:mass2p4} the corresponding masses.  These correlation functions fall to zero much faster, and so we can only go out to t=2 before statistical noise dominates.  This state should be heavy enough that we can extract our estimate from the first column of effective masses.  Also shown in the last column of the mass table is the overlap with the corresponding $0^{++}$ operator.  These overlaps should tell us how well our operator is coupling to the J=2.  The lowest estimate of the mass is 1.63(06); the next lowest is 1.68(05).  These states have an overlap of about 1.5\%.  The third lowest estimate is 1.79(07), again with overlap of 1.5\%.  This amount of overlap is not sufficient to lower a mass of 1.8 down to 1.6:
\beq (1.5 \times 1.2 + 98.5 \times 1.8)/100 = 1.79 \; .\eeq
We might be wary of accepting the estimate coming from $\alpha=.25$, as the only correlation function not hitting the statistical noise at t=2 is the size 4 operator giving us our lowest mass.
Nonetheless, these are all in the $2\sigma$ ballpark of 1.80(8), even if a little low.  Thus, we do conclude that this operator is coupling effectively to the J=2.  Perhaps future calculations with better statistics will overcome this minor discrepancy.

\begin{table}[!h]
\centering
\begin{tabular}{|c|ccc|} \hline
t = & 0 & 1 & 2  \\\hline

size  &   & \multicolumn{2}{c}{$\alpha=.1$}  \\ \hline
3 & 1 & 8.0769e-02 & 1.1711e-02 \\
4 & 1 & 9.3026e-02 & 1.8498e-02 \\
5 & 1 & 8.0629e-02 & 7.3778e-03 \\
6 & 1 & 9.8041e-02 & 3.5999e-03 \\\hline

  &   & \multicolumn{2}{c}{$\alpha=.15$}  \\ \hline
3 & 1 & 8.4990e-02 & 1.0426e-02 \\
4 & 1 & 1.2780e-01 & 3.0552e-02 \\
5 & 1 & 9.8406e-02 & 1.3467e-02 \\
6 & 1 & 1.0109e-01 & 1.6209e-03 \\\hline

  &   & \multicolumn{2}{c}{$\alpha=.2$}   \\ \hline
3 & 1 & 8.6963e-02 & 5.0222e-03 \\
4 & 1 & 1.8721e-01 & 4.8227e-02 \\
5 & 1 & 1.3880e-01 & 2.6834e-02 \\
6 & 1 & 9.9971e-02 &-4.0202e-03 \\\hline

  &   & \multicolumn{2}{c}{$\alpha=.25$}   \\ \hline
3 & 1 & 1.2540e-01 &-9.1018e-03 \\
4 & 1 & 1.9512e-01 & 8.5575e-03 \\
5 & 1 & 1.6615e-01 &-2.0638e-05 \\
6 & 1 & 1.2756e-01 &-6.6520e-04 \\\hline

\end{tabular}
\caption{Correlation functions for $2^{++}$ after torelon removal.}
\label{table:corr2p4}
\end{table}

\begin{table}[!h]
\centering
\begin{tabular}{|c|ccc|c|} \hline
t =  & & 1 & 2 & $|<\CO_2 | \CO_0>|$ \\\hline

size  & &  \multicolumn{2}{c}{$\alpha=.1$} & \\\hline
3 & &  2.52(08) & 1.93(05) & .005(4) \\
4 & &  2.37(07) & 1.62(03) & .007(3) \\
5 & &  2.52(09) & 2.39(07) & .008(6) \\
6 & &  2.32(06) & -        & .006(7) \\\hline

  & &  \multicolumn{2}{c}{$\alpha=.15$} & \\\hline
3 & &  2.47(08) & 2.10(50) & .005(4) \\
4 & &  2.06(06) & 1.43(19) & .011(3) \\
5 & &  2.32(08) & 1.99(42) & .011(5) \\
6 & &  2.29(06) & -        & .007(7) \\\hline

  & &  \multicolumn{2}{c}{$\alpha=.2$}  & \\\hline
3 & &  2.44(08) & 2.85(1.15)& .004(4) \\
4 & &  1.68(05) & 1.36(13)  & .016(4) \\
5 & &  1.97(08) & 1.64(23)  & .014(5) \\
6 & &  2.30(06) & -         & .008(6) \\\hline

  & &  \multicolumn{2}{c}{$\alpha=.25$} &  \\\hline
3 & &  2.08(07) & -         & .009(6) \\
4 & &  {\bf{1.63(06)}} & 3.13(1.11)& .015(7) \\
5 & &  1.79(07) & -         & .015(6) \\
6 & &  2.06(07) & -         & .010(4) \\\hline

\end{tabular}
\caption{Masses for $2^{++}$ after torelon removal.  The last column is the overlap with the $0^{++}$.}
\label{table:mass2p4}
\end{table}

\clearpage

\section{J = 4: The acid test}

Finally, we come to the operator we have all been waiting for:  the J=4.  Our hypothesis is that earlier lattice calculations have misidentified the pseudoscalar as being J=0 when it in fact has J=4 (and hence is not a pseudoscalar).  The state in question is the $0^{-+}$ with a mass of 2.10(33), from~\cite{Teper:1997tq}.  Our technique as developed so far only produces PC quantum numbers of ++.  However, gauge theory in D=2+1 dimensions posses the curious property that in general states with $J \neq 0$ come in degenerate parity doublets .  Thus, while we might measure the mass of a state $4^{++}$, we may infer that there exists a state $4^{-+}$ with equivalent mass.

The correlation functions for J=4 are presented in Table~\ref{table:corr4p4}, and the corresponding masses in Table~\ref{table:mass4p4}.  Again, the overlap with the $0^{++}$ is shown in the last column of the mass table.  This column is the one we are most interested in.  Whatever operator has the lowest number here is our best estimate for the $4^{++}$.  Scanning down these numbers, one state jumps out, the size 6 at $\alpha=.2$.  This state has an overlap of less than 1\%!  The next best operators have overlaps in the 5\% range, too much for us to consider their masses as accurate.  The mass of this state is 2.13(04).  Let us emphasize that this error is the statistical error resulting from the jackknife analysis, and does not include the unknown systematic effects, which may be far greater.  Nonetheless, it is pleasing to have such a tight statistical error resulting from a new technique of calculation.
While this operator couples to the $4^{++}$ state, we invoke parity doubling to claim that there exists a state with quantum numbers $4^{-+}$ at the same mass.  Thus, we claim that the conventional assignment of $0^{-+}$ to the state measured with a mass of 2.10(33) is mistaken, and that this state really posses spin 4.  Conventional lattice gauge theory needs to reevaluate how spin is assigned within the various symmetry channels in light of this result.

\begin{figure}[!h]
\centering
\includegraphics[width=.6\textwidth]{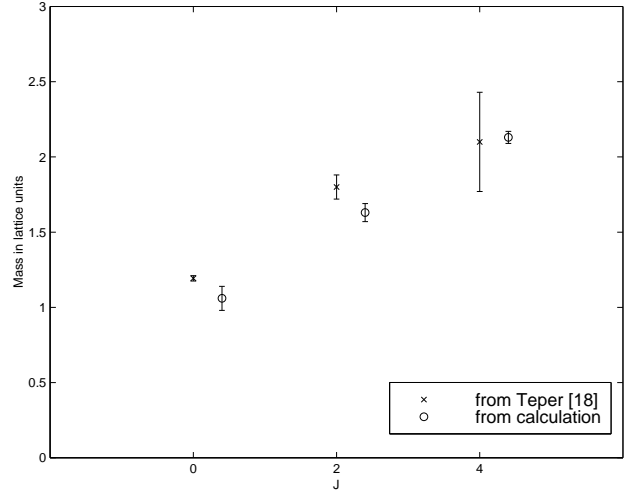}
\caption{The results of the lattice calculation compared to the masses from Teper~\cite{Teper:1997tq}.}
\label{fig:latres}
\end{figure}

These final results are shown in Figure~\ref{fig:latres}.

\begin{table}[!t]
\centering
\begin{tabular}{|c|ccc|} \hline
t =  & 0 & 1 & 2  \\\hline

size  &   & \multicolumn{2}{c}{$\alpha=.1$}  \\ \hline
3 & 1 & 8.4252e-02 & 5.2959e-03 \\
4 & 1 & 1.1608e-01 & 1.7138e-02 \\
5 & 1 & 6.4503e-02 & 8.6513e-03 \\
6 & 1 & 9.8620e-02 & 1.7948e-02 \\\hline

  &   & \multicolumn{2}{c}{$\alpha=.15$}  \\ \hline
3 & 1 & 6.5045e-02 &-2.7807e-03 \\
4 & 1 & 1.1552e-01 & 1.7367e-02 \\
5 & 1 & 8.1668e-02 & 1.7576e-02 \\
6 & 1 & 1.0721e-01 & 2.0704e-02 \\\hline

  &   & \multicolumn{2}{c}{$\alpha=.2$}   \\ \hline
3 & 1 & 5.0137e-02 &-1.0371e-02 \\
4 & 1 & 1.3613e-01 & 2.6607e-02 \\
5 & 1 & 1.3583e-01 & 3.8166e-02 \\
6 & 1 & 1.1937e-01 & 2.4707e-02 \\\hline

  &   & \multicolumn{2}{c}{$\alpha=.25$}   \\ \hline
3 & 1 & 6.9668e-02 &-1.7408e-02 \\
4 & 1 & 1.2066e-01 & 7.5657e-03 \\
5 & 1 & 1.2103e-01 & 1.7527e-02 \\
6 & 1 & 1.2769e-01 & 4.7764e-05 \\\hline

\end{tabular}
\caption{Correlation functions for $4^{++}$ after torelon removal.}
\label{table:corr4p4}
\end{table}

\begin{table}[!h]
\centering
\begin{tabular}{|c|ccc|c|} \hline
t = & & 1 & 2 & $|<\CO_4 | \CO_0>|$ \\\hline

size  & &  \multicolumn{2}{c}{$\alpha=.1$} & \\\hline
3 & &  2.47(08) & -        & .39 \\
4 & &  2.15(05) & 1.91(28) & .35 \\
5 & &  2.74(09) & 2.01(99) & .27 \\
6 & &  2.32(04) & 1.70(53) & .13 \\\hline

  & &  \multicolumn{2}{c}{$\alpha=.15$} & \\\hline
3 & &  2.73(10) & -        & .27 \\
4 & &  2.16(06) & 1.89(27) & .24 \\
5 & &  2.51(07) & 1.54(46) & .17 \\
6 & &  2.23(04) & 1.64(47) & .05 \\\hline

  & &  \multicolumn{2}{c}{$\alpha=.2$}  & \\\hline
3 & &  2.99(12) & -        & .11 \\
4 & &  1.99(05) & 1.63(21) & .11 \\
5 & &  2.00(05) & 1.27(18) & .14 \\
6 & &  {\bf{2.13(04)}} & 1.58(39) & .007 \\\hline

  & &  \multicolumn{2}{c}{$\alpha=.25$} &  \\\hline
3 & &  2.66(08) & -        & .03 \\
4 & &  2.11(07) & 2.77(93) & .38 \\
5 & &  2.11(05) & 1.93(28) & .05 \\
6 & &  2.06(05) & -        & .49 \\\hline

\end{tabular}
\caption{Masses for $4^{++}$ after torelon removal.  The last column is the overlap with the $0^{++}$.}
\label{table:mass4p4}
\end{table}

\chapter{Conclusions} \label{chap:conclusion}

As we have seen, pure gauge theory does possess a rich structure, even without the complication of quarks.  Understanding the properties of the pure gauge field is vital if we are going to reach a deeper understanding of our universe.  These gauge fields can produce quite interesting bound states:  glueballs, gluelumps, and even glueknots.  Even when restricting our attention to D=2+1, these glueballs in Flatland can intrigue and surprise us.  Reaching towards a better comprehension of glueballs gives us insight into the structure of both mesons and baryons.

The original model of Isgur and Paton was quite successful in explaining meson structure.  Its application to glueballs predated any experimental evidence for such states.  Nonetheless, its rough predictions agreed well with the lattice data available at the time.  While the extent of lattice calculations has improved greatly, not much work had been done to refine the model.  Indeed, the most interesting refinements, such as the inclusion of a dissipating string tension and the possibility of the k-string, have required the input of research conducted within the last two years.  Now, with a wealth of lattice data spanning several values of N from 2 to $\infty$, in both 2 and 3 dimensions, we can test the predictions of these extensions against the lattice data.

Several refinements to the original model have been proposed.  The inclusion of an effective elasticity accords with the evidence coming from other forms of flux tubes and promotes the L\"{u}scher correction term from a constant to a free parameter for closed flux loops.  Various mechanisms to produce states with negative charge conjugate eigenvalue have been developed, from a simple direct mechanism of flux orientation reversal to more complicated mechanisms requiring the existence of higher topological contributions or of more exotic types of gauge field flux tube.  The properties of this flux tube at very small distances, where the idea of an infinitesimally thin flux tube must break down, has been incorporated.  While still non-relativistic in character, these models are quite successful at reproducing the structure of the lattice spectrum.  The best of these models, developed in Chapter~\ref{chap:kstring}, incorporates all these features:  the k-string model with dissipating flux tube and effective elasticity.  It can match the lattice spectrum to well within $\pm$5\% across a range of spins and excitations.

One feature of all these models is that they disagree spectacularly with one particular state, traditionally assigned quantum numbers $0^{-+}$.  No amount of parameter manipulation can bring that state into agreement without ruining the agreement across the rest of the spectrum, simply because any flux tube model must have several costly, ie heavy, phonons to produce a spin cancellation while providing negative parity.  However, they all make a prediction that there exists a state with quantum numbers $4^{-+}$ at just around that state's mass.  Hence we are led to the hypothesis:  ``The lattice state traditionally labelled by $0^{-+}$ really possesses a spin of 4.''  Such an hypothesis cannot be automatically discounted owing to the mod4 spin ambiguity of operators on a cubic lattice.

To test this hypothesis, a new approach to constructing operators on a cubic lattice is developed, starting in Chapter~\ref{chap:lattice}.  This approach borrows from techniques of Greens function inversion already used in fermionic calculations, but applies the inversion to the full lattice, not just to all paths emanating from one particular site.  The numerical cost of this technique is more expensive than traditional methods, but the potential to overcome cubic group spin ambiguities makes its pursuit worthwhile.  As this technique receives further exploration, methods to tame the numerics will be developed, as in Section~\ref{sect:hermLU}.

The result:  there {\it{does}} exist a state with quantum numbers $4^{-+}$ precisely at the mass which the lattice community calls the $0^{-+}$.  This is a beautiful result, confirming by a lattice calculation an unconventional hypothesis coming from a dynamical model of flux tube glueballs.  The agreement is remarkable, comparing a mass of 2.13(04) with a mass of 2.10(33) respectively.  At the end of the day, such close agreement strengthens both aspects of the hypothesis, giving much credibility to the flux tube model of glueballs as well as inspiring further research into the method of Greens function inversion.

\renewcommand{\chaptermark}[1]{\markboth{Appendix
\ \thechapter.\ #1}{}}
\renewcommand{\sectionmark}[1]{\markright{\thesection.\ #1}}
\appendix

\chapter{Bayesian Methods of Inferential Statistics} \label{chap:bayes}

When comparing a model to a set of data points, the most appropriate method to use is that of inferential statistics~\cite{Bretthorst:1988,Sivia:1996}. This method provides means to estimate the values of the parameters that are most likely, given data points to match, as well as to compare the relative likelihood of several competing models.
What we are given is a set of data \{D\} of lattice measurements from which
we are to infer the most probable values $\{\alpha_i\}$ of the
parameters of the underlying theory.  
What we actually compute is the probability distribution function as a function in multi-dimensional parameter space.
The joint distribution of $X$ and
$Y$ is given by Bayes Theorem as
\beq
pr(X|Y,I) = {{pr(Y|X,I) \times pr(X|I)}\over{pr(Y|I)}} ,
\eeq
where $X$ and $Y$ are propositions conditioned on the background
knowledge $I$.  When these variables represent our hypothesis and
data, we have
\beq
pr(\mathit{hypothesis}|\mathit{data},I) \propto
pr(\mathit{data}|\mathit{hypothesis},I) \times pr(\mathit{hypothesis}|I) .
\eeq
The lhs is called the ``posterior'' and is the function we wish to
maximize.  The first factor on the rhs is called the ``likelihood'' of
the data given the hypothesis (and $I$).  The second factor is the
``prior'', and is often a source of confusion for beginning
Bayesians.  Physicists often assign a uniform prior (sometimes when
not appropriate!), reducing the problem to the familiar one of maximum
likelihood parameter estimation.  However, the Principle of
Insufficient Reason, as stated by Bernoulli, can prescribe more useful
priors in most circumstances.  The omitted factor, the ``evidence'',
is only a normalizing constant when peforming parameter estimation,
evaluated by normalizing the posterior, but reappears when we become
interested in model selection.

For uniform priors, the method reduces to the familiar one of maximum likelihood estimation~\cite{Press:1992}.  The likelihood, under certain assumptions, can be taken to be $\chi^2$, a nonnegative quantity.  As the probability is defined as the exponential of $\chi^2$, we often work with the log of the posterior, which would simply be our likelihood function.

For future reference, we state the means by which we remove unwanted
propositions from a joint distribution.  Continuous marginalization is
given by
\beq
pr(X|I) = \int_{-\infty}^{+\infty} pr(X,Y|I)dY
\eeq
which goes over in the trivially ultimate discrete case to
\beq
pr(X|I) = pr(X,Y|I) + pr(X,\obar Y|I) .
\eeq
When the posterior is sharply peaked around a unique maximum, we may
make a local quadratic approximation to determine the uncertainty in a
fitted parameter by fitting a local Gaussian to the peak.  This case
corresponds to the usual notion of confidence intervals.  When the
posterior is not sharply peaked, or even multinodal, such a gross
approximation is not adequate.  Even so, in this case we may still define
Bayesian intervals to represent our state of knowledge of the
parameters given the data at hand.

\chapter{Review of the Isgur-Paton Model in Two Spatial Dimensions} \label{chap:IPreview}

While perturbative QCD works demonstrably well for high-momentum
scattering experiments, the phenomenology of hadrons lies beyond
perturbative methods.  Quark models, on the other hand, give a
reasonably good accounting of the hadronic spectrum.  Full QCD,
though, must have a spectrum more complicated than the quark model, as
the self-coupling of the gluonic sector predicts bound states devoid
of any quarks.  These states, called ``glueballs'' or ``gluelumps'',
can be inferred from the spectrum of the pure Yang-Mills theory
corresponding to quarkless (``quenched'') QCD.  Two avenues for
approach present themselves: weak-coupling models or strong-coupling
models.  Weak-coupling models are epitomized by the MIT-bag model~\cite{Juge:1998nd},
where ``constituent gluons'' are postulated, in analogy with the
familiar ``constituent quarks'', to live within a confining potential
bag.  While these models can reproduce some features of the Yang-Mills
spectrum, many issues are unresolved.  Additionally, these models
actually violate the ``no-hair'' theorem~\cite{Robson:1980}, at least at the
semi-classical level.  Turning to strong-coupling models, Isgur and
Paton developed a model~\cite{Isgur:1985bm} where the gluonic degrees of freedom condense
into tubes of chromoelectric flux between static quarks.  The
quarkless extension of this point of view envisages pure glue states
as a closed loop of flux--imagine annihilating the $q$ and $\obar q$
of a meson leaving the flux tube behind as ring of flux.  Extensions
to this naive flux tube model are possible, with higher representation
flux and more complicated topologies allowed, as long as flux
junctions are properly treated and final states remain gauge
invariant.

Without the presence of the static quarks to provide a natural
adiabatic limit, Isgur and Paton supposed that the low-lying spectrum
will be described by the simplest state:  a closed loop of fundamental
flux.  A detailed application of the Isgur-Paton model to the case of
D=2+1 and SU(2) was performed in~\cite{Moretto:1993tm}.  The flux tube is described by
polar coordinates $(r,\theta)$.  For small fluctuations about a circle
of radius $r_0$, we expand the radial variable in Fourier modes which
we identify as phonons.  These phonons contribute to the potential
energy of the Schroedinger equation for the radial flux tube when we
apply an \textit{ad hoc} Born-Oppenheimer approximation to absorb the
``fast'' phonon modes.  The resulting Hamiltonian is diagonalized to
give the spectrum.  The traditional Isgur-Paton model contains a
cutoff parameter $f$, as well as the string tension $\sigma$ which is
set equal to unity.

For SU(N$\geq$3) the flux tube has a direction, so a 
circular string will have one of two orientations.
These we shall label by $L$ and $R$. In 2 space dimensions 
one cannot change the orientation of a loop by a rotation
so there is no link between the rotational and charge
conjugation properties of a state (in contrast to the
case in 3 space dimensions). Moreover the model provides
no means by which an $L$ loop will transform into an
$R$ loop. Thus the states which are
labelled by the orientation, ${\cal O}=L$ or $R$, of the
loop, by the phonon occupation numbers, $\{n^\pm_m\}$,
and by the radial quantum number, $n_R$, are energy
eigenstates in the model and form a complete set.
We may write them as $|{\cal O}; \{n_m^\pm\} ; n_R \rangle$. 
Since the theory is invariant under charge conjugation 
and parity we can also label the energy eigenstates by these 
more physical quantum numbers. What are these states?

The action of charge conjugation, $\cal C$, is to reverse the 
orientation of the glue loop without changing its modes 
in any way:
\beq
|L; \{n_m^\pm\} ; n_R\ \rangle
\stackrel{\cal C}{\rightarrow} |R; \{n_m^\pm\} ; n_R \rangle.
\label{2.8}
\eeq
Moreover for all quantum numbers a state and its charge conjugate
are different. Thus when we apply $1 \pm {\cal C}$ to one of our energy
eigenstates we get a pair of non-null degenerate states of
opposite quantum numbers. Hence the masses in the $C=+$ and $C=-$
sectors are identical. It is clear that this degeneracy simply 
follows from the fact that there is no mixing in the model 
between loops of opposite orientation.

Turning now to parity, $P$, we recall that
in 2 space dimensions it transforms $(x,y) \to (x,-y)$ 
(up to a rotation) and so changes the sign of angular
momentum. Thus it will reverse the helicity of each phonon.
It is also clear that it  reverses the orientation of a closed
flux loop. Thus
\beq
|L; \{n_m^\pm\} ; n_R \rangle
\stackrel{\cal P}{\rightarrow} |R; \{n_m^\mp\} ; n_R \rangle.
\label{2.9}
\eeq
These two states are different but degenerate. Thus when 
we apply $1 \pm {\cal P}$ to form the two parity eigenstates
we get a pair of non-null degenerate states. That is
to say, the mass spectra in the $P=+$ and $P=-$ 
sectors are identical. For $J \not= 0$ this degeneracy 
is in fact quite general, in 2 space dimensions. However
the general argument breaks down for $J=0$ and here it
is the special features of the model that lead to such a
degeneracy. 

\begin{figure}[t]
\centering
\subfigure[$\psi = |R; n_2^+ \rangle$]{
	\label{fig:pctrans:a}
	\includegraphics[width=4cm]{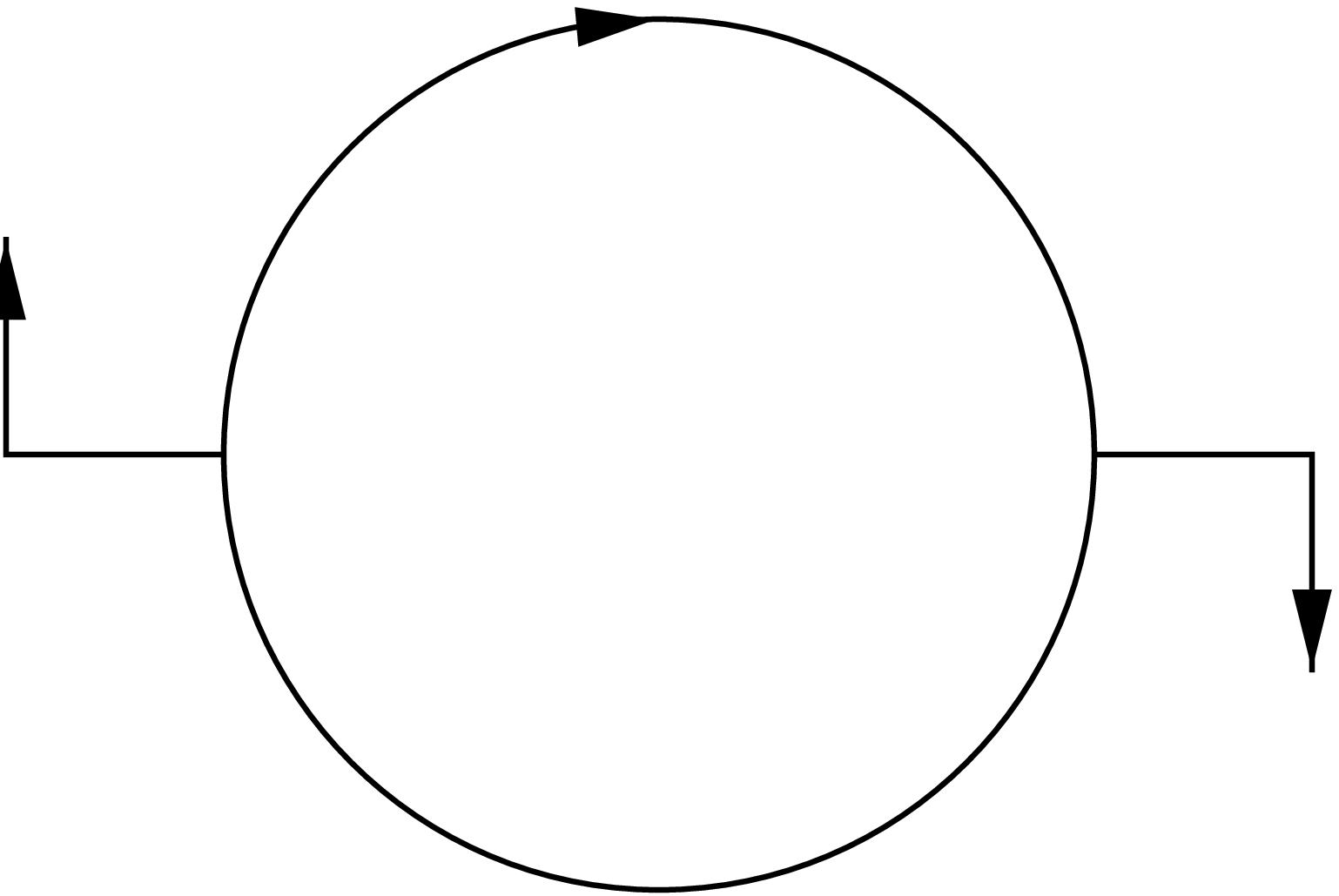}}
\subfigure[$P \psi = |L; n_2^- \rangle$]{
	\label{fig:pctrans:b}
	\includegraphics[width=4cm]{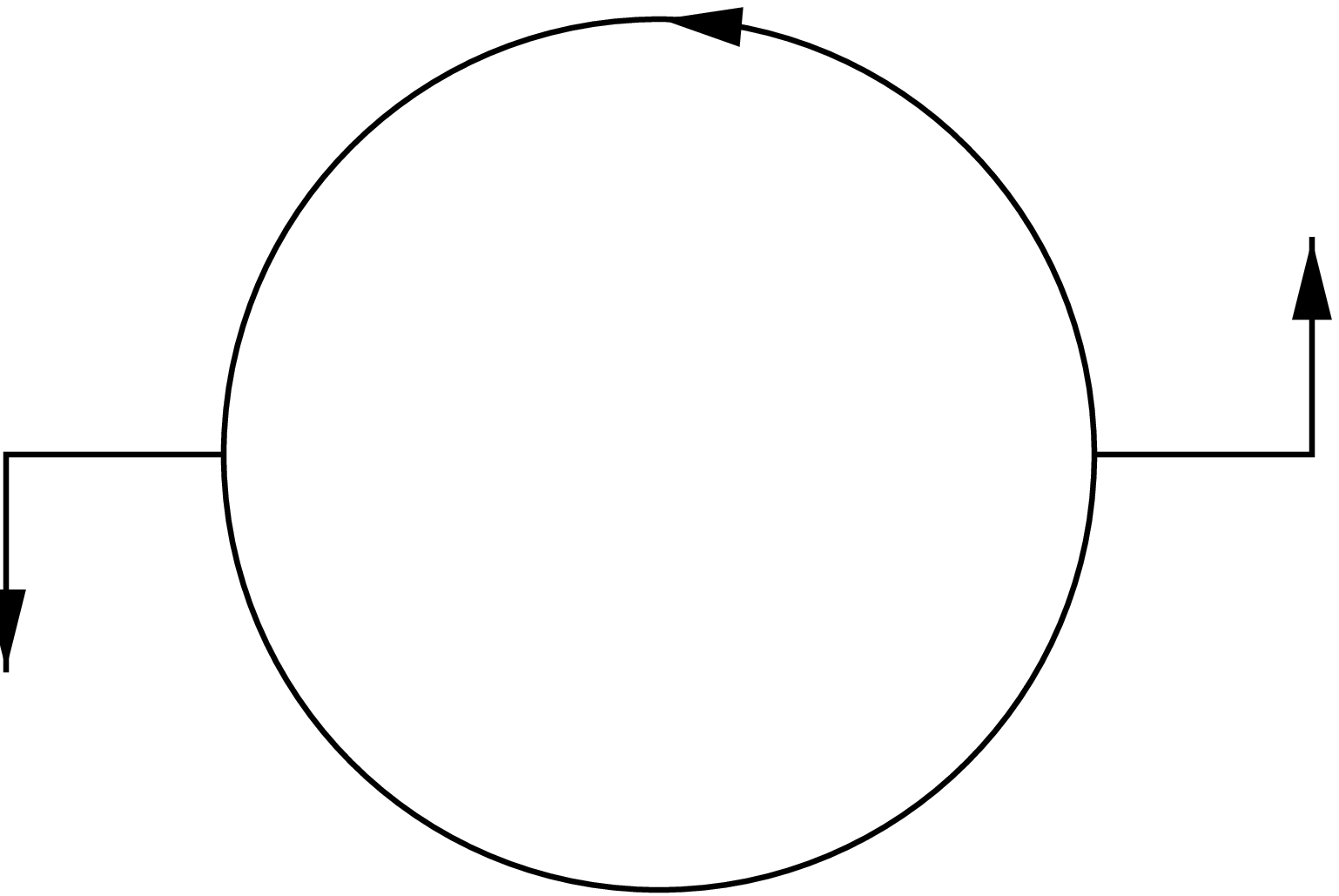}}
\subfigure[$C \psi = |L; n_2^+ \rangle$]{
	\label{fig:pctrans:c}
	\includegraphics[width=4cm]{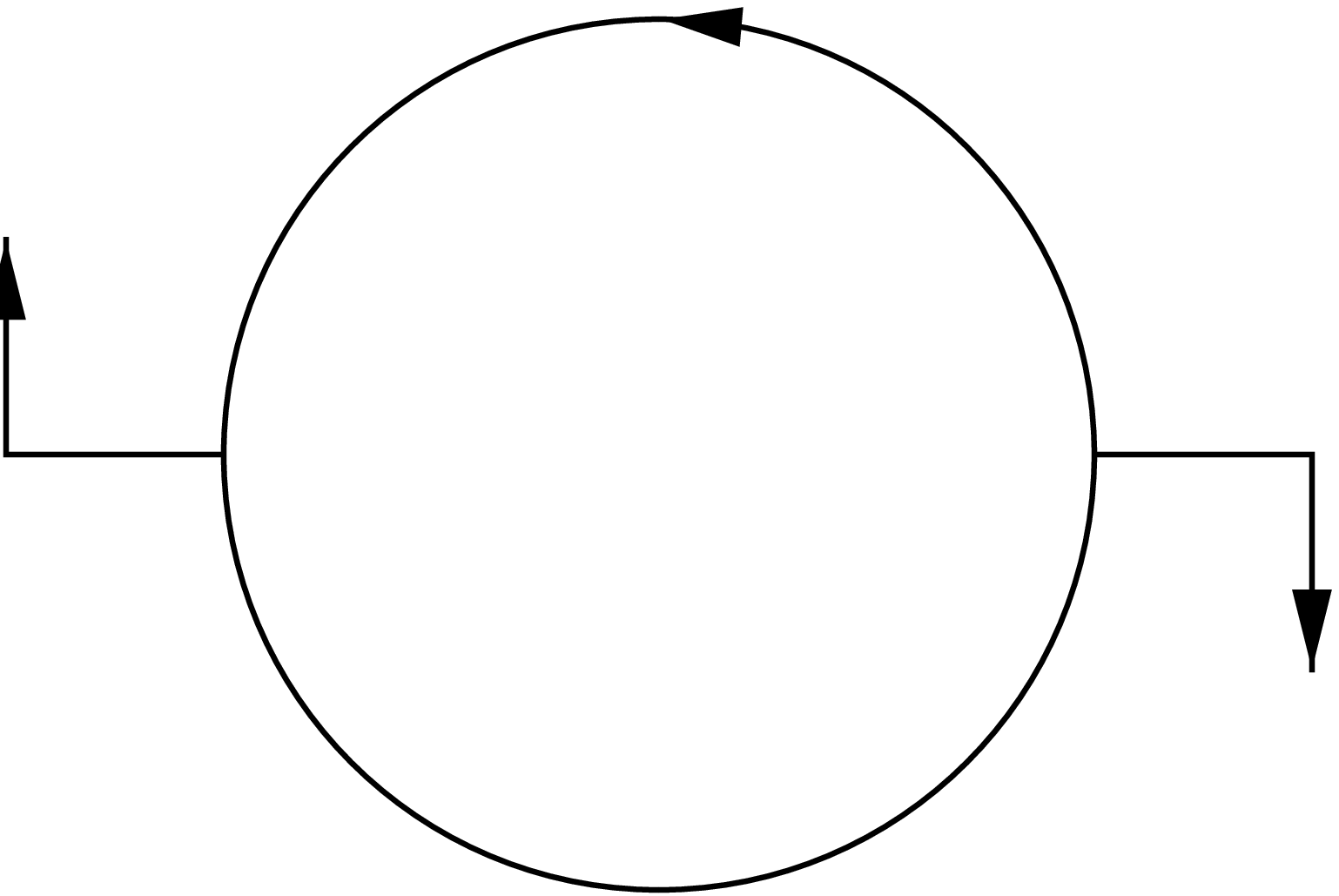}}
\caption{Transformation under parity and charge conjugaion.}
\label{fig:pctrans}
\end{figure}

A pictorial representation of the action of these transformations 
is given in Figure~\ref{fig:pctrans}.

From the above we see that an energy eigenstate definite 
$P$ and $C$ can be written (up to a normalisation) as
\beq 
\Psi_{P,C} = ( |L; n_m^\pm \rangle + \eta_P |R; n_m^\mp
\rangle ) + \eta_C ( |R; n_m^\pm \rangle + \eta_P |L; n_m^\mp \rangle ) 
\label{2.10}
\eeq 
where $\eta_P = \pm1$ is the eigenvalue for parity
and $\eta_C = \pm1$ is the eigenvalue for charge conjugation.
(We have abbreviated our notation for the states in an obvious 
way - in particular by suppressing the radial quantum 
number, $n_R$, which is unchanged by both $P$ and $C$.)
For example, the lowest lying $J=2$ state has a single $m=2$ 
phonon,  and so the $P=+, C=-, J=2$ ground state is
\beq 
\psi_{J=2} = ( |L; n_2^+=1 \rangle + |R; n_2^-=1 \rangle )
- (|R; n_2^+=1 \rangle + |L; n_2^-=1 \rangle )
\label{2.11}
\eeq

From Equation~\ref{2.10} we can readily infer some
qualitative features of the SU($N\geq$3) mass spectra:

{\noindent}1) The very lightest states are those with no phonons 
at all (and with $n_R=0$). There are two such states, the $0^{++}$ 
and the $0^{--}$, and they will be degenerate.

{\noindent}2) Since each of the component states in Equation~\ref{2.10}
is itself an energy eigenstate and they are all degenerate, this
implies that for a given $J$ the four sectors labelled by the
four combinations of $P,C$ are degenerate except where one of
the linear combinations is null. The latter can only occur if the
phonon content is identical for both helicities, 
$\{n_m^+\} = \{n_m^-\}$, and this can only occur for $J=0$.
So, for example, if we take the state with 
no phonons at all and try to form a $0^{-+}$ state degenerate
with the the  $0^{++}$ ground state we find, from Equation~\ref{2.10},
\beq \psi_{0^{-+}} = ( |L; 0 \rangle - |R; 0 \rangle )
+ (|R; 0 \rangle - |L; 0 \rangle ) \equiv 0 \eeq
It is easy to see that the lightest $0^{-+}$ arises for $n_4^+ =1, n_2^2=2$
and so is a (very) heavy state.

All the above is for SU($N\geq$3). The SU(2) spectrum 
has no $C=-$ sector which, in the string model, is
embodied in the fact that there is no orientation on the flux
loop. So if we identify $L$ and $R$ in the above equations
we obtain expressions for the energy eigenstates. Once
again we have parity doubling for $J \neq 0$ and 
for that part of the $J=0$ spectrum that contains 
differing positive and negative phonon contents.
The following table summarizes the assignment of spin and parity quantum number$J^P$ available in D=2+1 for the Isgure-Paton model.

\begin{table}[!h]
\centering
\begin{tabular}{|c|l|} \hline
M & $J^P$ \\\hline
0 & $0^+$ \\
2 & $2^\pm$ \\
3 & $3^\pm$ \\
4 & $0^+$,$4^\pm$ \\
5 & $1^\pm$,$5^\pm$ \\
6 & $0^+$,$2^\pm$,$6^\pm$ \\
7 & $1^+$,$3^\pm$,$7^\pm$ \\
8 & $0^\pm$,$2^\pm$,$4^\pm$,$8^\pm$ \\
9 & $1^\pm$,$3^\pm$,$5^\pm$,$9^\pm$ \\
  & \it{etc.} \\\hline
\end{tabular}
\caption{Allowable quantum numbers for each value of M.}
\label{table:MtoJ}
\end{table}

\chapter{Programs Used for the Calculations} \label{chap:programs}

All calculations were performed using the \textsc{Matlab} programming environment.  Some advantages of \textsc{Matlab} include the ease with which complex algorithms may be altered and its more mathematical programming environment.  The inherent matrix data type lends itself well to prototyping logical algorithms without having to change large swaths of dedicated C or Fortran.  Also, the mathematics behind the calculation are more easily seen using \textsc{Matlab} programs.  The crucial disadvantage is speed.  While \textsc{Matlab} does run remarkably fast, its inherent overhead slows down production calculation.  It simply cannot compete with highly optimized and parallelized code.  For the purposes of these calculations, it was the appropriate tool to use, but for any future work (on the lattice calculations), optimized C is definitely the way forward.  Note, the SuperLU code~\cite{Li:1997} was modified slightly to create a .mex file which handles complex values. 

\clearpage

\begin{figure}
\centering
\includegraphics[scale=.8]{IPprog1.epsi}
\caption{Programs used for the extended Isgur-Paton model.}
\end{figure}
\clearpage

\begin{figure}
\centering
\includegraphics[scale=.8]{IPprog2.epsi}
\caption{Programs used for the extended Isgur-Paton model.}
\end{figure}
\clearpage

\begin{figure}
\centering
\includegraphics[scale=.8]{lattprog1.epsi}
\caption{Programs used for the lattice calculation.}
\end{figure}
\clearpage

\begin{figure}
\centering
\includegraphics[scale=.8]{lattprog2.epsi}
\caption{Programs used for the lattice calculation.}
\end{figure}
\clearpage

\begin{figure}
\centering
\includegraphics[scale=.8]{lattprog3.epsi}
\caption{Programs used for the lattice calculation.}
\end{figure}
\clearpage

\begin{figure}
\centering
\includegraphics[scale=.8]{lattprog4.epsi}
\caption{Programs used for the lattice calculation.}
\end{figure}
\clearpage

\begin{figure}
\centering
\includegraphics[scale=.8]{lattprog5.epsi}
\caption{Programs used for the lattice calculation.}
\end{figure}
\clearpage

\begin{figure}
\centering
\includegraphics[scale=.8]{lattprog6.epsi}
\caption{Programs used for the lattice calculation.}
\end{figure}
\clearpage


\end{document}